\newcommand{\sect}{Section~}
\newcommand{\sects}{Sections~}

\newcommand{\kms}{{km$~\!$s$^{-1}$}}
\newcommand{\ang}{{\AA}}
\newcommand{\ston}{S/N}

\newcommand{\msol}{\mathcal{M}_\odot}

\newcommand{\lksol}{\mathcal{L}_\odot^K}
\newcommand{\sdu}{$\msol~\!$pc$^{-2}$}

\newcommand{\muu}{mag arcsec$^{-2}$}

\newcommand{\oiii}{[O{\scshape$~\!$iii}]}
\newcommand{\oiiinb}{O{\scshape$~\!$iii}}
\newcommand{\hone}{H{\scshape$~\!$i}}
\newcommand{\halp}{H$\alpha$}

\newcommand{\none}{[N{\scshape$~\!$i}]}

\newcommand{\nii}{[N{\scshape$~\!$ii}]}
\newcommand{\niinb}{N{\scshape$~\!$ii}}
\newcommand{\sii}{[S{\scshape$~\!$ii]}}
\newcommand{\siinb}{S{\scshape$~\!$ii}}
\newcommand{\mgi}{Mg{\scshape$~\!$i}}

\newcommand{\fei}{Fe{\scshape$~\!$i}}
\newcommand{\molh}{H$_2$}

\newcommand{\gaz}{\theta}
\newcommand{\pa}{\phi_0}

\newcommand{\ikin}{i_{\rm kin}}
\newcommand{\itf}{i_{\rm TF}}

\newcommand{\vhi}{V_{\mbox{\rm \footnotesize H{\scshape i}}}}
\newcommand{\vhel}{V_{\rm hel}}
\newcommand{\vflow}{V_{\rm flow}}
\newcommand{\vsys}{V_{\rm sys}}

\newcommand{\vlos}{V_{\rm LOS}}

\newcommand{\vt}{V_\theta}

\newcommand{\vrotproj}{V_{\rm rot}^{\rm proj}}
\newcommand{\vrot}{V_{\rm rot}}
\newcommand{\vmax}{V_{\rm max}}
\newcommand{\vflat}{V_{\rm flat}}
\newcommand{\vc}{V_{\rm c}}

\newcommand{\vstar}{V_{\ast}}
\newcommand{\vdiskgas}{V^{\rm disk}_{\rm gas}}
\newcommand{\vdiskstar}{V^{\rm disk}_\ast}

\newcommand{\vbulgestar}{V^{\rm bulge}_\ast}
\newcommand{\vhalo}{V_{\rm halo}}
\newcommand{\vdm}{V_{\rm DM}}

\newcommand{\vgas}{V_{\rm gas}}

\newcommand{\vbary}{V_{\rm b}}

\newcommand{\sgas}{\sigma_{\rm gas}}
\newcommand{\sstar}{\sigma_{\ast}}
\newcommand{\sobs}{\sigma_{\rm obs}}

\newcommand{\sinst}{\sigma_{\rm inst}}

\newcommand{\sbs}{\sigma_{\rm beam}}

\newcommand{\sigr}{\sigma_R}
\newcommand{\sigp}{\sigma_{\gaz}}
\newcommand{\sigz}{\sigma_z}

\newcommand{\dad}{\delta_{\rm AD}}

\newcommand{\sddisk}{\Sigma_{\rm dyn}}
\newcommand{\sdbary}{\Sigma_{\rm b}}
\newcommand{\sds}{\Sigma_{\rm \ast}}
\newcommand{\sdhi}{\Sigma_{\mbox{\rm \footnotesize H{\scshape i}}}}

\newcommand{\sdmh}{\Sigma_{\rm H_2}}
\newcommand{\sdg}{\Sigma_{\rm gas}}

\newcommand{\rhodm}{\rho_{\rm DM}}
\newcommand{\rhobulge}{\rho_{\rm bulge}}
\newcommand{\rhobary}{\rho_{\rm b}}

\newcommand{\mtot}{\mathcal{M}^{\rm tot}_{\rm dyn}}
\newcommand{\mtotbary}{\mathcal{M}^{\rm tot}_{\rm b}}
\newcommand{\mtotbar}{\mathcal{M}^{\rm tot}_{\rm bar}}

\newcommand{\mdisk}{\mathcal{M}_{\rm disk}}

\newcommand{\mhalodm}{\mathcal{M}^{\rm halo}_{\rm DM}}

\newcommand{\mhi}{\mathcal{M}_{\mbox{\hone}}}
\newcommand{\mht}{\mathcal{M}_{\mbox{\molh}}}

\newcommand{\mb}{\mathcal{M}_{\rm b}}
\newcommand{\mdm}{\mathcal{M}_{\rm DM}}

\newcommand{\mldyn}{\Upsilon_{\rm dyn}}

\newcommand{\mldynkdisk}{\Upsilon_{{\rm dyn},K}^{\rm disk}}

\newcommand{\mls}{\Upsilon_\ast}

\newcommand{\mlsk}{\Upsilon_{\ast,K}}

\newcommand{\mlsksps}{\Upsilon_{\ast,K}^{\rm SPS}}

\newcommand{\mlskdisk}{\Upsilon_{\ast,K}^{\rm disk}}

\newcommand{\Fd}{\mathcal{F}_{\rm disk}}
\newcommand{\Fdisk}{\mathcal{F}_{\ast}^{\rm disk}}

\newcommand{\Fbary}{\mathcal{F}_{\rm b}}

\newcommand{\dct}{{\it DC3}}
\newcommand{\imi}{I_{24\mu{\rm m}}}
\newcommand{\ico}{I_{\rm CO}\Delta V}

\newcommand{\xco}{X_{\rm CO}}
\newcommand{\iha}{i_{\mbox{\scriptsize \halp}}}
\newcommand{\iot}{i_{\mbox{\scriptsize \oiii}}}
\newcommand{\ist}{i_{\ast}}
\newcommand{\mumg}{\mu_{\mbox{\scriptsize \mgi}}}
\newcommand{\muha}{\mu_{\mbox{\halp}}}
\newcommand{\mujhk}{\mu_{J\!H\!K}}
\newcommand{\jhk}{$J\!H\!K$}
\newcommand{\qtg}{Q_{\rm T,gas}}
\newcommand{\qts}{Q_{{\rm T,}\ast}}
\newcommand{\qrg}{Q_{\rm R,gas}}
\newcommand{\qrs}{Q_{{\rm R,}\ast}}
\newcommand{\qr}{Q_{\rm R}}
\newcommand{\kl}{k_{\lambda}}

\newcommand\azmean[1]{\overline{#1}}
\newcommand\allmean[1]{\langle #1 \rangle}

\newcommand{\dustt}{Schechtman-Rook et al., {\it in prep}}
\newcommand{\haIt}{Swaters et al., {\it in prep}}
\newcommand{\haIIt}{Andersen et al., {\it in prep}}
\newcommand{\pV}{Paper~V}

\documentclass[apj,iop,numberedappendix]{emulateapj}
\RequirePackage{calc}
\usepackage{natbib}
\usepackage{url}
\shortauthors{Westfall et al.}
\shorttitle{The DiskMass Survey. IV.}

\defcitealias{2010ApJ...716..198B}{Paper~I}
\defcitealias{2010ApJ...716..234B}{Paper~II}
\defcitealias{2011ApJS..193...21W}{Paper~III}
\defcitealias{2008AJ....136.2782L}{L08}
\defcitealias{2001ApJ...563..694V}{V01}
\defcitealias{1986RSPTA.320..447V}{vAS86}
\defcitealias{2003ApJS..149..289B}{B03}
\defcitealias{2009MNRAS.400.1181Z}{Z09}
\begin{document}

\title{The DiskMass Survey. IV. The Dark-Matter-Dominated Galaxy UGC 463 }

\author{Kyle B. Westfall\altaffilmark{1,2}, Matthew A.
Bershady\altaffilmark{3}, Marc A. W. Verheijen\altaffilmark{1}, David R.
Andersen\altaffilmark{4}, Thomas P. K. Martinsson\altaffilmark{1}, Robert A.
Swaters\altaffilmark{5}, \& Andrew Schechtman-Rook\altaffilmark{3}}

\altaffiltext{1}{Kapteyn Astronomical Institute, University of Groningen,
Landleven 12, 9747 AD Groningen, the Netherlands}

\altaffiltext{2}{National Science Foundation (USA) International Research
Fellow}

\altaffiltext{3}{Department of Astronomy, University of Wisconsin-Madison, 475
N. Charter St., Madison, WI 53706, USA}

\altaffiltext{4}{NRC Herzberg Institute of Astrophysics, 5071 West Saanich Road,
Victoria, BC V9E 2E7, Canada}

\altaffiltext{5}{National Optical Astronomy Observatory, 950 North Cherry Ave,
Tucson, AZ 85719, USA }

\email{westfall@astro.rug.nl}

\begin{abstract}

We present a detailed and unique mass budget for the high-surface-brightness
galaxy UGC 463, showing it is dominated by dark matter (DM) at radii beyond one
scale length ($h_R$) and has a baryonic-to-DM mass ratio of approximately 1:3
within 4.2$h_R$.  Assuming a constant scale height ($h_z$, calculated via an
empirical oblateness relation), we calculate dynamical disk mass surface
densities from stellar kinematics, which provide vertical velocity dispersions
after correcting for the shape of the stellar velocity ellipsoid (measured to
have $\sigp/\sigr=1.04\pm0.22$ and $\sigz/\sigr=0.48\pm0.09$).  We isolate the
stellar mass surface density by accounting for all gas mass components and find
an average $K$-band mass-to-light ratio of
$0.22\pm0.09(ran)~^{+0.16}_{-0.15}(sys)~\msol/\lksol$;
\citeauthor{2009MNRAS.400.1181Z}\ and \citeauthor{2003ApJS..149..289B}\ predict,
respectively, 0.56 and 3.6 times our dynamical value based on
stellar-population-synthesis modeling.  The baryonic matter is submaximal by a
factor of $\sim3$ in mass and the baryonic-to-total circular-speed ratio is
$0.61^{+0.07}_{-0.09}(ran)~^{+0.12}_{-0.18}(sys)$ at 2.2$h_R$; however, the disk
is globally stable with a multi-component stability that decreases
asymptotically with radius to $Q\sim2$.  We directly calculate the circular
speed of the DM halo by subtracting the baryonic contribution to the total
circular speed; the result is equally well described by either a
Navarro-Frenk-White halo or a pseudo-isothermal sphere.  The volume density is
dominated by DM at heights of $|z|\gtrsim1.6h_z$ for radii of $R\gtrsim h_R$.
As is shown in follow-up papers, UGC 463 is just one example among nearly all
galaxies we have observed that contradict the hypothesis that
high-surface-brightness spiral galaxies have maximal disks.

\end{abstract}


\section{Introduction}
\label{sec:intro}

A primary goal of modern extragalactic astronomy is to reduce the complex,
stochastic process of galaxy formation into a few fundamental physical
parameters.  Such a goal appears tractable given the tight scaling relations
exhibited by galaxies over a large dynamic range in observed properties, which
to first order may be tied to a single physical characteristic
\citep{2008Natur.455.1082D}.  For example, measures of galaxy size, luminosity,
and a virialized dynamical quantity (such as the circular velocity in
rotationally supported systems and velocity dispersion in pressure-dominated
systems) demonstrate strong covariance.  Correlations among galaxy properties
are found in two-dimensional scatter plots \citep[e.g.,][]{2007ApJ...671..203C,
2010ApJ...715..606N, 2011ApJ...726...77S}, lines through multi-dimensional space
\citep[e.g.,][]{2011ApJ...726..108T}, and more complex, multi-dimensional
manifolds \citep[e.g.,][]{2008ApJ...682...68Z}.  Empirical and theoretical
understanding of these relations over cosmic time \citep[as in,
e.g.,][]{2011MNRAS.410.1660D} are critical.

Two long-standing scaling relations are the Tully--Fisher relation
\citep[][hereafter the TF relation]{1977A&A....54..661T} --- the correlation
between the rotation velocity of spiral galaxies and their total luminosity ---
and the Fundamental Plane \citep[FP;][]{1987ApJ...313...42D,
1987ApJ...313...59D} --- the plane relating the size, surface brightness, and
velocity dispersion of elliptical galaxies.  These fundamental relations are
strongly linked to mass:  The baryonic TF (BTF) relation
\citep{2000ApJ...533L..99M, 2005ApJ...632..859M}, created by replacing total
luminosity with total baryonic mass, exhibits less scatter than the nominal TF
relation over a wide range of luminosity and spiral type.  The mass-based FP
\citep{2007ApJ...665L.105B}, incorporating the total
(baryonic$+$dark-matter[DM]) mass surface density instead of surface brightness,
also exhibits lower scatter than its luminosity-based counterpart.  It is
interesting that the residuals are reduced for both the BTF and mass-based FP
relation despite the exclusion of DM from the former.  The tightness of the BTF
implies that either DM is a rather negligible mass component or there exists a
strict proportionality, in both relative amplitude and distribution, between DM
and baryonic mass in spiral galaxies.  The former is incompatible with our
current understanding of gravity and the current paradigm of hierarchical
disk-galaxy formation \citep[see, e.g.,][]{1980MNRAS.193..189F,
1997ApJ...482..659D, 1998MNRAS.295..319M, 2011MNRAS.410.1391A}, and the latter
is tantamount to the discomforting disk-halo conspiracy\footnote{
The observed fine tuning of the relative fraction and distribution of baryonic
and DM mass required to produce a total rotation curve that is dominated by
baryonic matter at small radii with a smooth transition to a roughly constant
rotation speed at large radii \citep[cf.][]{1991AJ....101.1231C,
2010A&A...519A..47A}.
} \citep[][hereafter \citetalias{1986RSPTA.320..447V}]{1986RSPTA.320..447V}.
One can begin to address this contentious issue by placing direct constraints on
the detail mass composition of galaxies.

Although there are multiple methods of measuring the total mass enclosed within
a given radius (e.g., dynamics, lensing), a robust decomposition of total mass
into fractional contributions from DM, stars, and the interstellar medium (ISM)
is non-trivial.  Measurements of the atomic-gas mass can be made directly using
21cm \hone\ emission, and molecular-gas mass can be estimated using CO emission.
However, stellar mass estimates depend on the calibration of stellar
mass-to-light ratios, $\mls$, via resolved stellar populations in the most
nearby (dwarf) galaxies or stellar-population-synthesis (SPS) modeling of
integrated light.  The latter remains substantially uncertain
\citep{2009ApJ...699..486C, 2010ApJ...708...58C, 2010ApJ...712..833C}.\footnote{
Here, the remarkable success of \citet{2005ApJ...632..859M} in reducing the
residuals in his BTF relation by using $\mls$ as derived from the
mass-discrepancy--acceleration relation \citep{2004ApJ...609..652M} is
noteworthy; however, it is possible that this is more reflective of the ability
of MOND \citep{1983ApJ...270..365M} to fit rotation curves and/or the disk-halo
conspiracy than it is of the absolute calibration of these $\mls$ measurements.
}

Rotation-curve mass decompositions provide upper limits on $\mls$ when one
adopts the ``maximum-disk'' hypothesis, the assumption that the rotation
velocity at the center is dominated by the luminous matter
\citepalias{1986RSPTA.320..447V}.  For example, \citet{2001ApJ...550..212B} used
the ``maximum-disk'' rotation curve decompositions made by \citet[][hereafter
\citetalias{2001ApJ...563..694V}]{2001ApJ...563..694V} to place limits on the
allowed $\mls$.  However, rotation curves cannot provide unique measurements of
$\mls$ as we have recently illustrated \citep[][hereafter
\citetalias{2010ApJ...716..198B}]{2010ApJ...716..198B}; inference of $\mls$
based on rotation-curve mass decompositions are unconstrained due to the
disk-halo degeneracy \citep{1985ApJ...295..305V}.

Given the uncertainty in SPS model zero-points and the disk-halo degeneracy, a
direct measurement of $\mls$ is needed.  Following the work of
\citet{1984ApJ...284L..35B}, \citet{1984ApJ...278...81V, 1986ApJ...303..556V},
and \citet{1993A&A...275...16B}, the DiskMass Survey
\citepalias[DMS;][]{2010ApJ...716..198B} aims to tackle this problem via
dynamical measurements of the mass surface density, $\sddisk$, of $\gtrsim$40
low-inclination, late-type galaxies.  Our measurements uniquely describe the
baryonic mass distributions and DM-halo density profiles, $\rhodm$, of each
galaxy within $\sim$3 disk scale lengths ($h_R$), thereby breaking the disk-halo
degeneracy and allowing for detailed calculations of disk-galaxy mass budgets.
In this paper, we focus on providing a detailed, initial example of these
calculations using UGC 463, located at equatorial (J2000.0) coordinates (RA,DEC)
= (00$^{\rm h}$43$^{\rm m}$32$\fs$5,+14$^{\rm d}$20$^{\rm m}$34$^{\rm s}$).  We
continue our series by summarizing the baryonic mass fractions in 29 additional
galaxies in Bershady~et~al.~(2011; hereafter \pV), {\it submitted}.

Here we summarize some salient properties of UGC 463: It is a well isolated
galaxy with a moderately-high extrapolated central disk surface brightness
\citepalias{2010ApJ...716..198B}, which is a factor of $\sim2$ above the mean
derived by \citet{1970ApJ...160..811F}.  It is of late type
\citepalias[SABc;][]{2010ApJ...716..198B} and demonstrates an interesting
three-arm spiral structure.  The SDSS $g$-band surface photometry demonstrates a
clear Type II surface-brightness profile, as defined by
\citet{1970ApJ...160..811F}, with a profile ``break'' at a radius of
$\sim15\arcsec$, well within the field-of-view (FOV) of our kinematic data; the
break becomes less pronounced toward longer wavelengths.  The disk is also
bright in the mid- and far-infrared {\it Spitzer} bands, suggestive of
significant star-formation activity and molecular gas mass.  In general, UGC 463
is unexceptional in its optical and near-infrared (NIR) color, size, and
luminosity; however, it is slightly redder and more luminous (in $M_K$) than
typical of galaxies in the DMS Phase-B sample \citepalias[as defined
in][]{2010ApJ...716..198B}.

Our study of UGC 463 is a detailed example in the use of our full suite of data
to produce quantities of fundamental relevance to the science goals of the DMS
\citepalias{2010ApJ...716..198B}, following much of the formalism developed in
\citet[][hereafter \citetalias{2010ApJ...716..234B}]{2010ApJ...716..234B}.
Given the large number of observational ingredients, we have relocated some
detailed information to future papers, which we refer to throughout our
discussion.  An outline of our paper is as follows: \sect \ref{sec:data}
presents all the data products.  We derive the on-sky geometric projection of
the disk using our two-dimensional kinematic data in \sect \ref{sec:geom},
including an extensive discussion of the inclination, $i$.  Based on this
projection geometry, we produce azimuthally averaged kinematic profiles and
beam-smearing corrections and discuss the axial symmetry of the galaxy in \sect
\ref{sec:spk+ppk}.  In \sect \ref{sec:disk}, we derive salient properties of the
disk including the shape of the disk stellar velocity ellipsoid (SVE), the disk
stability, mass surface densities of all baryonic components, and dynamical
mass-to-light ratios.  In \sect \ref{sec:mass}, we produce a detailed mass
budget for UGC 463 out to 15 kpc ($\sim 4.2$ scale lengths); this analysis
relies on a traditional rotation-curve mass decomposition but uses our direct
measurements of $\mls$.  Having established the mass distribution of all the
baryonic components, \sect \ref{sec:mass} also presents the DM-halo density and
enclosed-mass distribution.  Therefore, \sects \ref{sec:data} --
\ref{sec:spk+ppk} are largely concerned with data handling, whereas \sects
\ref{sec:disk} -- \ref{sec:mass} produce the scientifically motivated
calculations that result from these data.  We summarize our study in \sect
\ref{sec:summary}.

We note here the nomenclature $\epsilon(x)$ signifies the measurement error in
$x$, $\azmean{x}$ is the azimuthal average of $x$, and $\allmean{x}$ is the
combined radial and azimuthal average of $x$.  When quoting two sets of errors
in any quantity (as done in the Abstract), the first and second set provide the
random and systematic uncertainties, respectively.

\section{Observational Data}
\label{sec:data}

\begin{figure*}
\centering
\leavevmode
\includegraphics[angle=-90,scale=1.15]{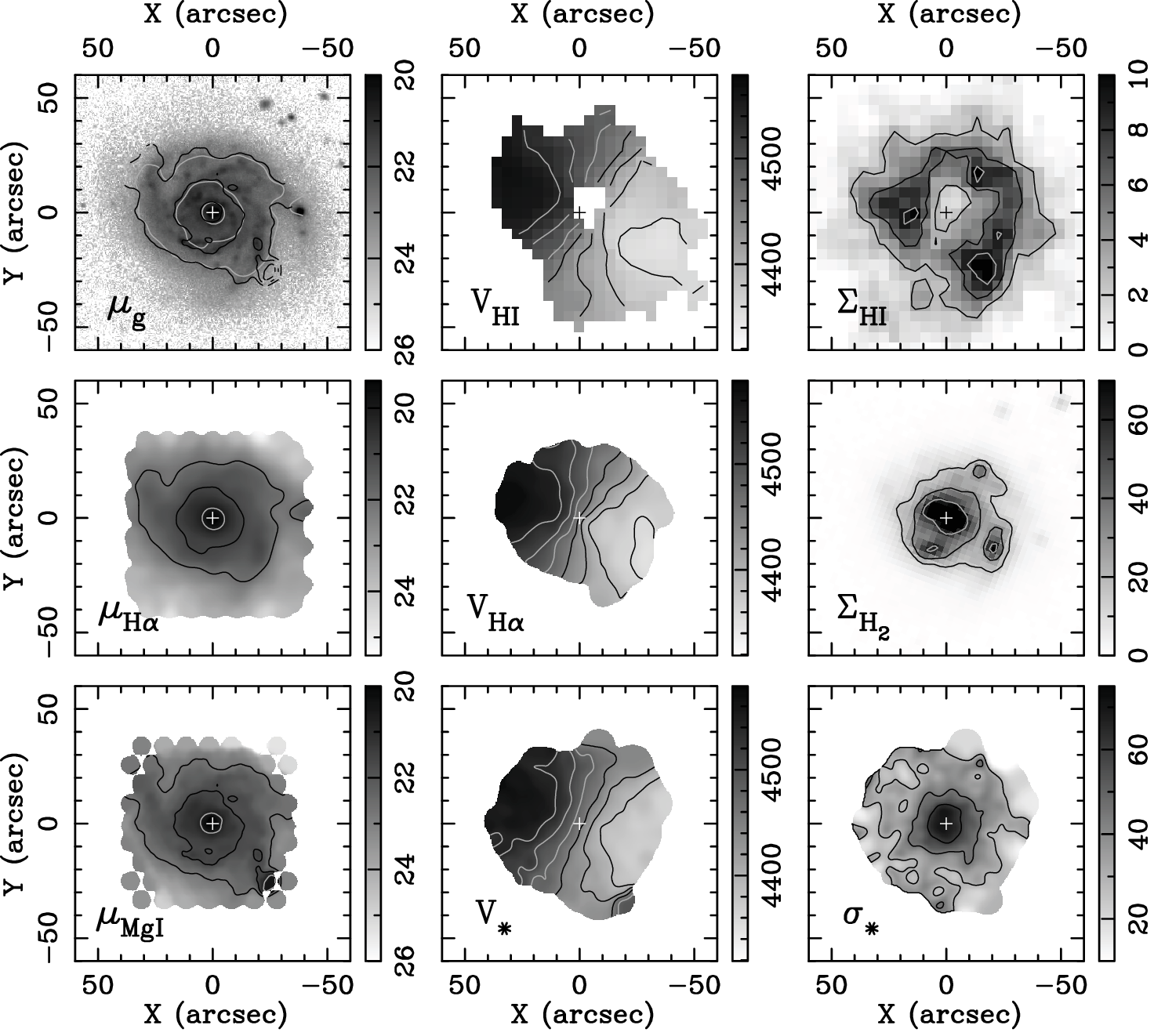}
\caption{
Two-dimensional data used in our study of UGC 463.  From top-to-bottom and
left-to-right, we provide the surface brightness in AB \muu\ for SDSS $g$-band
($\mu_g$), SparsePak \halp-region ($\muha$), and SparsePak$+$PPak \mgi-region
($\mumg$); LOS velocity in \kms\ for \hone\ ($\vhi$), \halp\ ($V_{{\rm
H}\alpha}$), and stars ($\vstar$); mass surface density in \sdu\ for \hone\
($\sdhi$) and \molh\ ($\sdmh$); and stellar velocity dispersion
($\sigma_{\ast}$) in \kms.  The galaxy center, as provided by NED, is marked by
either a black or white plus sign.  All images have the same spatial scale and
have a sky-right orientation.  Contour levels are: $\mu_{{\rm H}\alpha}$ --- 22,
21, 20 \muu; $\mumg$ --- 22.7, 21.7, 20.7 \muu; $\vhi$, $V_{{\rm H}\alpha}$,
$\vstar$ --- 4360, 4390, 4420, 4450, 4480, 4520, 4540 \kms; $\sdhi$ --- 3, 6, 9
\sdu; $\sdmh$ --- 15, 30, 60 \sdu; and $\sigma_{\ast}$ --- 30, 45, 60 \kms.
Gray contours are used to ease visibility with respect to the background
grayscale image.  The contours of $\mu_{{\rm H}\alpha}$ ({\it gray}) and $\mumg$
({\it black}) are overplotted on the $\mu_g$ image for comparison.
}
\label{fig:maps}
\end{figure*}

The DMS has collected an extensive suite of data to reach our science goals, as
described in \citetalias{2010ApJ...716..198B}.  We draw upon a large fraction of
those observations specific to UGC 463 for use in this paper.  Table
\ref{tab:data} provides a list of all the data products used herein, their
observational source, the year of the relevant observations, a reference to the
section containing a description of each dataset, and (when available) a
reference to papers containing more detailed information.  On-sky maps of much
of the relevant data products are provided in Figure \ref{fig:maps}; see \sect
\ref{sec:maps} and Appendix \ref{app:maps} for a full description of how these
maps were generated.

\begin{deluxetable}{ l r r r r }
\tabletypesize{\small}
\tablewidth{0pt}
\tablecaption{Observational Data}
\tablehead{ Product & Source & Year & Section & Ref\tablenotemark{a}}
\startdata
\halp\ kinematics           & SparsePak         & 2002 &   \ref{sec:spkha}, \ref{sec:gaskin}  & 1,2     \\
\oiii\ kinematics           & SparsePak         & 2006 &    \ref{sec:oiii}, \ref{sec:gaskin}  & 3       \\
                            & PPak              & 2004 &    \ref{sec:oiii}, \ref{sec:gaskin}  & 4       \\
stellar kinematics          & SparsePak         & 2006 & \ref{sec:spkstar}, \ref{sec:starkin} & 3       \\
                            & PPak              & 2004 & \ref{sec:ppkstar}, \ref{sec:starkin} & 4       \\
$g$, $r$, $i$ photometry    & SDSS              & 2003 &                     \ref{sec:optnir} & \nodata \\
$K$, \jhk photometry        & 2MASS             & 2000 &                    \ref{sec:optnir}  & \nodata \\
\hone\ aperture-synthesis   & VLA               & 2005 &                       \ref{sec:vla}  & 4       \\
24-$\mu$m photometry        & {\it Spitzer}     & 2007 &                    \ref{sec:24phot}  & \nodata
\enddata
\tablenotetext{a}{{\bf References:} 1. \haIt; 2. \haIIt; 3. \citet{KBWPhD}; 4. \citet{TPKMPhD}}
\label{tab:data}
\end{deluxetable}

\subsection{Distance}
\label{sec:dist}

The distance to UGC 463 is used to calculate: (1) the total absolute $K$-band
magnitude, $M_K$, providing an inclination measurement via inversion of the TF
relation (\sect \ref{sec:itf}); and (2) the disk scale height based on a
measured scale length in kpc \citepalias[see equation 1
from][]{2010ApJ...716..234B}.  In \sect \ref{sec:ikin}, we find $\vsys =
4460\pm1$ \kms, consistent with $\vsys = 4452\pm9$ \kms\
\citep{1999ApJS..121..287H} provided by NED.\footnote{
The NASA/IPAC Extragalactic Database, operated by the Jet Propulsion Laboratory,
California Institute of Technology, under contract with the National Aeronautics
and Space Administration.
}  Applying the 104 \kms\ flow correction \citep{2000ApJ...529..786M}, we
calculate a flow-corrected velocity of $\vflow = 4356\pm52$ \kms, where we have
taken half the flow correction as its error \citepalias{2010ApJ...716..234B}.
Using $H_0 = 73 \pm 5$ km s$^{-1}$ Mpc$^{-1}$ for Hubble's constant
\citep[provided by NED, cf.][]{2009ApJ...699..539R, 2011ApJS..192...16L}, we
calculate a flow-corrected distance of $D = \vflow/H_0 = 59.67 \pm 0.01 \pm
4.15$ Mpc; the systematic error is dominated by the uncertainty in $H_0$.

\subsection{Optical and Near-Infrared Emission}
\label{sec:optnir}

We use archival $g$-, $r$-, and $i$-band data obtained from the Sloan Digital
Sky Survey \citep[SDSS;][]{2000AJ....120.1579Y} and $J$-, $H$-, and $K$-band
data obtained from the Two-Micron All-Sky Survey
\citep[2MASS;][]{2006AJ....131.1163S} to produce surface-brightness profiles and
large-aperture total magnitudes.  Photometric measurements are in AB magnitudes
for SDSS data and Vega-based magnitudes for 2MASS data.  SDSS and 2MASS images
are, respectively, $10\farcm2 \times 13\farcm8$ and $8\farcm3 \times 17\farcm1$
with UGC 463 well separated from the frame edges.

\subsubsection{Surface Photometry}
\label{sec:sbprof}

Given the basic image reduction and photometric calibration provided by SDSS and
2MASS, our surface photometry is primarily concerned with sky-background
subtraction and masking sources other than UGC 463.

Source catalogs have been created for each band using Source
Extractor.\footnote{
\url{http://www.astromatic.net/software/sextractor}
}  Each catalog has been visually inspected and pruned of erroneous source
identifications, such as along meteor streaks or diffraction spikes; these
features are masked from our final results by including pseudo-sources in our
catalog.  We have created a master catalog for the region surrounding UGC 463 by
merging the catalogs from all bands, identifying sources detected in multiple
bands.

Using the {\it IRAF}\footnote{
{\it IRAF} (Image Reduction and Analysis Facility) is distributed by the
National Optical Astronomy Observatory, which is operated by the Association of
Universities for Research in Astronomy, Inc., under cooperative agreement with
the National Science Foundation.
} task {\tt imsurfit}, we determine the sky background of each image by fitting
a Legendre polynomial surface (with cross terms) to each image where all sources
and artifacts are replaced, initially, by a ($\pm3$) sigma-clipped mean of the
image.  After the lowest-order surface fit (2-$\times$2-order), masked regions
are replaced by the fitted surface values as the fit order is increased.  By
inspection, we find there is little improvement in the sky flatness when using
surfaces of more than 9-$\times$9-order (terms up to $x^8$), and higher order
fits begin to introduce artificial structure.  The backgrounds of SDSS images
are generally well-behaved, whereas the 2MASS $H$- and $K$-band data exhibit
significant background structure.

Surface photometry has been performed on each image after subtracting the sky
background and masking all artifacts and sources, except for UGC 463.  Source
masking is forced to be identical in every band.  From these masked images, we
perform elliptical aperture photometry over a range of radii, each with an
aspect ratio and orientation coinciding with the derived geometry of the disk
discussed in \sect \ref{sec:inclination}.  Given the shallow depth of the 2MASS
data, we have also produced a \jhk\ surface-brightness profile using the
unweighted sum of the $J$, $H$, and $K$ surface-brightness profiles.  This
extends the NIR surface-brightness profile to larger radii.  The ``\jhk\
bandpass'' has an effective band-width of $\delta\lambda/\lambda = 0.62$ and a
Vega zero-point of 1062 Jy.

\begin{deluxetable*}{ c c c c c c c }
\tabletypesize{\small}
\tablewidth{0pt}
\tablecaption{Exponential Disk Scale Lengths and Scale Heights}
\tablehead{ & \colhead{$R_{\rm max}$} & \colhead{$\mu_0$} & \colhead{$h_R$} & \colhead{$h_R$} & \colhead{$q$} & \colhead{$h_z$} \\ \colhead{Band} & \colhead{(arcsec)} & \colhead{(mag arcsec$^{-2}$)} & \colhead{(arcsec)} & \colhead{(kpc)} & \colhead{($h_R/h_z$)} & \colhead{(kpc)} }
\startdata
 $g$ & 70 & $20.16\pm0.07$ & $12.0\pm0.5$ & $3.5\pm0.1\pm0.2$ & $8.1\pm0.3^{+2.0}_{-1.7}$ & $0.43\pm0.03^{+0.11}_{-0.08}$ \\
 $r$ & 70 & $19.58\pm0.06$ & $12.2\pm0.4$ & $3.5\pm0.1\pm0.2$ & $8.1\pm0.2^{+2.0}_{-1.7}$ & $0.44\pm0.02^{+0.11}_{-0.09}$ \\
 $i$ & 70 & $19.29\pm0.05$ & $12.6\pm0.3$ & $3.6\pm0.1\pm0.3$ & $8.2\pm0.2^{+2.1}_{-1.7}$ & $0.44\pm0.02^{+0.11}_{-0.09}$ \\
 $K$ & 55 & $16.66\pm0.09$ & $11.9\pm0.3$ & $3.4\pm0.1\pm0.2$ & $8.0\pm0.2^{+2.0}_{-1.6}$ & $0.43\pm0.02^{+0.10}_{-0.08}$ \\
\jhk & 80 & $17.18\pm0.05$ & $12.3\pm0.3$ & $3.6\pm0.1\pm0.2$ & $8.1\pm0.2^{+2.0}_{-1.7}$ & $0.44\pm0.02^{+0.11}_{-0.09}$
\enddata
\tablenotetext{~}{{\bf Notes.} Columns are: (1) photometric band; (2) maximum
fitted radius, the minimum radius is always $R_{\rm min} = 15\arcsec$; (3) best-fitting
central surface brightness; (4) best-fitting scale length; (5) scale length in
kpc; (6) oblateness calculated using equation 1 from
\citetalias{2010ApJ...716..234B}; and (7) scale height.}

\label{tab:expdisk}
\end{deluxetable*}

\begin{figure}
\epsscale{1.1}
\plotone{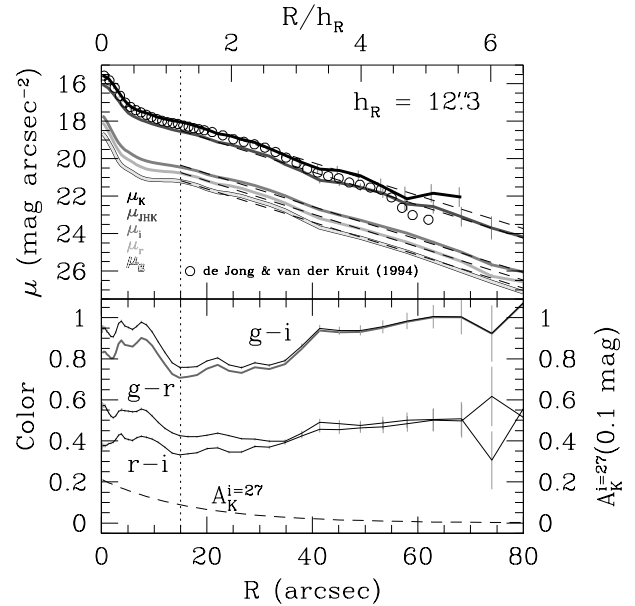}
\caption{
Optical and NIR photometry of UGC 463, corrected for Galactic extinction; SDSS
measurements are in AB magnitudes and 2MASS measurements are in Vega magnitudes.
{\it Top} --- From top-to-bottom, the 2MASS $K$-, combined 2MASS \jhk-, SDSS
$i$-, SDSS $r$-, and SDSS $g$-band surface-brightness profiles are plotted as
solid lines; the bands are differentiated by grayscale as given by the key.
Exponential-disk fits with a fixed scale length of $h_R = 12\farcs3$ are shown
as dashed lines (compare to the results in Table \ref{tab:expdisk} when $h_R$ is
allowed to be free for each band); the vertical dotted lines marks the innermost
radius included in the fit.  The $K$-band data from \citet{1994A&AS..106..451D}
are plotted as open circles.  {\it Bottom} --- SDSS $g-r$, $r-i$, and $g-i$
colors; the thick gray line is the $g-i$ color after correcting for internal
extinction.  The estimate of the internal $K$-band extinction, $A^{i=27}_K$,
calculated in \sect \ref{sec:dustext} is also provided as a dashed line.
}
\label{fig:sbprof}
\end{figure}

Figure \ref{fig:sbprof} provides all the surface photometry used in our present
study of UGC 463.  We apply Galactic extinction corrections\footnote{
\url{http://irsa.ipac.caltech.edu/applications/DUST/}
} of $A_g = 0.357$, $A_r = 0.242$, $A_i = 0.173$, and $A_K = 0.034$; no
extinction correction is applied to $\mujhk$ (see \sect \ref{sec:muk}).  We also
plot the UGC 463 $K$-band photometry from \citet{1994A&AS..106..451D}, which is
in very good agreement with our 2MASS photometry at $R < 40\arcsec$.   Table
\ref{tab:expdisk} provides the result of fitting an exponential disk to all
bands, including the \jhk\ data, demonstrating marginal change in the
best-fitting scale length and no evident trend with wavelength.  The Table
provides the radii over which the exponential disk is fit; the minimum radius is
always $15\arcsec$ in order to avoid non-exponential features seen near the
galaxy center.

\subsubsection{$K$-Band Surface Brightness Profile}
\label{sec:muk}

We apply two corrections to $\mujhk$ and use it as the primary NIR
surface-brightness measurement for our study of UGC 463: (1) a color correction
to produce a more accurate $K$-band surface-brightness profile at large radius and
(2) an instrumental-smoothing correction.  The color correction compares $\mu_K$
and $\mujhk$ from Figure \ref{fig:sbprof}, such that it includes the Galactic
extinction correction to $\mujhk$.  We find $\mu_K - \mujhk = -0.47\pm 0.03$ at
$R<40\arcsec$ with a maximum deviation from the mean of 0.05 magnitudes, which
is at most $\sim\epsilon(\mu_K)$.  The NIR color gradients are small over this
radius, with $\Delta(J-H) \lesssim 0.2$ and $\Delta(J-K) \lesssim 0.1$, and
roughly consistent with no color gradient to within the photometric error.

Our instrumental-smoothing correction effectively performs a one-dimensional
bulge-disk decomposition of $\mujhk$.  We assume the central light concentration
intrinsically follows a \citet{1963BAAA....6...41S} profile.  After first
subtracting an exponential surface-brightness profile fitted to $\mujhk$ data
between $10\arcsec< R<33\arcsec$, we model the central light concentration
within $R<8\arcsec$ by a S{\'e}rsic profile convolved with a Gaussian kernel,
assuming the latter is a good approximation for {\it all} instrumental effects.
We find a best-fitting S{\'e}rsic index of $n=1.5$ and an effective (half-light)
radius of $R_e = 1\farcs7$.  Random errors in this modeling are approximated by
the root-mean-square (RMS) difference between the measured and modeled central
light profile.  We assume that the ``intrinsic disk surface brightness'' is the
remainder of the profile after subtracting the model of the central-light
concentration, the ``intrinsic central light concentration'' is the S{\'e}rsic
component of the model, and the ``intrinsic $\mujhk$ profile'' is the sum of
these two components; our instrumental-smoothing correction then consists of the
difference between the measured and ``intrinsic'' $\mujhk$ profile.  We adopt a
conservative 50\% systematic error in this correction due to the inherent
uncertainties in the true parameterization of the central light concentration
and instrumental-smoothing kernel.  Although relevant to our assessment of any
central mass concentration in our calculations of $\sddisk$ (\sect
\ref{sec:sdd}), $\mlskdisk$ (\sect \ref{sec:ml}), and the mass budget (\sect
\ref{sec:mass}), our instrumental-smoothing correction is immaterial to the
fundamental conclusions of our paper.

Hereafter, we refer to the corrected \jhk\ measurement as $\mu^{\prime}_K$;
$h_R$ always refers to the measurement based on this profile unless otherwise
stated.  The fully corrected $\mu^{\prime}_K$ profile is shown in, e.g., Figure
\ref{fig:sdgas}.

\subsubsection{Total Apparent $K$-Band Magnitude}
\label{sec:mk}

In \sect \ref{sec:itf}, we use $M_K$ to calculate the inverse-TF inclination of
UGC 463 based on the TF relations derived by \citetalias{2001ApJ...563..694V}.
Therefore, we calculate the total apparent magnitude, $m_K$, using the 2MASS
$K$-band data according to the procedure used by
\citetalias{2001ApJ...563..694V}:  We measure a roughly isophotal magnitude at
$\mu_K = 21.5$ \muu\ (the rough surface-brightness limit occurring at $R
\lesssim 55\arcsec$) and extrapolate to infinity based on the fitted exponential
disk ($h_R = 12\farcs3\pm0\farcs3$).  We find $m_K(R\leq55\arcsec) =
9.45\pm0.01$ and a correction of -0.09 mag for the extrapolation of the disk.
Accounting for Galactic-extinction ($A_K = 0.034$), we find $m_K = 9.32\pm0.02$.

\subsubsection{Internal Dust Extinction}
\label{sec:dustext}

Left uncorrected, $\mu^{\prime}_K$ may overestimate our dynamical mass-to-light
ratios (\sect \ref{sec:ml}) due to internal dust extinction in $K$-band,
$A^i_K$.  Using the dust-slab model from \citet{1985ApJS...58...67T}\footnote{
See also discussion in \citetalias{2001ApJ...563..694V} and
\citetalias{2010ApJ...716..234B}.
} and $i=27\arcdeg\pm 2\arcdeg$ (\sect \ref{sec:inclination}), we calculate the
function $A^{i=27}_{K}(R)$ as shown in Figure \ref{fig:sbprof}, which has a
maximum of $A^{i=27}_{K}(R=0) = 0.02\pm 0.01$ at the galaxy center.  Assuming
$A_V/E(B-V) = 3.1$ \citep{1989ApJ...345..245C}, we calculate $A_K/E(g-i) =
0.18\pm0.02$, which we use to apply an internal reddening correction to our
$g-i$ color when comparing our dynamical $\mls$ measurements to those predicted
by SPS modeling.  Figure \ref{fig:sbprof} plots the uncorrected $g-r$, $r-i$,
and $g-i$ colors, as well as our internal-reddening-corrected $(g-i)_0$ color.
Compared to more realistic radiative-transfer modeling including a clumpy ISM
and spiral structure, these simple dust-slab model predictions tend to
overestimate both the level of dust extinction and reddening toward high
inclination; the predictions are more reasonable at the low inclination
appropriate for UGC 463 (\dustt).  Given the marginal extinction in $K$-band and
reddening of $g-i$, the simpler model is sufficient for our purposes.

\subsubsection{Dust Emission}
\label{sec:dustem}

We note here that the $K$-band also contains emission from hot dust and
polycyclic aromatic hydrocarbons; however, based on a preliminary modeling of
the spectral energy distribution of our full suite of NIR and {\it Spitzer}
imaging, we expect this to be no more than a 3\% contribution, which is
immaterial to the conclusions of this paper.

\subsection{Ionized-Gas Kinematics}

We primarily use ionized-gas kinematics obtained by the SparsePak\footnote{
Mounted on the 3.5-meter WIYN telescope, a joint facility of the University of
Wisconsin-Madison, Indiana University, Yale University, and the National Optical
Astronomy Observatories.
} \citep{2004PASP..116..565B, 2005ApJS..156..311B} and PPak\footnote{
Mounted with PMAS on the 3.5-meter telescope at the Calar Alto Observatory,
operated jointly by the Max-Planck-Institut für Astronomie (MPIA) in Heidelberg,
Germany, and the Instituto de Astrofísica de Andalucía (CSIC) in Granada, Spain.
} \citep{2004AN....325..151V, 2005PASP..117..620R, 2006PASP..118..129K}
integral-field units (IFUs), augmented by our \hone\ observations, to produce
the total rotation curve of UGC 463.  These ionized-gas data also provide
measurements of the gas velocity dispersion, which we use to correct the gas
rotation speed to the circular speed (\sect \ref{sec:vccorr}).  Below, we
briefly describe the IFU data available for UGC 463 and our extraction of the
ionized-gas kinematics.

\subsubsection{\halp, \niinb, \& \siinb\ Spectroscopy}
\label{sec:spkha}

SparsePak integral-field spectroscopy (IFS) of UGC 463 was obtained on the
nights of UT 02 January 2002 and UT 20 October 2002, following the setup
provided for the \halp\ region as listed in Table 1 of
\citetalias{2010ApJ...716..198B}.  We obtained four pointings during the January
run and an additional three pointings during the October run.  The pointings
nominally followed the 3-pointing dither pattern designed to fully sample the
$72\arcsec\times 71\arcsec$ FOV of SparsePak
\citep{2004PASP..116..565B};\footnote{
\url{http://www.astro.wisc.edu/~mab/research/sparsepak/}
} the fourth pointing during the January 2002 run was a repeat of the center
pointing.  We obtained 2$\times$15-minute exposures for each pointing.  Each
exposure pair is combined, before extraction of the spectra, while
simultaneously removing cosmic rays.  Spatially overlapping fibers among the
seven pointings are not combined but treated individually throughout our
analysis.  Further details of the reduction of these data (basic image
reduction, spectral extraction, wavelength calibration, and sky and continuum
subtraction) are provided by \haIt, largely following methods described in
\citet{2006ApJS..166..505A} with continuum-subtraction techniques described in
\citet{2005ApJS..156..311B}.  The RMS difference between the catalogued and
measured line centroids for the ThAr lines used in our wavelength calibration,
i.e. the ``wavelength calibration error,'' is typically $\lesssim 0.1$ \kms.

We also measure the instrumental dispersion, $\sinst$, as a function of
wavelength for all spectra, using the ThAr emission lines from our calibration
lamp spectra \citepalias{2010ApJ...716..234B}.  The intrinsic widths of the ThAr
features are negligible such that the second moment of these lines is equivalent
to $\sinst$ to good approximation.  After identifying a set of appropriate
(unblended) lines from the calibration spectrum, we fit single Gaussian
functions to each line using the same code described by
\citet{2008ApJ...688..990A} to fit the emission-line features in our galaxy
spectra (see also \haIt).  For each fiber, we fit a quadratic Legendre
polynomial to $\sinst(\lambda)$, which is used to interpolate $\sinst$ at any
wavelength.  The average instrumental broadening across the full spectral range
for all fibers is 13 \kms.

\subsubsection{\oiiinb\ Spectroscopy}
\label{sec:oiii}

Our optical continuum spectra in the \mgi\ region taken with both SparsePak and
PPak --- described in \sects \ref{sec:spkstar} and \ref{sec:ppkstar},
respectively --- have sufficient spectral range to include the \oiii
$\lambda$5007 emission feature.  Therefore, we also use these lines as tracers
of the ionized-gas kinematics.  No adjustment of the continuum-data reduction
recipe was needed to accommodate the proper handling of the emission features.
Instrumental dispersions are calculated as described in \sects \ref{sec:spkstar}
and \ref{sec:ppkstar} for the SparsePak and PPak data, respectively.

\subsubsection{Kinematic Measurements}
\label{sec:gaskin}

Ionized-gas kinematics are measured for all available emission lines.  Following
\citet[][see also \citealt{2006ApJS..166..505A}]{2008ApJ...688..990A}, both
single and double Gaussian line profiles are fitted in a 20\ang\ window centered
around each line.  All Gaussian fits have been visually inspected to ensure each
emission line was fitted properly.  Velocities ($cz$) of each atomic species are
calculated separately using the wavelength of the Gaussian centroid.  Of all
fitted line profiles, 27\% are better fit by a double Gaussian profile
\citep[][\haIIt]{2008ApJ...688..990A}; in these cases, a single component is
used to measure the line-of-sight (LOS) velocity.  Ionized-gas velocity
dispersions, $\sgas$, also use a single component and are corrected for the
measured instrumental line width.

For our \halp-region spectroscopy, we combine all available velocity and
velocity dispersion measurements (any combination of the \nii$\lambda$6548,
\halp, \nii$\lambda$6583, \sii$\lambda$6716, and \sii$\lambda$6731 lines) into
an error-weighted mean velocity for each fiber.  Due to the large uncertainties
in $\sgas$ for lines other than \halp, we only include measurements with
$\epsilon(\sgas) < 3$ \kms\ in the combined value.  In this spectral region, both
velocities and $\sgas$ are dominated by the \halp\ line measurements due to the
higher $\ston$ of these lines; hereafter, we refer to these kinematics as
``\halp'' kinematics, despite their inclusion of other ionized atomic species.

In addition to the consistency check among measurements made by SparsePak and
PPak, the \oiii\ kinematics provide a useful comparison with the \halp\ results
(see \sect \ref{sec:spk+ppk}).  However, the \halp\ line generally provides
higher quality kinematics and velocity fields: the line $S/N$ is higher on
average and the filling factor of the kinematic measurements in the disk of UGC
463 is more uniform.  We eventually combine all ionized-gas kinematics into a
single, axisymmetric set of measurements (\sect \ref{sec:rcsdisp}); however, it
is useful to keep in mind this distinction between the merit of the \halp\ and
\oiii\ kinematic data.

\subsection{Stellar Kinematics}

Our optical continuum spectra are at the heart of this paper and, in fact, our
entire survey.  UGC 463 is among a set of 19 galaxies in (and roughly half of)
our Phase-B sample that have both SparsePak and PPak continuum spectroscopy near
the \mgi\ triplet.  This intentional duplication provides an internal
consistency check of our stellar kinematics using two different instruments, and
we find excellent agreement among the observations (\sect \ref{sec:spk+ppk}).
Here, we provide information concerning our observations and our derivation of
the LOS stellar kinematics.

\subsubsection{SparsePak Spectroscopy}
\label{sec:spkstar}

SparsePak IFS of UGC 463 was obtained using the \mgi-region setup as listed
in Table 1 of \citetalias{2010ApJ...716..198B} --- primarily targeting \fei\ and
\mgi\ stellar-atmospheric absorption lines.  The observation and reduction of
these data are described by \citet{KBWPhD}; we review the salient details here.

UGC 463 was observed on consecutive nights during a single run from UT $23-25$
September 2006.  Four, six, and four 45-minute exposures were taken during the
three nights of observation, respectively.  No dithering of the pointing was
applied between exposures; the original pointing was repeated to the best of our
ability for each night.  Exposures taken within a given night have been combined
into a single image, using an algorithm that simultaneously rejects cosmic rays,
and reduced (basic image reduction, spectral extraction, wavelength calibration,
and sky-subtraction) on a night-by-night basis.  The basic reduction procedures
are nearly the same as that used for the \halp\ data (\sect \ref{sec:spkha}).
The wavelength calibration errors are typically $\lesssim 0.1$ \kms.  Error
spectra have been calculated in a robust and parallel analysis.  The
repeatability of the pointing across the three nights of observation was good to
less than one arcsecond, determined by \citet{KBWPhD} by forcing the kinematic
centers of all the SparsePak \halp\ and \oiii\ data to follow from the same
on-sky kinematic geometry.  Thus, the extracted spectra from each night have
been combined on a fiber-by-fiber basis and weighted by the spectral
$(\ston)^2$.  The weighting is relevant due to changing conditions; moon
illumination increased for each night (from $1-7$\%) and a number of exposures
($\sim30$\%) suffered from variable transparency losses due to passing clouds,
particularly during the second night of observation.  The combined spectra, from
all 10.5 hours of integration, are analyzed in \sect \ref{sec:starkin} to
measure stellar kinematics for UGC 463.

\begin{figure}
\epsscale{1.1}
\plotone{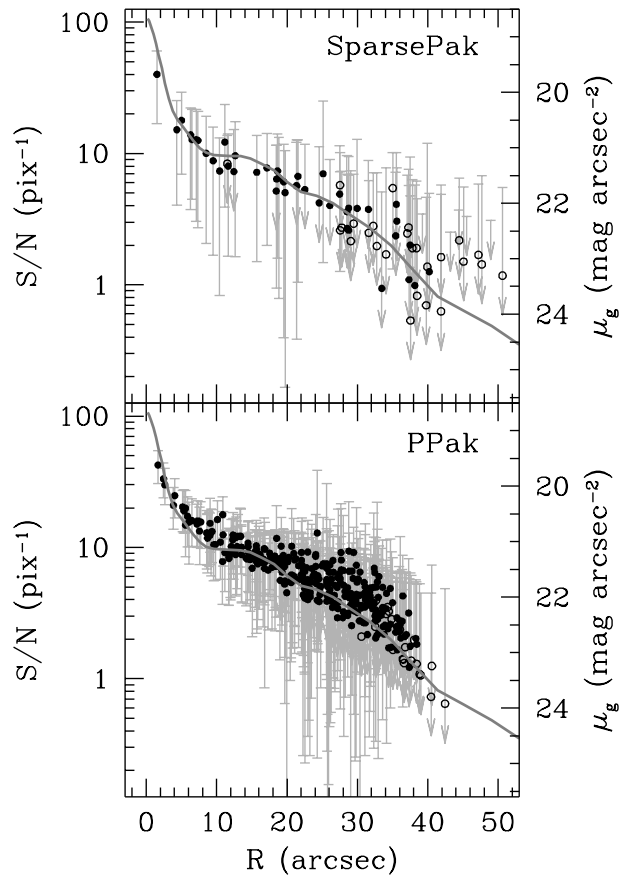}
\caption{
Mean $\ston$ measurements for both our SparsePak ({\it top}) and PPak ({\it
bottom}) \mgi-region IFS of UGC 463.  Data are plotted as a function of radius
and against the SDSS $g$-band profile ({\it gray line}; \sect \ref{sec:sbprof}).
Bars provide the range in $\ston$ of all pixels over the full spectrum; upper
limit arrows are used in cases where $\ston$ drops below zero due to the random
errors in sky subtraction.  Spectra that have yielded stellar kinematics are
plotted by filled points and as open circles otherwise.
}
\label{fig:sn}
\end{figure}

Using our calculated error spectra, Figure \ref{fig:sn} plots the mean $\ston$
of our SparsePak IFS as a function of radius against the $g$-band
surface-brightness profile, demonstrating that we have measured stellar
kinematics for fibers with $\mu_g\sim 22.5$ at $\ston\sim 3$.  The scale
translation between mean $\ston$ and $\mu_g$ assumes the data are
detector-limited ($\ston\propto$ flux), which is only an approximation for our
data.

Finally, we estimate $\sinst$ for both the template and galaxy spectra for use
in measuring instrumental-broadening corrections (\sect \ref{sec:sinst}).  We
use the same approach as described for the \halp-region spectroscopy in \sect
\ref{sec:spkha} for both the galaxy and template observations.  The average
instrumental broadening for these galaxy spectra is 11 \kms.

\subsubsection{PPak Spectroscopy}
\label{sec:ppkstar}

PPak IFS of UGC 463 was obtained using the \mgi-region setup as listed in Table
1 of \citetalias{2010ApJ...716..198B}.  Compared to the SparsePak optical
continuum spectra, individual PPak spectra have approximately 1.5 times the
spectral range, 0.7 times the spectral resolution, and 0.6 times the on-sky
aperture per fiber; however, PPak contains 4.4 times as many fibers as SparsePak
in its main fiber bundle.  These data are described by \citet{TPKMPhD}, which
includes all \mgi-region data taken by PPak.  Here, we briefly review the
acquisition and reduction of the data specifically for UGC 463.

We obtained two consecutive 1-hour exposures targeting UGC 463 on each of three
nights from UT $13-15$ November 2004, yielding six hours of total on-target
integration.  Flexure corrections have been applied on an exposure-by-exposure
basis, requiring spectral extraction, wavelength calibration, and sky
subtraction to be performed on each exposure individually. The RMS wavelength
calibration error is typically of the same order as that found for the SparsePak
observations ($\sim0.1$ \kms).  Error spectra have been calculated in a robust
and parallel analysis.  For UGC 463, no pointing offsets were detected; the CCD
of the guide camera of the PMAS spectrograph allows for accurate reacquisition
of our target galaxies on subsequent nights.  All six spectra for a given fiber
have been combined using weights depending on the instrumental resolution and
the spectral $\ston$ as described by \citet{TPKMPhD}.  

Figure \ref{fig:sn} provides the mean $\ston$ of the PPak spectra, alongside
that of the SparsePak spectra.  Despite the shorter integration time and smaller
fiber aperture, the PPak data have slightly higher mean $\ston$, an effect of both
the lower spectral resolution and the better efficiency of PMAS over the WIYN
Bench Spectrograph.\footnote{
Our data were taken before the completion of the upgrade to the WIYN Bench
Spectrograph that improved its overall efficiency by a factor of $\gtrsim 2$
\citep{2008SPIE.7014E..15B, 2010SPIE.7735E.240K}.
}

Instrumental broadening measurements are performed differently for PPak data
than described above for SparsePak data.  PPak observations provide simultaneous
calibration spectra in 15 fibers, evenly distributed along the pseudo-slit among
the galaxy spectra.  These spectra have proven critical for applying the
necessary flexure corrections \citep{TPKMPhD}.  Moreover, they provide {\it
simultaneous} measurements of the instrumental dispersion, obtained by fitting
Gaussian functions to the ThAr emission lines.  \citet{TPKMPhD} has fit a
quadratic Legendre polynomial {\it surface} to these measurements of $\sinst$
for each object frame.  We use this description of $\sinst$ to calculate the
instrumental-dispersion corrections for both the ionized-gas and stellar
kinematics measured from the PPak spectra.  The mean instrumental dispersion
across all fibers and all spectral channels is 17 \kms\ for our PPak data of UGC
463.

\subsubsection{Raw Kinematic Measurements}
\label{sec:starkin}

We use the Detector-Censored Cross-Correlation (\dct) software presented in
\citet[][hereafter \citetalias{2011ApJS..193...21W}]{2011ApJS..193...21W} to
determine spatially resolved stellar kinematics in UGC 463 from both our
SparsePak and PPak spectra.  Based on a preliminary analysis
\citepalias{2010ApJ...716..234B}, we find a K1 III star provides a minimum
template-mismatch error of $\sim 5$\% in the observed velocity dispersion
($\sobs$) with no systematic trend in radius.  In the future, we can improve
upon this by using composite templates; however, a 5\% template-mismatch error
in $\sobs$ is satisfactory for the present study.

As per our survey protocol \citepalias{2010ApJ...716..198B}, we have observed
template stars in the \mgi\ region using both SparsePak and PPak.  For the
analysis here, we specifically use the K1 III stars HD 167042 and HD 162555 for
our SparsePak and PPak stellar kinematics, respectively; Table \ref{tab:tpl}
presents salient information regarding the template spectra.  Template star
observations are performed under nominally the same spectrograph/telescope
configuration as for our galaxy data.  For both SparsePak and PPak, template
stars are observed by drifting the star through the full FOV, yielding many
spectra that are combined to provide high-$\ston$ templates; the final $\ston$
of each template is provided in Table \ref{tab:tpl}.

\begin{deluxetable*}{ c c r c c c c c c }
\tabletypesize{\footnotesize}
\tablewidth{0pt}
\tablecaption{Stellar Templates}
\tablehead{ & & & & & \multicolumn{4}{c}{Physical Quantities\tablenotemark{a}} \\ \cline{6-9} & \colhead{Spectral} & & & \colhead{$\ston$} & \colhead{$\vhel$} & \colhead{$T_{\rm eff}$} & \colhead{$\log g$} & \colhead{[Fe/H]} \\ \colhead{HD} & \colhead{Type} & \colhead{Instrument} & \colhead{UT Date} & \colhead{(pix$^{-1}$)} & \colhead{(\kms)} & \colhead{(K)} & \colhead{(cm s$^{-2}$)} & \colhead{(dex)} }
\startdata
167042 & K1 III & SparsePak & 2001-06-09 & 0.4$\times10^3$ & $-18.01\pm 0.17$ & 4878 & 2.74 & -0.11 \\
162555 & K1 III &      PPak & 2007-01-15 & 1.1$\times10^3$ & $-14.84\pm 0.20$ & 4660 & 2.72 & -0.21
\enddata
\tablenotetext{a}{Measurements of $\vhel$ are from \citet{2005A&A...430..165F};
remaining data are from \citet{2004ApJS..152..251V}.}
\label{tab:tpl}
\end{deluxetable*}

We fit any galaxy--template cross-correlation (CC) function that peaks within a
few hundred \kms\ of the systemic velocity of UGC 463 (as recorded by NED)
regardless of the $\ston$.  For each fiber, we adopt a Gaussian broadening
function, and we use a cubic Legendre polynomial to minimize continuum
differences between the broadened template and the fitted galaxy spectrum.
Additionally, we mask the \oiii$\lambda$5007 and \none\ ($\lambda$5198 and
$\lambda$5200) nebular emission regions from both the template and galaxy
spectra.  For the PPak data, the redshifted \oiii$\lambda$4960 line is also
visible; however, \dct\ masks the CC to a rest-wavelength range common to both
the galaxy and template spectrum, thereby automatically masking this line.  Each
CC fit has been visually inspected to insure the proper peak was considered by
the fitting algorithm and that any unexpected artifacts --- poorly removed sky
lines and/or cosmic-ray detections --- were masked.  Based on this inspection,
spectra have been refit as necessary.  Our stellar kinematic analysis follows
the expectations derived for random errors in \citetalias{2011ApJS..193...21W}.
As assessed via $\chi^2$ and the velocity shift with respect to spatially
neighboring fibers, we find reasonable fits to spectra with mean $\ston$
approaching unity, albeit with large errors.  Systematic errors should be
negligible for velocity measurements at all $\ston$, and they should be
$\lesssim20$\% in $\sobs$ at $\ston\gtrsim2$; systematic errors are always
smaller than the calculated random error \citepalias{2011ApJS..193...21W}.

\subsubsection{Instrumental-broadening Corrections}
\label{sec:sinst}

We correct our observed stellar kinematics for the system response function by
considering the following two separable components: (1) The broadening of the
intrinsic absorption-line widths due to the spectrograph optics, accounted for
using an ``instrumental-broadening'' correction, $\delta\sinst$; and (2) The
smearing of the intrinsic surface-brightness, velocity, and velocity dispersion
distributions by the response of the atmosphere$+$telescope system, accounted
for using a ``beam-smearing'' correction, $\sbs$.  The final LOS dispersion is
$\sstar^2 = \sobs^2 - \delta\sinst^2 - \sbs^2$ \citepalias{2010ApJ...716..234B}.
Unlike $\sbs$, $\delta\sinst$ is independent of the on-sky geometry and
intrinsic kinematic structure of the observed galaxy; therefore, we calculate
$\delta\sinst$ here.  We calculate $\sbs$ before combining our SparsePak and
PPak kinematics in \sect \ref{sec:spk+ppk} using the projection geometry derived
in \sect \ref{sec:geom}.

Each CC is used to compare template and galaxy absorption-line shapes such that
$\delta\sinst$ is determined by the difference in $\sinst$ measured for the
template and galaxy spectrum; we calculate $\delta\sinst$ following Appendix A
of \citetalias{2011ApJS..193...21W} using our measurements of $\sinst$ for both
the template and galaxy spectra.  We adopt a 4\% error in $\delta\sinst$
\citepalias{2010ApJ...716..234B}, which is marginal when compared to
$\epsilon(\sobs)$.  These corrections differ rather dramatically between
SparsePak and PPak; however, in both cases, $\delta\sinst^2$ is typically small.
Corrections to $\sobs$ --- i.e., the ratio $(\sobs^2 -
\delta\sinst^2)^{\frac{1}{2}} / \sobs$ --- are $\lesssim4$\% and $\lesssim20$\%
for, respectively, 90\% and 99\% of all measurements; a few measurements have
rather large corrections due to dispersion measurements of $\sobs < 10$ \kms,
which are likely erroneously low \citepalias{2011ApJS..193...21W}.  We always
find $\delta\sinst < \epsilon(\sinst)$.

\subsection{Atomic-Gas Content}
\label{sec:vla}

As part of our general survey strategy \citepalias{2010ApJ...716..198B}, we have
obtained 21cm aperture-synthesis imaging for the DMS Phase-B sample.  These
data measure neutral hydrogen (\hone) surface densities ($\sdhi$) and extend the
rotation-curve measurements of each galaxy; the ionized-gas kinematics can be
limited by the FOV of our \halp\ spectroscopy and/or the extent of the \halp\
emission in the disk.  For UGC 463, we obtained 2.3 hours of on-source
integration using the Very Large Array (VLA); observations were taken in the C
configuration yielding a synthesized beam of $14\farcs7\times12\farcs9$ and a
velocity resolution of 10.5 \kms.  In the end, these data provide only a
marginal radial extension of the ionized-gas rotation curve of UGC 463.  The
acquisition and reduction of these data is fully described by \citet{TPKMPhD}.

\begin{figure}
\epsscale{1.15}
\plotone{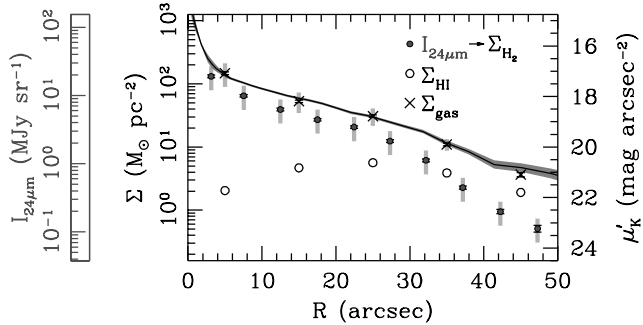}
\caption{
Mass surface density measurements for \hone\ (via direct observation; {\it open
circles}) and \molh\ (via an indirect calculation; {\it gray-filled points}),
and all gas including helium and metals ({\it crosses}).  The gray-filled points
also provide the $\imi$ profile as delineated by the ordinate to the far left.
Errors (random in black and systematic in gray) shown for $\sdmh$ and $\sdg$ are
discussed in the text; errors in $\sdhi$ are of order the size of the plotted
symbol.  As delineated by the right ordinate, we overplot $\mu^{\prime}_K$
(\sect \ref{sec:muk}) as the solid line with the error region plotted in dark
gray.
}
\label{fig:sdgas}
\end{figure}

The two-dimensional \hone\ mass-surface-density map and velocity field are
presented in Figure \ref{fig:maps}.  The azimuthally averaged measurements of
$\sdhi(R)$ for UGC 463 are presented in Figure \ref{fig:sdgas}; we adopt
$\epsilon(\sdhi) = 0.1\sdhi$.  As is typical of late-type spiral galaxies we
find a decrease in the \hone\ mass surface density toward the galaxy center; the
peak surface density of $\sdhi= 5.6$ \sdu\ occurs at $R=1.9 h_R$.  Although the
spatial resolution of our \hone\ column-density map is a factor of $\gtrsim3$
larger than our optical IFU data, we do not attempt to match the resolution of
these two data sets; such a correction to the azimuthally averaged total mass
surface density is negligible for the purposes of this paper.

\subsection{Molecular-Gas Content}
\label{sec:hmol}

Our nominal survey scope does not include direct observations of the molecular
content of our galaxy sample via, e.g., the $^{12}$CO ($J=1\rightarrow0$)
emission line --- henceforth all discussion of ``CO emission'' refers to this
emission feature unless noted otherwise.  Until we obtain such data, estimation
of the molecular content in DMS galaxies relies on available literature data
and/or inference from other suitable tracers for which data is available.

For our characterization of the molecular content of UGC 463, we use our
24$\mu$m {\it Spitzer} imaging to produce a rough approximation of the CO
surface-brightness distribution.  Numerous studies exist demonstrating a
correlation between the infrared luminosity of star-forming galaxies (dominated by
thermal dust emission) and their molecular-gas content as traced by CO emission.
For example, the integrated infrared luminosity based on {\it IRAS}
observations\footnote{
\url{http://irsa.ipac.caltech.edu/IRASdocs/iras.html}
} is well-correlated with the integrated CO flux \citep{1991ARA&A..29..581Y}.
Moreover, \citet{2007MNRAS.380.1313B} note a similar dependence of the
distribution of 24$\mu$m emission on morphological type as was noted by
\citet{1995ApJS...98..219Y} for the molecular gas traced by CO emission.  At
spatially resolved scales, \citet{2006A&A...456..847P} have studied the
correlation between CO ($\ico$) and 24$\mu$m emission ($\imi$) in a set of 6
nearby spiral galaxies to find $\imi \propto (\ico)^{0.9\pm0.1}$.  Correlations
between the CO and 8$\mu$m emission have also been discussed
\citep{2006ApJ...652.1112R, 2010MNRAS.402.1409B}; however, we prefer to focus on
the correlation between $\ico$ and $\imi$ as the latter should be dominated by
warm dust emission and be less dependent on the fraction of dust in the form of
polycyclic aromatic hydrocarbons \citep[see, e.g.,][]{2007ApJ...657..810D}.

In \sect \ref{sec:24phot}, we describe the procedure used to measure $\imi(R)$
in UGC 463.  In \sect \ref{sec:spitztoco}, we detail the calibration of our
$\imi$-to-$\ico$ surface-brightness relation using data made available by
\citet[][hereafter \citetalias{2008AJ....136.2782L}]{2008AJ....136.2782L}.
Finally, in \sect \ref{sec:xco}, we convert from $\imi$ to $\ico$ and then
calculate the molecular mass surface density, $\sdmh$, using the traditional
$X$-factor, $\xco$.  This latter step dominates the systematic error in our
estimation of the molecular content of UGC 463.

\subsubsection{24$\mu$m {\it Spitzer} Photometry}
\label{sec:24phot}

The survey strategy for all our {\it Spitzer} observations are provided in \sect
6.2.3 of \citetalias{2010ApJ...716..198B}.  In general, 24$\mu$m images
collected for the DMS demonstrate significant background structure, due to both
detector effects and intrinsic structure in the Galactic ISM, with fluctuations
on angular scales close to that of our galaxies.  To account for these
fluctuations, we mask out all statistically significant sources, including a
substantial radial region surrounding UGC 463, and create a
$47\arcsec\times47\arcsec$ boxcar-smoothed background image.  Masked regions are
iteratively filled by the boxcar smoothing, effectively interpolating the sky
background and its gross structure, across all detected sources.  We simply
subtract this smoothed image from our 24$\mu$m image of UGC 463 and use the
result to calculate the 24$\mu$m surface-brightness profile.

Our background-subtraction procedure has been carefully assessed to ensure that
the low-surface-brightness extent of UGC 463 has not been systematically
over-subtracted.  Preliminary tests with UGC 463 and other galaxies in our
survey demonstrate that our $\imi$ profiles become strongly affected by the
sky-subtraction errors at a source intensity below $\imi < 0.05$ MJy sr$^{-1}$
($\sdmh < 0.21$ \sdu\ in Figure \ref{fig:sdgas}).  UGC 463 is the third
brightest 24$\mu$m emitter in our entire sample, meaning that this surface
brightness limit falls outside the radial region relevant to this paper.  Our
measured 24$\mu$m surface-brightness profile uses elliptical apertures following
a geometry identical to that used for the optical and NIR photometry in \sect
\ref{sec:sbprof}.  Figure \ref{fig:sdgas} provides the 24$\mu$m
surface-brightness profile and the result of its conversion to $\sdmh$,
according to the discussion in the next two sections.  The random errors in our
$\imi$ measurements incorporate a constant 4\% calibration error
\citep{2007PASP..119..994E} and a sky-subtraction error estimated by the change
in $\imi$ introduced by a factor of two change in the smoothing-box size; the
latter results in 1\% and 10\% sky-subtraction errors at $R\sim30\arcsec$ and
$\sim47\arcsec$, respectively.  

\subsubsection{24$\mu$m-to-CO Surface Brightness Calibration}
\label{sec:spitztoco}

We use measurements of both CO and 24$\mu$m emission provided by
\citetalias{2008AJ....136.2782L} (see their Table 7) to measure the correlation
between $\imi$ and $\ico$.  Twelve of the 23 galaxies studied by
\citetalias{2008AJ....136.2782L} include both CO and 24$\mu$m observations;
however, four of those galaxies (NGC 2841, NGC 3627, NGC 4736, and NGC 5194) are
listed in NED as having either LINER or Seyfert activity, unlike UGC 463.
Therefore, we calibrate $\imi/\ico$ using only the remaining
eight galaxies, hereafter the ``$\imi/\ico$ subsample.''

The quantities provided by \citetalias{2008AJ....136.2782L} are
matched-resolution, azimuthally averaged radial profiles of $\Sigma_{\rm H_2}$
(based on CO emission and a value for $\xco$) and the contribution of embedded
star formation (determined from the 24$\mu$m surface brightness) to the total
star-formation-rate surface density.  We revert these quantities to $\ico$ (in K
\kms) and $\imi$ (in MJy sr$^{-1}$) using equations A2 and D1 from
\citetalias{2008AJ....136.2782L}.  All eight galaxies in the $\imi/\ico$
subsample were observed as part of the HERACLES Survey
\citep{2009AJ....137.4670L}, observing only the $^{12}$CO($J=2 \rightarrow 1$)
emission line, where \citetalias{2008AJ....136.2782L} adopt a line ratio of
$^{12}$CO($J=2 \rightarrow 1$)/$^{12}$CO($J=1 \rightarrow 0$) = 0.8.  The CO
surface brightness has been determined by integrating the emission profile over
the full line width and converting the flux units per beam to Kelvin using the
Rayleigh-Jeans limit.  The 24$\mu$m fluxes are determined from surface
photometry of {\it Spitzer} imaging data obtained by the SINGS Survey
\citep{2003PASP..115..928K}.

\begin{figure}
\epsscale{0.9}
\plotone{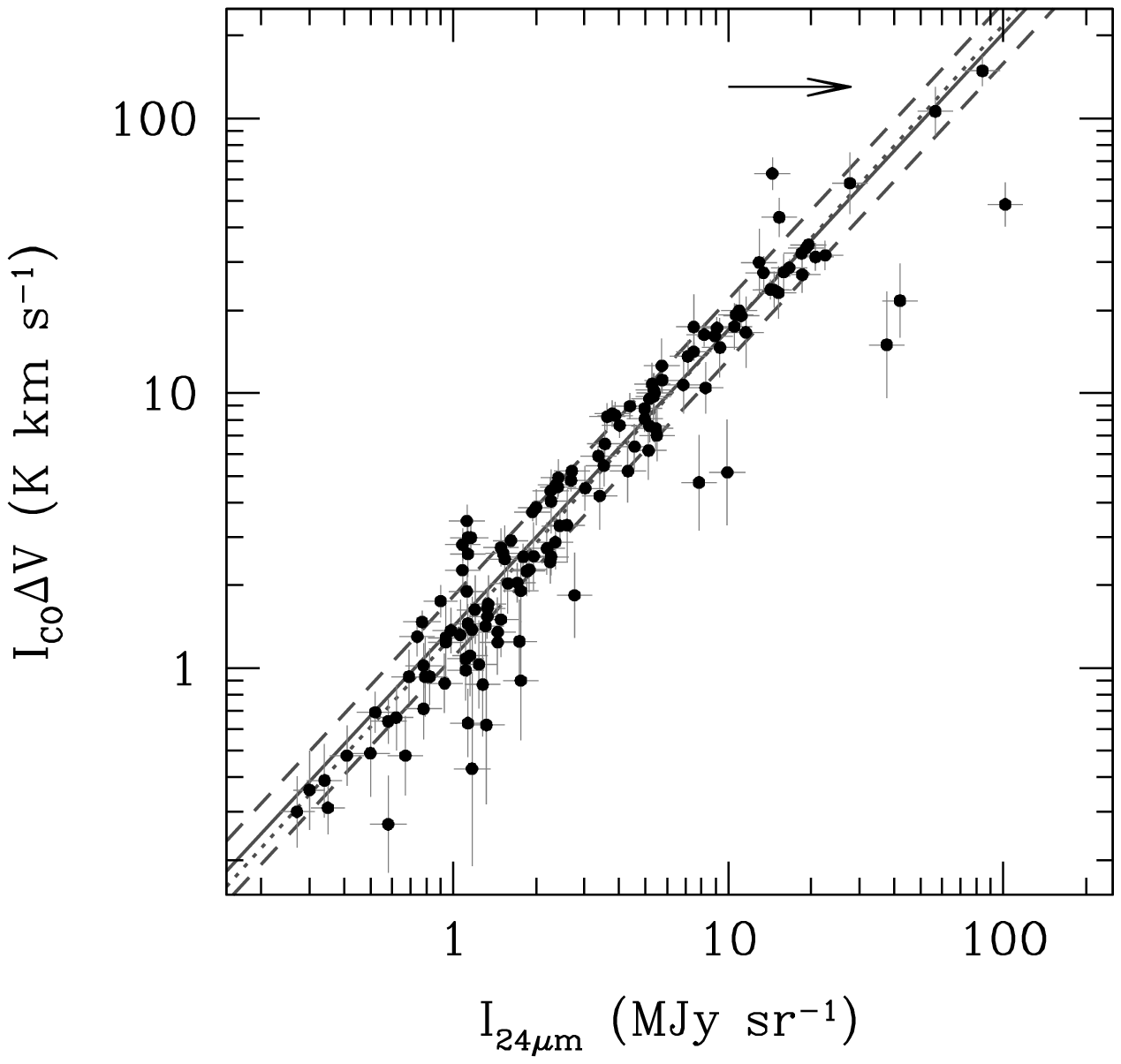}
\caption{
Correlation between azimuthally averaged values of the 24$\mu$m and CO surface
brightness from \citetalias{2008AJ....136.2782L} for the $\imi/\ico$ subsample
(see text).  The best-fit linear correlation (equation \ref{eq:24_co_fit}) is
given by the solid line; the dashed lines illustrate the weighted standard
deviation about the fit.  The dotted line has the slope determined by
\citet{2006A&A...456..847P} using a different galaxy sample with different CO
data.  The arrow at the top of the plot indicates the peak $\imi$ measured for
UGC 463.
}
\label{fig:cocal}
\end{figure}

Figure \ref{fig:cocal} presents the data for the $\imi/\ico$ subsample
regardless of the galaxy or radial region from which it has been measured.
Table 7 from \citetalias{2008AJ....136.2782L} is used to calculate
$\epsilon(\ico)$ directly, whereas we adopt a uniform $\epsilon(\imi) =
0.15\imi$ due to insufficient information; we expect this $\epsilon(\imi)$ to be
an upper limit.  We fit a power-law relationship between $\imi$ and $\ico$ to
all available data, incorporating errors in both coordinates \citep[\sect 15.3
of][]{NR3}, finding a best fit of
\begin{equation}
\log\left[ \frac{\ico}{\rm K\ km\ s^{-1}}\right] = \left(1.08\ \log\left[
\frac{\imi}{\rm MJy\ sr^{-1}}\right] + 0.15\right),
\label{eq:24_co_fit}
\end{equation}
with a weighted standard deviation of $\pm0.11$ dex, in good agreement with the
previous result from \citet{2006A&A...456..847P}.  Thus, given that UGC 463 has
physical parameters that are comparable to the $\imi/\ico$ subsample, our
calibration is expected to estimate $\ico$ for this galaxy to within $\sim30$\%.

\subsubsection{H$_2$ Mass Surface Density}
\label{sec:xco}

Using equation \ref{eq:24_co_fit}, we convert our sky-subtracted 24$\mu$m image
of UGC 463 to a CO surface brightness map.  Subsequently, we calculate the H$_2$
mass surface density using the traditional conversion factor, $\xco$, following
\begin{eqnarray}
\left[\frac{\sdmh}{\msol\ {\rm pc}^{-2}}\right] & = & 1.6 \left[\frac{\ico}{\rm K\
km\ s^{-1}}\right] \times \nonumber \\
 & & \left[\frac{\xco}{\rm 10^{20}\ cm^{-2}\ (K\ km\
s^{-1})^{-1}}\right] \cos i,
\label{eq:sdh2}
\end{eqnarray}
where $i$ is the galaxy inclination and $\xco$ is the ratio of the H$_2$ column
density to the CO line strength.  This use of $\xco$ to calculate $\sdmh$ is a
common procedure for calculating the molecular-gas content of external galaxies;
however, it may suffer from substantial systematic error.

A large number of studies have been devoted to measuring $\xco$ both in our own
Galaxy and within the Local Group.  Empirical and theoretical studies suggest
$\xco$ likely depends on multiple physical parameters, such as metallicity,
radiation field, gas mass surface density, and density structure
\citep{1996PASJ...48..275A, 1999ApJ...515...89M, 2002A&A...384...33B,
2006MNRAS.371.1865B, 2011arXiv1104.4118N}.  Moreover, direct measurement of
$\xco$ is observationally challenging: For example, the assumption of virial
equilibrium and the finite spatial resolution of giant molecular clouds, in even
Local Group galaxies, may both lead to inflated values of $\xco$
\citep{2007prpl.conf...81B, 2008ApJ...686..948B}; see
\citet{2008ApJ...686..948B} for a more general review of $\xco$ measurements.
Keeping these complications in mind, our analysis here adopts a simple approach:
Combining the Galactic measurement of $\xco = 1.8 \pm 0.3$ from
\citet{2001ApJ...547..792D} with the measurements for M31 ($\xco = 3.6 \pm 0.3$)
and M33 ($\xco = 2.6 \pm 0.4$) from \citet{2008ApJ...686..948B}, we find a mean
and range of $\xco = (2.7 \pm 0.9) \times 10^{20}$ cm$^{-2}$ (K \kms)$^{-1}$.
We assume this value to be representative of UGC 463, given that the Milky Way,
M31 and M33 are arguably the only spiral galaxies with well-resolved
observations of giant molecular clouds or associations from which robust
measurements of $\xco$ can be made.

Figure \ref{fig:maps} provides the 24$\mu$m image, converted to $\sdmh$ in units
of $\msol\ {\rm pc}^{-2}$ using the $\imi/\ico$ calibration from \sect
\ref{sec:spitztoco} and assuming $\xco = 2.7\times10^{20}$ cm$^{-2}$ (K
\kms)$^{-1}$.  Figure \ref{fig:sdgas} provides the azimuthally averaged surface
density profile $\sdmh(R)$ using the 24$\mu$m surface-brightness profile from
\sect \ref{sec:24phot}.  Errors in $\sdmh$ are plotted separately for random and
systematic components; the former includes errors from the $\imi$ calibration
and photometry and the inclination, whereas the latter includes the $\imi/\ico$
calibration error and range in $\xco$.  This estimate of $\sdmh(R)$ agrees with
the expectation that \molh\ is concentrated toward the ``hole'' in the \hone\
mass surface density, also seen in Figure \ref{fig:maps} and studies of other
galaxies \citepalias[e.g.,][]{2008AJ....136.2782L}.

Out to 15 kpc (4.2 $h_R$), we find $\mht/\mhi = 3.2$, a value that is reasonable
with respect to direct CO and \hone\ studies in the literature.   In particular,
\citet{1989ApJ...347L..55Y} find a range of 0.2 and 4.0 for, respectively, late-
and early-type spiral galaxies, comparable to the range measured by the more
recent COLD GASS survey \citep{2011MNRAS.415...32S}.  However, despite having a
{\it total} $\mht/\mhi$ that is decreased by $\sim10$\% compared to the
measurement within 15 kpc \citep{TPKMPhD}, UGC 463 is more rich in molecular gas
than the mean $\mht/\mhi$ calculated by \citet{2011MNRAS.415...32S} for the COLD
GASS survey by approximately twice the standard deviation.  Similarly, by
combining our 24$\mu$m imaging, the $\imi/\ico$ and $\xco$ values derived
herein, and the \hone\ data presented by \citet{TPKMPhD} for 24 galaxies in the
DMS, we find a mean of $\mht/\mhi = 0.48$, consistent with the
\citet{2011MNRAS.415...32S} measurement after accounting for the difference in
their adopted $\xco$; UGC 463 has the maximum value of $\mht/\mhi$ and 21 of 24
galaxies have $\mht/\mhi < 1$.  Therefore, consideration of the molecular mass
component in UGC 463 is relatively more important to our dynamical $\mls$
measurements and the baryonic mass budget than the majority of galaxies in the
DMS.  In \sect \ref{sec:ml}, we discuss both the total dynamical mass-to-light
ratios as well as stellar-mass-only measurements to illustrate the effects of
the gas-mass corrections.

\subsection{Total Gas Content}
\label{sec:totgas}

Figure \ref{fig:sdgas} provides a calculation of the total gas content of UGC
463, $\sdg =1.4(\sdhi+\sdmh)$; the factor of 1.4 accounts for the helium and
metal fraction.  For this calculation, measurements of $\sdmh$ have been
interpolated to the radii of the $\sdhi$ measurements; different interpolation
schemes are used in subsequent sections.  The random and systematic error have
been separated as discussed above for $\sdmh$.  Figure \ref{fig:sdgas}
demonstrates a strong correspondence between $\mu^{\prime}_K$ and $\sdg$.
Therefore, if $\mu^{\prime}_K$ is a reasonable tracer of the stellar mass, the
correspondence in Figure \ref{fig:sdgas} suggests that the radial distribution
of the stellar mass is roughly equivalent to that of the {\it total} gas mass.

\section{On-Sky Geometric Projection}
\label{sec:geom}

Geometric projection parameters are a fundamental consideration for the DMS.  In
particular, accurate inclinations are required to decompose $\sstar$ into the
vertical component, $\sigz$, using the measured axial ratios of the SVE (\sect
\ref{sec:sve}), and to produce the deprojected rotation curve, $\vrot(R)$, used
in our mass decomposition.  Errors in $\sigz$ and $\vrot$ have opposite trends
with inclination such that intermediate inclinations ($25\arcdeg \leq i \leq
35\arcdeg$) are preferred and, therefore, used to select optimal galaxies for
our survey (Papers I and II).

We present a detailed discussion of our determination of the geometric
projection of UGC 463 below.  We determine the inclination in \sect
\ref{sec:inclination}; we measure the on-sky pointing of each observation in
\sect \ref{sec:pointing}; and, in \sect \ref{sec:maps}, we discuss the
two-dimensional maps presented in Figure \ref{fig:maps} created using our final
pointing geometry.

\subsection{Inclination}
\label{sec:inclination}

We measure inclination using two methods: (1) ``kinematic inclinations''
($\ikin$) are determined by modeling an observed velocity field by a circularly
rotating disk and (2) ``inverse-Tully--Fisher inclinations'' ($\itf$) are
determined by inverting the TF relation \citep{1995ApJ...447...82R}.  Kinematic
measurements are used in {\it both} inclination estimates, but in different
ways:  For $\itf$, one measures a fiducial velocity from the rotation curve
\citepalias[e.g., $V_{\rm max}$ or $V_{\rm flat}$;][]{2001ApJ...563..694V} {\it
in projection} and compares with the inclination-corrected rotation speed
predicted by the TF relation for a known absolute magnitude.  In
contradistinction, $\ikin$ measurements are independent of any distance or
photometric measurement, instead determined by minimizing the difference between
measured and model isovelocity contours. 

We demonstrated in \citetalias{2010ApJ...716..234B} that the combination of
$\ikin$ and $\itf$ is ideal for minimizing the errors at low (using $\itf$) and
high (using $\ikin$) inclination; the two methods produce roughly equivalent
errors at $i \sim 30\arcdeg$.  A comparison of $\ikin$ and $\itf$ allows for an
internal assessment of the accuracy and precision of each.  We present
measurements of $\ikin$ and $\itf$ for UGC 463 below, and find that the
measurements are consistent at 1.2 times the combined error, which is
satisfactory for our purposes.  A statistically rigorous combination of the two
inclination estimates is derived by Andersen \& Bershady, {\it in prep};
however, here we simply produce the error-weighted mean value $i =
27\arcdeg\pm2\arcdeg$, which is used in our analysis in \sect \ref{sec:pointing}
and thereafter.  One can also estimate inclination via eccentricity measurements
of isophotal contours; however, this method is particularly poor at low
inclination and for galaxies that have significant outer-disk spiral structure,
as is true of UGC 463 (Figure \ref{fig:maps}).  Nevertheless, we calculate a
mean isophotal inclination of $27\arcdeg\pm 3\arcdeg$ by combining
Source-Extractor eccentricity measurements in the SDSS $g$, $r$, and $i$ bands
and the 2MASS $J$, $H$, and $K$ bands; this photometric measurement is easily
consistent with our adopted inclination based on kinematic measurements.

\subsubsection{Kinematic Inclination}
\label{sec:ikin}

We use the method described in \citet{2008ApJ...688..990A} to measure kinematic
inclinations; see also \citet{2003ApJ...599L..79A}.  The strength of this method
is in its simultaneous use of the full two-dimensional information in our
observed velocity fields.  However, it assumes a single set of geometric
projection parameters for the entire disk (a ``one-zone'' model), assumes that
all motion is purely circular rotation in the disk plane, and adopts a
parameterization for the projected rotation curve; therefore, one must justify
these assumptions.

Based on edge-on galaxies, literature studies have repeatedly found that warps
in non-interacting galaxies only influence disk morphology at large radii.
Highlighting two recent studies, \citet{2007A&A...466..883V} have shown that gas
disks (as traced by HI) typically begin to warp at approximately the outer
truncation radius of the stellar disk and \citet{2009MNRAS.396..409S} have shown
that stellar-disk warps occur at $R>3h_R$.  Thus, we expect no warping within
the FOV of our IFS of UGC 463, and only a marginal warping of our HI data.
Indeed, our \hone\ data only begin to show a position-angle warp for the last
measured radial bin \citep[$R=45\arcsec$;][]{TPKMPhD}.  For our IFU data,
post-analysis of the velocity-field residuals demonstrates little to no radial
dependence of the geometric parameters, as determined by translating velocity
residuals with respect to our nominal model (as developed in this and subsequent
sections) into model-parameter residuals.  This is done by holding all but one
parameter fixed and adjusting the free parameter until the velocity residual is
nearly or identically zero.  We find no correlation between the parameter
residuals and radius, with the possible exception of the position angle.  There
is some indication of a positive slope in position angle with radius; however,
the magnitude of the position angle change is small (less than $5\arcdeg$ over
the full radial range) and the significance of the slope is marginal.  This
means that the use of a radially dependent position angle is only marginally
justified and, more importantly, inconsequential to our measurement of the
rotation curve.  Therefore, a one-zone velocity-field model provides an adequate
description of our optical kinematic data.

As briefly noted in \sect \ref{sec:maps}, the isovelocity contours in our
velocity fields appear to show slight non-circular motions.  These motions are
most prevalent for the \halp\ data where some coherent structure is seen in the
velocity-field residual map, particularly along the spiral-arm to the south-west
of the galaxy center (on the approaching side of the velocity field).  These
coherent residuals likely represent streaming motions along this spiral arm
toward the galaxy center given their spatial correlation to the photometric
feature and the sign of the residual.  The magnitude of this streaming is less
than 10 \kms\ along the LOS (less than 25 \kms\ in the disk plane) and the
covering fraction of all non-circular motions is small.  Therefore, we expect
that our best-fitting velocity-field model should suffer only marginally from
these motions, particularly given the benefits afforded the one-zone model in
this respect.

The parameterization of $\vrotproj(R) = \vrot(R)\sin i$ does not adversely bias
the derived geometric parameters: \citet{2003ApJ...599L..79A} and
\citet{2008ApJ...688..990A} have chosen a hyperbolic tangent ($\tanh$) function
--- a simple two-parameter model that enforces an asymptotically flat rotation
curve.  Although inappropriate for the rare declining rotation curve in the DMS
sample, kinematic inclinations derived for such galaxies using a more
appropriate parameterization \citep[e.g.,][]{1997AJ....114.2402C} are within the
formal errors of those derived using a $\tanh$ model.

Given our highly sampled and high-quality velocity fields of UGC 463, here we
use a step function to define $\vrotproj(R)$, effectively fitting a set of
co-planar ``rings'' with constant rotation speed.  We fit up to 13 rings, each
with a width of $3\arcsec$ (approximately the diameter of a single PPak fiber)
such that rotation-speed gradients within each ring are small, except for
possibly the central ring.  The final ring includes all data at $R>36\arcsec$
and may be omitted for some tracers if no data exist at these radii.  Despite
our use of the term ``ring'' here, we emphasize that this fit is not a typical
tilted-ring fit given that we are defining only a single set of geometric
parameters.

We measure independent kinematic inclinations for the three kinematic data sets
provided by the SparsePak \halp\ data and the PPak \oiii\ and stellar data.  All
geometric and rotation-curve parameters are fit simultaneously, with one
exception: \citet{TPKMPhD} has used reconstructed continuum images to determine
the morphological center of UGC 463 relative to the PPak fibers, to which we
affix the dynamical center when modeling these data.  Greater detail regarding
our velocity-field fitting approach is provided in Appendix \ref{app:ikin},
including a full description of which measurements are omitted from
consideration during the fit.  However, Appendix \ref{app:ikin} is primarily
focused toward an assessment of the optimal data-weighting scheme for modeling
the velocity field of UGC 463.  Therein, we use bootstrap simulations \citep[see
\sect 15.6.2 of][]{NR3} to produce inclination probability distributions based
on four different weighting schemes.  We thereby demonstrate that we obtain the
most correspondent inclinations among the different data sets by adopting
weights defined by the derivative of the model LOS velocity, $\vlos$, with
respect to the inclination, i.e.\ $\partial\vlos/\partial i$.  These weights
approximately follow a $\sin^2(2\gaz)$ function in azimuth and a direct
proportionality in radius; therefore, data with the most leverage on the fitted
inclination \citep[at approximately $\pm45\arcdeg$ from the major
axis;][]{2003ApJ...599L..79A} have the highest weight.  The best-fitting
inclination, position angle, and systemic velocity for each tracer are given in
Table \ref{tab:kgeom}; bootstrap simulations are used to calculate the 68\%
confidence intervals.  The results provided for our \hone\ data from
\citet{TPKMPhD} are based on traditional tilted-ring fitting
\citep{1989A&A...223...47B}.

\begin{deluxetable}{ l c c c }
\tabletypesize{\small}
\tablewidth{0pt}
\tablecaption{Kinematic Geometry}
\tablehead{ & \colhead{$\ikin$} & \colhead{$\pa$} & \colhead{$\vsys$} \\ \colhead{Data Set} & \colhead{(deg)} & \colhead{(deg)} & \colhead{(\kms)} }
\startdata
\halp & $24.1^{+4.5}_{-2.1}$  & $68.8^{+0.3}_{-0.3}$ & $4458.6^{+1.0}_{-0.5}$ \\
\oiii & $26.5^{+4.6}_{-3.9}$  & $68.8^{+0.4}_{-0.7}$ & $4460.4^{+0.6}_{-0.8}$ \\
Stars & $25.5^{+4.7}_{-10.8}$ & $68.4^{+0.6}_{-0.7}$ & $4461.3^{+0.6}_{-0.8}$ \\
\hone &               \nodata & $68.8\pm 1.5$        & $4459.5\pm 1.5$        \\
\hline
Mean  & $25.1\pm2.5$          & $68.8\pm0.3$         & $4460\pm1$
\enddata
\label{tab:kgeom}
\end{deluxetable}

The geometric parameters listed in Table \ref{tab:kgeom} for each dynamical
tracer are in general agreement; the systemic velocities exhibit the most
statistically significant differences.  Such differences are likely due to
systematic errors in the heliocentric velocities of the template stars and/or
shifts in the pointing center.  In any case, these shifts are small and
irrelevant to our analysis of the mass distribution in UGC 463.  Using the half
width of the 68\% confidence interval from Table \ref{tab:kgeom} as the error,
we calculate error-weighted means of $\ikin = 25\fdg1 \pm 2\fdg5$ and $\pa =
68\fdg8\pm0\fdg3$.  The unweighted mean value $\vsys = 4460\pm1$ \kms\ has been
used in \sect \ref{sec:dist} to calculate the distance to UGC 463.

\subsubsection{Inverse Tully--Fisher Inclination}
\label{sec:itf}

Following the discussion in \citetalias{2010ApJ...716..234B}, inverse-TF
inclinations are calculated according to
\begin{equation}
\itf = \sin^{-1}\left[ 2\vrotproj\ {\rm dlog}\left(\frac{c_{1,\lambda} -
M_\lambda}{c_{2,\lambda}}\right)\right],
\label{eq:itf}
\end{equation}
where $c_{1,\lambda}$ and $c_{2,\lambda}$ are, respectively, the zero-point and
slope of the TF relation in wavelength band $\lambda$ and $M_{\lambda}$ is the
total absolute magnitude.  Combining $D=59.67 \pm 4.15$ Mpc (the error here is
the quadrature sum of the random and systematic error from \sect
\ref{sec:dist}), $m_K = 9.32\pm0.02$ (\sect \ref{sec:mk}), and a ${\mathcal
K}$-correction of 0.035 mag \citep{1995AJ....109...87B}, we find $M_K =
-24.59\pm0.15$ for UGC 463.  We use the $K$-band TF relations derived by
\citetalias{2001ApJ...563..694V} to calculate $\itf$ based on measurements of
the projected rotation speed.

\begin{figure*}
\epsscale{1.1}
\plotone{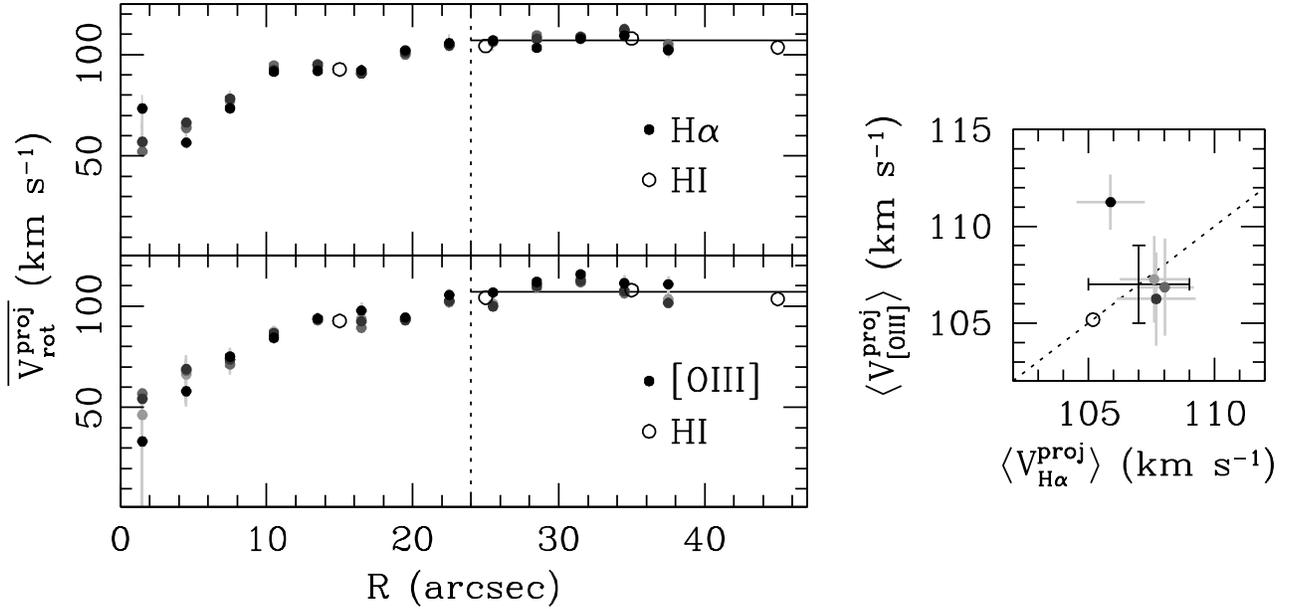}
\caption{
The projected rotation curves for all UGC 463 gas tracers determined by four
different weighting schemes (see Appendix \ref{app:ikin}): from light-gray to
black, the weights are uniform, error-based, from $\cos\gaz$, and from
$\partial\vlos/\partial i$.  {\it Left} --- The projected \halp\ (filled points
in the top panel), \oiii\ (filled points in the bottom panel), and \hone\ (open
points) rotation curves.  Measurements of $\langle\vrotproj\rangle$ are made at
$R>24\arcsec$, delineated by the vertical dotted line.  The mean
$\langle\vrotproj\rangle$ for all tracers is plotted as a solid horizontal line.
{\it Right} --- Measured $\langle\vrotproj\rangle$ for the \halp- (abcissa) and
\oiii-emitting (ordinate) gas for all four weighting schemes; the dotted lines
gives the 1:1 relation between the two tracers; $\langle\vrotproj\rangle$ for
\hone\ is plotted as an open circle on this relation.  The black error bars show
$\langle\vrotproj\rangle = 107\pm 2$ determined using all three gas tracers.
}
\label{fig:rc_itf}
\end{figure*}

We measure the projected rotation-curve for all gas tracers in UGC 463 for use
in calculating $\itf$; stellar measurements are not considered due to
significant asymmetric drift (\sect \ref{sec:rcsdisp}).  Figure \ref{fig:rc_itf}
presents $\vrotproj$ for the \halp\ and \oiii\ data resulting from all four
weighting schemes implemented in Appendix \ref{app:ikin}; \hone\ measurements
are directly from \citet{TPKMPhD}.  It also provides the error-weighted mean
measurements $\allmean{\vrotproj}$ for data at $R>24\arcsec$.  No beam-smearing
(\sect \ref{sec:beamc}) or pressure (\sect \ref{sec:vccorr}) corrections have
been applied; these are negligible considerations for the measurement of
$\allmean{\vrotproj}$.  The \halp\ rotation curve exhibits less dependence on
the applied weighting than does the \oiii\ rotation curve; however, they both
compare well with each other and with the \hone\ rotation curve, regardless of
the weighting scheme.  The \halp\ data, in particular, appear to asymptote at
$R>24\arcsec$; hence this radial region is chosen for measuring
$\allmean{\vrotproj}$.  Table \ref{tab:vsini} provides $\allmean{\vrotproj}$ for
each tracer; measurements of $\allmean{\vrotproj}$ from \halp\ and \oiii\ are
the unweighted mean of the results from all weighting schemes.  Using all
tracers, we find a mean and standard deviation of $\allmean{\vrotproj} = 107\pm
2$ \kms.

\begin{deluxetable}{ l c }
\tabletypesize{\small}
\tablewidth{0pt}
\tablecaption{Projected Rotation Speed}
\tablehead{ & \colhead{$\langle\vrotproj\rangle$} \\ \colhead{Data Set} & \colhead{(\kms)} }
\startdata
\halp & $107\pm 1$ \\
\oiii & $108\pm 2$ \\
\hone & $105\pm 2$ \\
\hline
Mean & $107\pm 2$
\enddata
\label{tab:vsini}
\end{deluxetable}

The calculation of $\itf$ and its uncertainty relies on two additional factors:
(1) the choice of the TF relation and (2) the estimation of its intrinsic
scatter.  \citetalias{2001ApJ...563..694V} created multiple samples based on a
rotation-curve- and asymmetry-based taxonomy, each sample yielding different TF
coefficients ($c_{1,\lambda}$ and $c_{2,\lambda}$) and intrinsic-scatter
estimates.  UGC 463 exhibits \hone\ and ionized-gas properties that are most
consistent with the ``RC/$FD$'' sample from \citetalias{2001ApJ...563..694V}
(see his Sections 4 and 5 for a detailed definition of this sample); UGC 463
fits within the definition of the ``RC/$FD$'' sample because its rotation curve
asymptotes to a nearly constant rotation speed and exhibits neither strong
rotation asymmetries (\sect \ref{sec:axisym}) nor signs of ongoing interaction.
Additionally, \citetalias{2001ApJ...563..694V} produced TF coefficients based on
two fiducial velocities, one measured at the rotation-curve peak ($\vmax$) and
the other where it ``flattened'' to a constant value ($\vflat$).  UGC 463
exhibits a well-defined $\vflat$ (Figure \ref{fig:rc_itf}); however, in most
cases $\vflat=\vmax$ for galaxies studied by \citetalias{2001ApJ...563..694V}.
The smallest {\it observed} scatter in the $K$-band TF relations derived by
\citetalias{2001ApJ...563..694V} (0.26 mag) was found by excluding the outlying
NGC 3992 measurements from the ``RC/$FD$'' sample (leaving measurements for 21
galaxies) and using $\vflat$ for the fiducial rotation measurement; this TF
relation is consistent with having zero {\it intrinsic} scatter.

Given the range in $c_{1,K}$ and $c_{2,K}$ for the ``RC/$FD$'' sample (with and
without NGC 3992 and based on either $\vmax$ or $\vflat$) from Table 4 of
\citetalias{2001ApJ...563..694V}, we find $28\arcdeg\leq \itf\leq 30\arcdeg$;
and we find $\epsilon(\itf) = 1\arcdeg$ and $2\arcdeg$ assuming, respectively,
0.0 and 0.2 magnitudes of intrinsic TF scatter, regardless of the assumed
$c_{1,K}$ and $c_{2,K}$.  Taking a mean across the four relevant TF relations
and assuming 0.2 magnitudes for the intrinsic TF scatter, we measure $\itf =
29\arcdeg\pm2\arcdeg$ for UGC 463; this measurement of $\itf$ is dominated by
systematic error with roughly equal contributions from the uncertainties in
$H_0$ and the TF relation.  Our conservative approach to measuring $\itf$ is
justified given that, at $M_K = -24.59\pm0.15$, UGC 463 is more luminous than
any galaxy considered by \citetalias{2001ApJ...563..694V}, far away from the
``pivot'' point of the fitted $K$-band TF relations ($M_K\sim -22$).

\subsection{Position Angle and Dynamical Center}
\label{sec:pointing}

The kinematic position angles derived in \sect \ref{sec:ikin} are very
consistent among all dynamical tracers; Table \ref{tab:kgeom} provides an
error-weighted mean value of $\pa = 68.8\pm0.3$, which is constant across the
optical disk to good approximation (as discussed in \sect \ref{sec:ikin}).  As
stated above, the PPak data affix the dynamical center to the morphological
center.

We determine the pointing of each SparsePak IFU observation relative to the
dynamical center by fitting the kinematic geometry (as in \sect \ref{sec:ikin})
with $i$ and $\pa$ fixed.  We {\it simultaneously} fit all kinematics measured
from our IFS, assuming all tracers are in co-planar rotation.  To do so, we
apply slight offsets to $\vsys$ according to the differences found in Table
\ref{tab:kgeom} such that all data can be forced to have the $\vsys$ measured
for the stars.  We also allow for asymmetric drift between the gas and stars by
simultaneously fitting different rotation curves to these components.  Finally,
we force the \oiii\ and stellar kinematics determined from the SparsePak
observations to have the same pointing center.  During the fitting procedure, we
omit velocity measurements based on the measurement error and the discrepancy
with the model, as described in Appendix \ref{app:ikin}, and weight according to
the velocity errors, as done in \citet{2003ApJ...599L..79A}.

\begin{deluxetable}{ l c c c c }
\tablewidth{0pt}
\tabletypesize{\small}
\tablecaption{Pointing Coordinates}
\tablehead{ & \colhead{$x_d$} & \colhead{$x_0$} & \colhead{$y_d$} & \colhead{$y_0$} \\ \colhead{Pointing} & \colhead{(arcsec)} & \colhead{(arcsec)} & \colhead{(arcsec)} & \colhead{(arcsec)} \\ \colhead{(1)} & \colhead{(2)} & \colhead{(3)} & \colhead{(4)} & \colhead{(5)} }
\startdata
\halp\ 02Jan02 p1 & 0.0 & $-0.1_{-0.3}^{+0.3}$ & 0.0 & $-0.5_{-0.4}^{+0.7}$ \\
\halp\ 02Jan02 p2 & 0.0 & $ 0.0_{-0.3}^{+0.3}$ & 0.0 & $-0.2_{-0.3}^{+0.7}$ \\
\halp\ 02Jan02 p3 & 0.0 & $ 0.5_{-0.2}^{+0.5}$ & 5.6 & $ 6.6_{-0.4}^{+0.6}$ \\
\halp\ 02Jan02 p4 & 4.9 & $ 6.6_{-0.5}^{+0.4}$ & 2.8 & $ 3.3_{-0.7}^{+0.7}$ \\
\halp\ 20Oct02 p1 & 0.0 & $ 0.5_{-0.3}^{+0.4}$ & 0.0 & $ 0.8_{-0.9}^{+0.6}$ \\
\halp\ 20Oct02 p2 & 0.0 & $ 1.3_{-0.2}^{+0.3}$ & 5.6 & $ 6.7_{-0.4}^{+0.5}$ \\
\halp\ 20Oct02 p3 & 4.9 & $ 5.9_{-0.2}^{+0.4}$ & 2.8 & $ 4.5_{-0.6}^{+0.4}$ \\
\mgi\  23Sep06    & 0.0 & $-1.5_{-0.3}^{+0.7}$ & 0.0 & $-0.2_{-1.4}^{+0.8}$ \\
\mgi\  PPak\tablenotemark{a} & 0.0 &         $-1.8\pm1.0$ & 0.0 &         $ 1.6\pm1.0$
\enddata
\tablenotetext{~}{{\bf Notes.}  Columns are: (1) pointing description; (2)
nominal RA dither position; (3) fitted RA position; (4) nominal DEC dither
position; and (5) fitted DEC position.  All coordinates are sky-right and
relative to the dynamical center.}
\tablenotetext{a}{The PPak coordinates are taken from \citet{TPKMPhD}.}
\label{tab:kcen}
\end{deluxetable}

Table \ref{tab:kcen} provides the resulting pointing coordinates relative to the
dynamical center for each IFS observation; as with the geometric quantities in
Table \ref{tab:kgeom}, errors are 68\% confidence limits determined using
bootstrap simulations.  Table \ref{tab:kcen} also provides the nominal
expectation for the pointings based on the dither pattern used during the
observations.  The kinematic fitting results are consistent with the dither
pattern, if allowing for $\sim1\arcsec$ systematic errors in the initial
pointing.  Moreover, reconstructed continuum images that use this pointing
geometry are in good agreement with direct images from SDSS (Figure
\ref{fig:maps}; \sect \ref{sec:maps}).

\subsection{Two-Dimensional Maps}
\label{sec:maps}

Five of the nine images in Figure \ref{fig:maps} have used an interpolation
algorithm to smooth over the interstitial regions of our IFS.  The continuum
surface-brightness maps of our IFS (labeled $\muha$ and $\mumg$) are determined
via a calibration to SDSS imaging data.  The detailed procedures used to both
perform the surface-brightness calibration and two-dimensional interpolation are
discussed in Appendix \ref{app:maps}.  These interpolated kinematic maps are
purely for illustration purposes, useful for qualitative assessments of our
registration of the dynamical center and a discussion of the two-dimensional
kinematic morphology; however, all quantitative analyses herein have been
performed using the direct fiber measurements, the IFU astrometric tables, and
our derived pointings.

The first column of Figure \ref{fig:maps} demonstrates the excellent agreement
among the reconstructed continuum images and the direct SDSS $g$-band image.
Indeed, the central contour of both $\muha$ and $\mumg$ directly overlap and are
centered on the NED-provided coordinate of UGC 463. The detailed spiral
structure is apparent in, particularly, the $\mumg$ image due to the small PPak
fibers.  The isovelocity contours of the gas data appear to exhibit streaming
motions associated with the spiral arm toward the south-west of the galaxy
center; this is less apparent in the stellar data.  Additionally, the effect of
asymmetric drift is seen in the stellar velocity field as the ``linearization''
of the isovelocity contours toward the galaxy center, which is due to a
shallower increase in the stellar rotation curve, corresponding to the steep
decrease in the stellar velocity dispersion, toward larger radius.  We further
explore the kinematic axisymmetry in \sect \ref{sec:axisym}.

\section{Azimuthally Averaged Kinematics}
\label{sec:spk+ppk}

Analyses in \sects \ref{sec:disk} and \ref{sec:mass} assume UGC 463 is axially
symmetric, considering only the azimuthally averaged kinematics that we derive
in the following subsections.  We apply beam-smearing corrections in \sect
\ref{sec:beamc} \citep[beam-smearing corrections for the \hone\ data are
described by][]{TPKMPhD} such that kinematic data from different
instruments can be combined.  In \sect \ref{sec:axisym}, we assess the degree of
dynamical symmetry by comparing approaching- and receding-side kinematics.
Finding no substantial asymmetries, we discuss the azimuthally averaged
kinematics in \sect \ref{sec:rcsdisp}.

\subsection{Beam-Smearing Corrections}
\label{sec:beamc}

Our beam-smearing corrections require a characterization of the beam profile,
the convolution of the point-spread function and the fiber aperture.  Since no
significant jitter was detected among or during the individual IFS observations,
the effective fiber aperture is given by the plate-scale (yielding $2\farcs7$
for PPak and $4\farcs7$ for SparsePak).  In Appendix \ref{app:maps}, we find
that the seeing of the SDSS imaging data --- $1\farcs5$ in $g$-band and
$1\farcs2$ in $r$-band --- is very close to the effective seeing of our
SparsePak IFS; \citet{TPKMPhD} provides a direct seeing measurement of
$1\farcs7$ for our PPak IFS.  Our beam-smearing corrections change negligibly
over the range of measured seeing; therefore, we simply adopt $1\farcs5$ seeing
to calculate all beam-smearing corrections.

Our approach to beam-smearing corrections \citepalias{2010ApJ...716..234B}
depends on comparing our UGC 463 data to models of the intrinsic
surface-brightness ($I$), velocity ($V$), and velocity-dispersion ($\sigma$)
distributions.  SDSS imaging data provide the model surface-brightness
distribution; $g$-band data are used for \mgi-region IFS and $r$-band are used
for \halp-region IFS.  We assume a polyex parameterization to model the
intrinsic rotation curve \citep{2002ApJ...571L.107G}.  The gas velocity
dispersion is assumed to be constant with radius and isotropic.  Only the \halp\
data are used to describe the velocity-dispersion profile; beam-smearing
corrections are marginally different if the \oiii\ dispersions are used instead.
For the stars, we adopt SVE axial ratios of $\alpha = \sigz/\sigr = 0.6$ and
$\beta = \sigp/\sigr$ as determined by the epicycle approximation (\sect
\ref{sec:sve}; equation \ref{eq:ea}).  The model radial profile for the
azimuthally averaged $\sstar$ ($\azmean{\sstar}$) combines an exponential
function with a cubic Legendre-polynomial perturbation at small radius; although
somewhat ad hoc, this form allows for deviations from a nominal exponential
while enforcing a well-behaved, exponential form at large radius.  For UGC 463,
deviations of $\azmean{\sstar}$ from an exponential form are small and
irrelevant to the calculated beam-smearing corrections.

Beam-smearing corrections are calculated as follows:  A fit to the uncorrected
data is used to generate a seed model of the intrinsic galaxy kinematics, which
is then ``observed'' by integrating a set of Gaussian line profiles, defined by
($I,V,\sigma$), discretely sampled over the beam profile of each fiber to create
a synthetic data set \citep{KBWPhD}.  The velocity and velocity-dispersion
corrections are the difference between this synthetic dataset and the model
value at the center of the fiber, and they are primarily correlated
with the velocity gradients across the fiber face.  The beam-smearing effects are
largest toward the galaxy center where the rotation curve is most steeply rising
and the azimuthal coverage of each fiber is largest.  The trend of the
correction is to {\it increase} the measured rotation speed and {\it decrease}
the measured velocity dispersion.  We converge to a set of beam-smearing
corrections iteratively by updating the model of the intrinsic galaxy
kinematics, done by fitting the corrected observational data, and minimizing the
difference between the observed and synthetic data sets.

Monte Carlo simulations demonstrate that the random errors in the beam-smearing
corrections are $\lesssim 10$\%; systematic errors, estimated by calculating
beam-smearing corrections using SVE-shape extrema, are typically much smaller.
Therefore, we adopt the quadrature sum of a 10\% random error and the estimated
systematic error for each fiber as the error in the beam-smearing correction;
the error is always 10\% for the gas data.  Although lower than the upper-limit
used in \citetalias{2010ApJ...716..234B}, this reduction in error has little
effect on the error budget.

The correspondence of the uncorrected \halp, \oiii, and \hone\ rotation curves
in Figure \ref{fig:rc_itf}, despite the factor of $\gtrsim3$ difference in the
beam size among the data sets, suggests beam-smearing corrections should be
small; this expectation is in agreement with our direct beam-smearing
calculations.  For the ionized gas data, corrections to $\vlos$ are less than 2
\kms\ for 93\% of the data, with a maximum correction of 14 \kms.  For the
stellar data, $\vlos$ corrections are less than 2 \kms\ for 84\% and 99\% of the
SparsePak and PPak data, respectively; the maximum correction is 7 \kms\ for
SparsePak and 8 \kms\ for PPak.  Corrections to $\sstar$ are less than 5\% for
91\% and 99\% of the SparsePak and PPak data, respectively; the maximum
correction is 41\% for SparsePak and 29\% for PPak.  Corrections to $\sstar$ are
typically less than $\epsilon(\sobs)$, with the only exceptions occurring near
the galaxy center.

\subsection{Axial Symmetry}
\label{sec:axisym}

\begin{figure*}
\epsscale{0.9}
\plotone{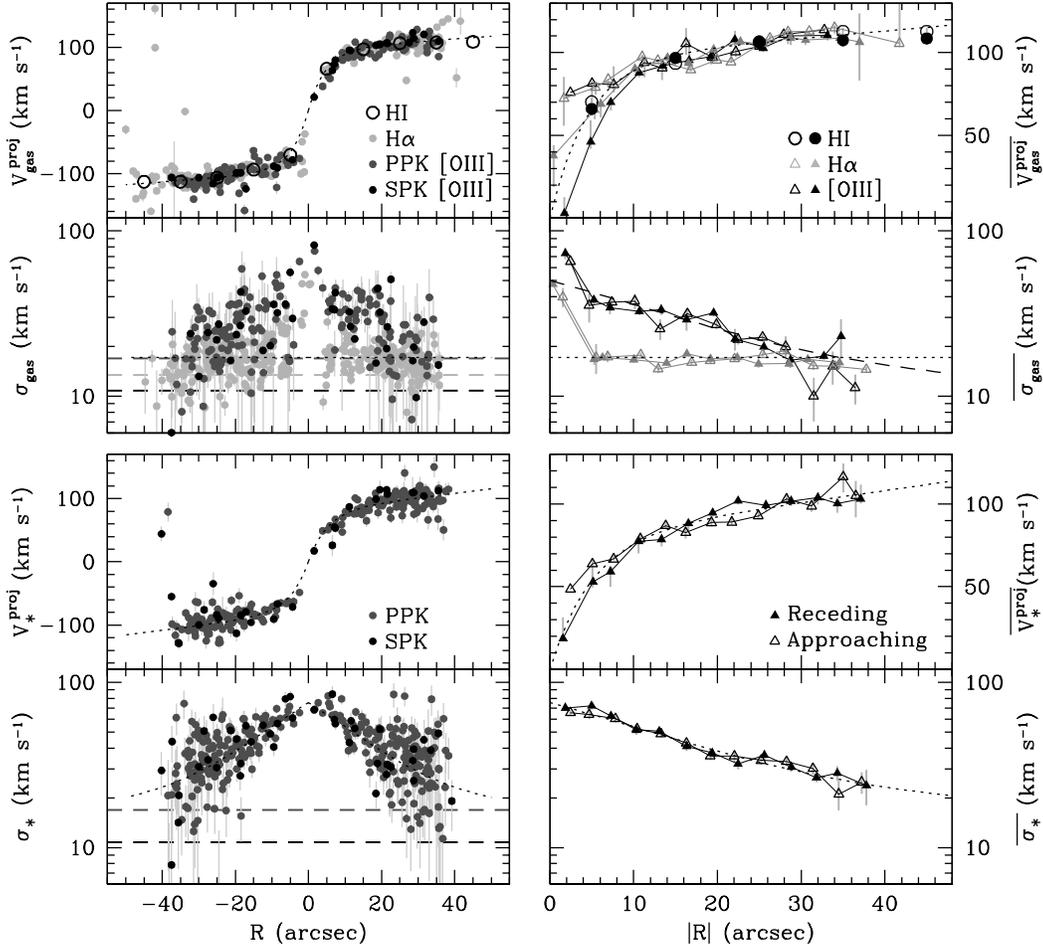}
\caption{
Corrected gaseous ($\vgas^{\rm proj}$, $\sgas$) and stellar ($\vstar^{\rm
proj}$, $\sstar$) kinematics for all tracers and instruments.  Panels to the
left present individual fiber measurements, whereas quantities in the right
panels are azimuthally averaged over the approaching (open symbols) and receding
(filled symbols) sides.  Horizontal dashed lines mark the mean instrumental
dispersion for the SparsePak \halp-region data (light gray), PPak \mgi-region
data (dark gray), and SparsePak \mgi-region data (black).  Model expectations
used to calculate the beam-smearing corrections are shown as dotted lines.  The
diagonal dashed line in the $\azmean{\sgas}$ panel is equal to $2/3$ of the
model $\azmean{\sstar}$ shown in the bottom-right panel.
}
\label{fig:beamc}
\end{figure*}

Figure \ref{fig:beamc} presents individual-fiber kinematics after correcting for
instrumental-broadening and beam-smearing, with point types indicating the
tracer and instrument.  For measurements located at in-plane azimuths within
$\pm60\arcdeg$ of the major axis, we deproject $\vlos$ to $\vrotproj$; rotation
velocities are plotted regardless of whether or not they were rejected from the
velocity-field fitting discussed in \sect \ref{sec:geom}.  Velocity-dispersion
data include all measurements made at any azimuth.  We find the kinematic
measurements from the different IFUs to be very well matched.  Data are
separated according to the approaching (negative radii) and receding sides.
Figure \ref{fig:beamc} also overlays mean quantities from either side of the
minor axis.  We determine $\azmean{\vrotproj}$ for the ionized gas and stars
using the velocity-field fitting procedure described in \sect \ref{sec:ikin},
with rejection and error-based weighting.  Errors in $\azmean{\vrotproj}$ are
68\% confidence limits calculated using bootstrap simulations.  The values of
$\azmean{\sgas}$ and $\azmean{\sstar}$ are error-weighted means.

The overlay of the binned data in Figure \ref{fig:beamc} from its two sides show
that UGC 463 exhibits little kinematic asymmetry, justifying our assumption of
axisymmetry in the following sections.  In detail, the ionized gas rotation
curves exhibit the strongest asymmetry at $R\lesssim5\arcsec$.  UGC 463 is
morphologically classified as an SABc galaxy \citepalias{2010ApJ...716..198B},
suggesting that this low-level asymmetry may be due to non-circular motions
imposed by the presence of a weak bar.  This kinematic asymmetry may also be
reflected in the stellar data at marginal significance.  The velocity-dispersion
profiles for both the gas and stars are very symmetric at all radii, more so
than the rotation velocities.

\subsection{Radial Kinematic Profiles}
\label{sec:rcsdisp}

\begin{figure}
\epsscale{1.1}
\plotone{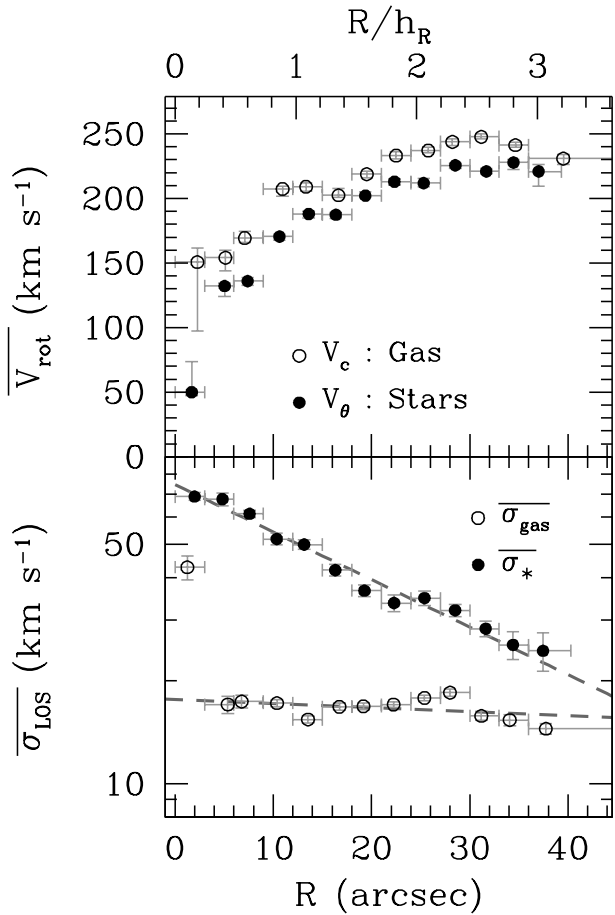}
\caption{
Kinematic profiles for UGC 463 averaged over all azimuth.  {\it Top} ---
Circular speed ({\it open circles}) and stellar tangential speed ({\it filled
points}), as described in the text.  {\it Bottom} --- Gas ({\it open circles})
and stellar ({\it filled points}) LOS velocity dispersions.  The dashed gray
lines are the best-fitting linear and exponential functions for the gas and
stellar data, respectively.
}
\label{fig:rcaxisym}
\end{figure}

Figure \ref{fig:rcaxisym} provides the azimuthally averaged kinematics analyzed
in \sects \ref{sec:disk} and \ref{sec:mass}, following the same procedure as
described in the previous section but over all azimuth.  Stellar kinematics
combine both SparsePak and PPak observations, and ionized-gas kinematics
incorporate all tracers from both instruments.  Rotation-velocity errors (68\%
confidence limits) are determined using bootstrap simulations, not by, e.g.,
considering the difference in rotation speed between the two sides of the
rotation curve; that is, we assume the disk contains no asymmetries such that
any asymmetries manifest themselves as an increased error in the measured
rotation speed via bootstrap simulations.  The ionized-gas rotation velocity at
the largest radius is averaged with the radially overlapping \hone\ measurement
to slightly extend the radial coverage.  The circular-speed curve provided in
Figure \ref{fig:rcaxisym} results from applying pressure corrections to the gas
rotation curve, as described in the next section.

Measurements of $\azmean{\sgas}$ include only the ionized-gas kinematics, not
the colder \hone.  Although the \oiii\ velocity dispersion is significantly
larger than the \halp\ velocity dispersion and decreasing with radius (Figure
\ref{fig:beamc}), the azimuthally averaged $\sgas$ is very nearly the same as
the \halp\ velocity dispersion.  This is because of the error-weighting and the
significantly higher quality of the \halp\ velocity dispersions.  We find
$\allmean{\sgas} = 16.6\pm1.1$ \kms\ when excluding the datum near the galaxy
center, which is nearly constant as a function of radius.  This result is
comparable to similar measurements made in other face-on spiral galaxies by
\citet{2006ApJS..166..505A} and, as shown by these authors, dominated by
turbulence given the expected thermal pressure.  Physically, the difference
between the \halp\ and \oiii\ dispersions may be related to the different
energetics involved in generating the two lines; for example, \oiii\ emission
may be naturally biased toward more turbulent regions of the ISM.  From Figure
\ref{fig:beamc}, we note that the \oiii\ dispersion is surprisingly well fit by
a radial profile following $2\azmean{\sstar}/3$ over a large radial range; it is
of interest to explore the reason for this relationship by comparing with other
galaxies.

\subsubsection{Circular-Speed Corrections}
\label{sec:vccorr}

\citet{2010ApJ...721..547D} derive
\begin{eqnarray}
\vc^2 & = & \vrot^2 - \sgas^2 \left(\frac{d\ln\sgas^2}{d\ln R} +
\frac{d\ln\sdg}{d\ln R}\right) \nonumber \\ & = & \vrot^2 + \delta_P \sgas^2,
\label{eq:vgcor}
\end{eqnarray}
where $\sgas$ is assumed to be dominated by turbulence and produced by an
isotropic gas velocity ellipsoid \citep[cf.][]{2009MNRAS.392..294A}.  We,
thereby, correct our gas rotation curve to the circular speed using our
measurements of the $\azmean{\sgas}$ in Figure \ref{fig:rcaxisym} and $\sdg$ in
Figure \ref{fig:sdgas}.  

One expects $\sgas$ and $\sdg$ to decrease with radius such that $\delta_P >
0$.\footnote{
When $\sgas \propto \sdg \propto R^{-1}$ ($\delta_P = 3$), equation
\ref{eq:vgcor} reduces to a similar equation used by \citet{2003ApJ...587L..19S}
following from virial theorem arguments.
}  \citet{2010ApJ...721..547D} propose exponential functions for use in
calculating $\delta_P$, which is appropriate for our $\sdg$ measurements;
however, we adopt the linear function plotted in Figure \ref{fig:rcaxisym} for
$\sgas$.  We calculate $0.2 \leq \delta_P \leq 3.3$ such that the circular-speed
corrections range from $0.4\leq (\vc-\vrot) \leq 1.6$ \kms, always below the
measurement error in the gas rotation speed.  If we treat the circular-speed
corrections independently for \oiii\ and \halp, we find the \halp-based and
\oiii-based circular speeds to be {\it more} consistent than if one correction
is applied to both.  However, the combined circular-speeds are roughly
independent of whether or not the \halp\ and \oiii\ data are treated separately.

\section{The Disk}
\label{sec:disk}

Using the data described above, we measure physical properties of the disk of
UGC 463.  In summary, we determine the kinematic scale length, $h_{\sigma}$,
defined as the $e$-folding length of $\azmean{\sstar}$ (\sect \ref{sec:hsigma});
we determine the shape of the SVE such that we can calculate $\sigz$ based on
our measurements of $\azmean{\sstar}$ (\sect \ref{sec:sve}); we determine the
total dynamical disk mass surface density, $\sddisk$, using equation 9 from
\citetalias{2010ApJ...716..234B} (\sect \ref{sec:sdd}); we calculate the stellar
mass surface density, $\sds$, by removing contributions to $\sddisk$ from
atomic- and molecular-gas (\sect \ref{sec:sds}); we calculate the stability of
the isolated gaseous and stellar disks, as well as a quantity for the
multi-component disk (\sect \ref{sec:Q}); and, finally, we measure the dynamical
and stellar mass-to-light ratios in $K$-band, $\mlskdisk$ (\sect \ref{sec:ml}). 

\subsection{Kinematic Scale Length, $h_{\sigma}$}
\label{sec:hsigma}

Measurements of $\azmean{\sstar}$ are well fit by an exponential profile; the
best-fit exponential has a central dispersion of $74.7\pm2.4$ \kms\ and
$h_{\sigma} = 31\farcs3\pm1\farcs8$ as plotted in Figure \ref{fig:rcaxisym}.
Given that $h_{\sigma} \sim 2.6 h_R$, either $\mls$ (or more appropriately the
dynamical disk mass-to-light ratio, $\mldyn$) or the scale height, $h_z$, may be
increasing exponentially with radius; the implied $e$-folding length is $h_R
h_{\sigma}/(h_{\sigma} - 2h_R) = 57\farcs5$.  Over the radial range of our data,
this suggests a factor of $\sim 2.3$ increase in either $\mldyn$
\citep[consistent with, e.g.:][]{2001ApJ...550..212B, 2009MNRAS.400.1181Z} or
$h_z$ \citep[compare with edge-on galaxy photometry from,
e.g.:][]{1981A&A....95..105V, 1997A&A...320L..21D, 2002A&A...390L..35N,
2002A&A...389..795B, 2006AJ....131..226Y, 2009MNRAS.396..409S}.  The effects of
radial variations in $\mldyn$ and/or $h_z$ are further discussed in \sect
\ref{sec:ml}.

\subsection{Stellar Velocity Ellipsoid Axial Ratios, $\alpha$ and $\beta$}
\label{sec:sve}

Calculations of disk mass surface density require measurements of $\sigz$.  We
obtain $\sigz$ by correcting $\azmean{\sstar}$ for the shape of the SVE, which
can be directly measured using stellar and ionized-gas kinematics
\citep{2003AJ....126.2707S, 2008MNRAS.388.1381N, KBWPhD}.  We define the two
axial ratios of the SVE to be $\alpha = \sigz/\sigr$ and $\beta = \sigp/\sigr$.
In the limit where the stellar orbits are nearly circular, the epicycle
approximation (EA) yields
\begin{equation}
\beta^2 = \beta_{\rm EA}^2 \equiv \frac{1}{2} \left(
\frac{\partial\ln\vt}{\partial \ln R} + 1\right),
\label{eq:ea}
\end{equation}
where $\vt$ is the tangential speed of the stars.  Also, assuming that UGC 463
is axially symmetric with an SVE that is always aligned with the cylindrical
coordinate axes, one can approximate
\begin{equation}
\dad \equiv \frac{(\vc^2-\vt^2)\sin^2i}{\azmean{\sstar^2}} \approx
\frac{\tan^2i}{\azmean{\gamma}\alpha^2} \left(\frac{4R}{h_{\sigma}} + \beta^2
- 1 \right)
\label{eq:dad}
\end{equation}
from the asymmetric drift (AD) equation \citep{2008gady.book.....B}, where we
use
\begin{equation}
\azmean{\gamma} \equiv \frac{\azmean{\sstar^2}}{\sigz^2\cos^2i} =
1+\frac{\tan^2i}{2\alpha^2}(1+\beta^2)
\label{eq:gamma}
\end{equation}
as defined in \citetalias{2010ApJ...716..234B} and $\alpha$ and $\beta$ are
constant over radial regions where $\partial\ln\azmean{\sstar^2}/\partial\ln R
\approx -2R/h_{\sigma}$.  Using these equations, each radially binned
measurement of $\vc$, $\vt$, and $\azmean{\sstar^2}$ provides a direct
measurement of $\beta_{\rm EA}$ and $\alpha = \alpha_{\rm AD}$, assuming that
the derivatives $\partial\ln\vt/\partial \ln R$ and $\partial
\ln\azmean{\sstar^2} /\partial\ln R$ do not strongly deviate from the
fitted-model expectations; for all measurements we assume $\azmean{\sstar^2} =
\azmean{\sstar}^2$.

\begin{figure*}
\epsscale{0.9}
\plotone{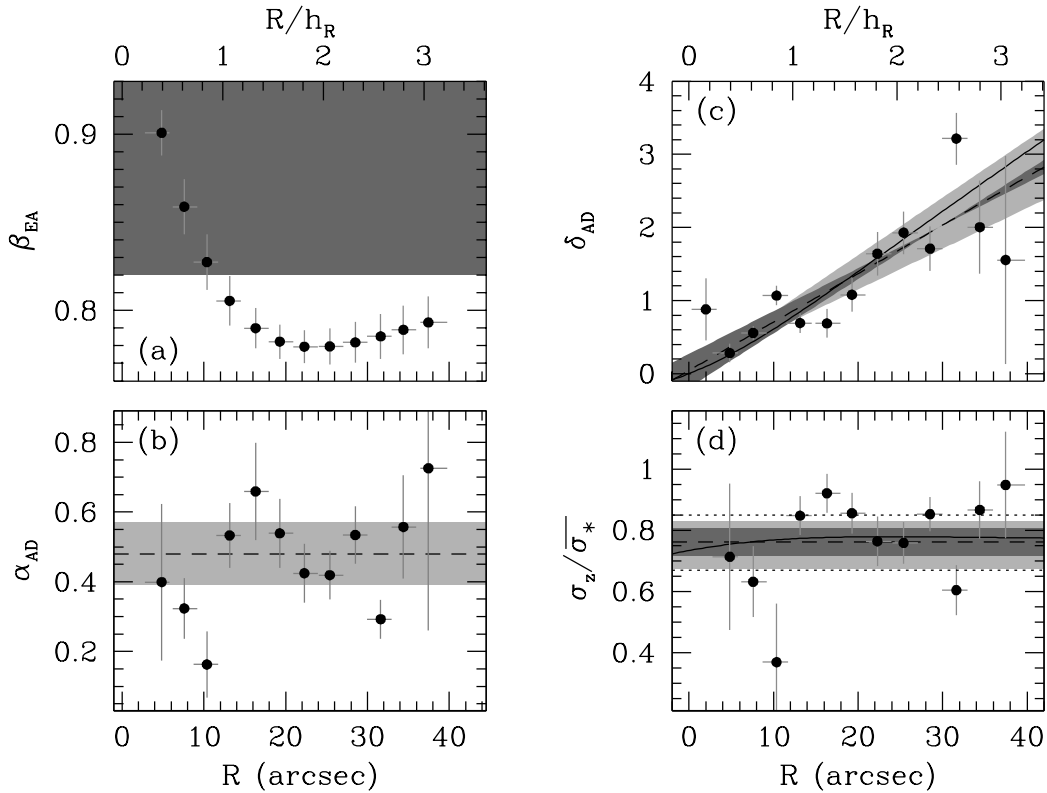}
\caption{
SVE results for UGC 463.  Dark- and light-gray regions, in any panel, show the
error range in $\beta = 1.04\pm0.22$ and $\alpha = 0.48\pm0.09$, respectively,
as fitted to the $\dad$ measurements in panel (c); the dashed line, in any
panel, results when these axial ratios are constant at all radii.  The solid
line assumes $\beta = \beta_{\rm EA}$ and $\alpha = 0.44\pm0.07$ (see text).
Panel (a) provides $\beta_{\rm EA}$, calculated assuming the model stellar
rotation curve from Figure \ref{fig:beamc}.  Panel (b) provides $\alpha_{\rm
AD}$ calculated assuming $\beta = \beta_{\rm EA}$ from panel (a) and
measurements of $\dad$ from panel (c).  Panel (d) provides measurements of
$\sigz/\azmean{\sstar}$ assuming $\alpha=\alpha_{\rm AD}$ and $\beta=\beta_{\rm
EA}$.  Dotted lines in panel (d) give the error range in $\sigz/\azmean{\sstar}
= 0.76\pm0.09$.
}
\label{fig:sve}
\end{figure*}

In Figure \ref{fig:sve}(a), we calculate $\beta_{\rm EA}$ for each radial ring
using the parameterized description of $\vt$ shown in Figure \ref{fig:beamc}.
Systematic errors of greater than 15\% are expected due to non-circular stellar
orbits \citep{1975ApJ...195..333V, 1991dodg.conf...71K, 1999AJ....118.1190D} and
non-axisymmetric streaming motions in the disk near spiral arms
\citep{2006MNRAS.373..197V,2008MNRAS.383..817V}; such effects will dominate the
$\lesssim 2$\% random error shown in Figure \ref{fig:sve}(a).  Adopting $\beta =
\beta_{\rm EA}$, Figure \ref{fig:sve}(b) provides $\alpha_{\rm AD}$ given the
observed values of $\dad$ in Figure \ref{fig:sve}(c).  Note that the calculation
of $\alpha_{\rm AD}$ near the galaxy center provides an imaginary result
($\alpha^2<0$) and is not shown in Figures \ref{fig:sve}(a) or \ref{fig:sve}(b).
Excluding this datum, the mean values are $\allmean{\alpha_{\rm AD}}=0.46$ and
$\allmean{\beta_{\rm EA}} = 0.81$.

We fit equation \ref{eq:dad} to our measurements of $\dad$ in Figure
\ref{fig:sve}(c) under the assumption that the SVE shape is constant (such that
$\dad$ is linear in radius).  We note that $\dad$ is inversely proportional to
$\alpha$, and $\dad$ pivots about $\dad = 2$ --- where $\dad$ is independent of
$\beta$ --- such that an increase in $\beta$ decreases the slope of $\dad(R)$.
Errors in $\alpha$ and $\beta$ are estimated using the quadrature sum of two
quantities: (1) the standard deviation returned by a Monte Carlo sampling of the
normal distributions $i=27\arcdeg\pm2\arcdeg$ and $h_{\sigma} =
31\farcs3\pm1\farcs8$; and (2) the mean of the error determined from a set of
500 bootstrap simulations performed for each instance of $i$ and $h_{\sigma}$.
We find $\alpha = 0.48\pm0.09$ and $\beta = 1.04\pm0.22$, shown as the dashed
line in Figure \ref{fig:sve};\footnote{
Note that we do not use the average nomenclature for these measurements of
$\alpha$ and $\beta$, unlike $\allmean{\alpha_{\rm AD}}=0.46$ and
$\allmean{\beta_{\rm EA}} = 0.81$.  These fitted parameters describe average
properties of the disk but they are not strictly averages of multiple
measurements.
} errors are shown as light- and dark-gray regions for $\alpha$ and $\beta$,
respectively.  If we force $\beta = \beta_{\rm EA}$ when fitting $\dad$, we find
$\alpha = 0.44\pm0.07$ with a fit shown by the solid line in Figure
\ref{fig:sve}.

Figure \ref{fig:sve}(a) shows that $\allmean{\beta_{\rm EA}} = 0.81$ and
$\beta=1.04\pm 0.22$ are different at the level of slightly more than the random
errors.  Although interesting, the relevance of this difference to our
subsequent analysis is primarily with respect to our conversion of
$\azmean{\sstar}$ to $\sigr$ (see \sect \ref{sec:Q}) and $\sigz$.  Figure
\ref{fig:sve}(d) provides individual measurements of the ratio
$\sigz/\azmean{\sstar} = (\azmean{\gamma}\cos^2i)^{-1/2}$ using $\alpha_{\rm
AD}$ and $\beta_{\rm EA}$.  Adopting $i=27\arcdeg\pm2\arcdeg$, $\alpha =
0.48\pm0.09$, and $\beta = 1.04\pm0.22$, we calculate the constant value
$\sigz/\azmean{\sstar} = 0.76\pm0.09$, which is identical to
$\allmean{\sigz/\azmean{\sstar}}$ calculated using the individual measurements
based on $\alpha_{\rm AD}$ and $\beta_{\rm EA}$ when excluding the datum near
the galaxy center.  Assuming $\alpha = 0.44\pm0.07$ and $\beta = \beta_{\rm EA}$
is insignificantly different from $\sigz/\azmean{\sstar} = 0.76\pm0.09$, as
shown by the solid line in Figure \ref{fig:sve}(d).  Moreover, fitting the
azimuthal variation in $\sstar/\azmean{\sstar}$ averaged over the full radial
range, akin to the analysis done by \citet{1997MNRAS.288..618G,
2000MNRAS.317..545G} and \citet{2003AJ....126.2707S}, provides $\beta\sim0.9$,
consistent with the error in our measurement based on $\dad$ and, therefore,
insignificant to the calculation of $\sigz/\azmean{\sstar}$.

\subsection{Dynamical Disk-Mass Surface Density, $\sddisk$}
\label{sec:sdd}

We calculate $\sddisk$ using equation 9 from \citetalias{2010ApJ...716..234B},
which is fundamentally $\sddisk = \sigz^2/\pi k G h_z$.  We calculate $\sigz$
using $\azmean{\sstar}$ and our SVE measurements above; the oblateness ratio
$q=h_R/h_z$, given by equation 1 in \citetalias{2010ApJ...716..234B}, provides
the vertical scale height, $h_z$, at the distance $D$; and the effect of the
assumed vertical mass distribution is quantified by $k$.
\citet{1988A&A...192..117V} calculate the value of $k$ for three vertical mass
distributions: exponential ($k=1.5$), sech ($k=1.71$), and sech$^2$ ($k=2$).  As
discussed in \citetalias{2010ApJ...716..234B}, our nominal approach is to adopt
a purely exponential disk as a reasonable approximation for the composite
(gas$+$stars) density distribution; thus, of the three density distributions
discussed by \citet{1988A&A...192..117V}, we are effectively maximizing the
measurement of $\sddisk$ and, hence, $\mlskdisk$.  In
\citetalias{2010ApJ...716..234B}, we suggested a 14\% systematic error in $k$
based on the range in $k$ among the exponential, sech, and sech$^2$
distributions.  For the case of UGC 463, we increase this to $k = 1.5\pm0.3$,
allowing for a relatively massive gas disk (low $k$) or spherical DM halo (high
$k$); see further discussion in \sects \ref{sec:mlsys} and \ref{sec:mass}.

\begin{figure}
\epsscale{1.1}
\plotone{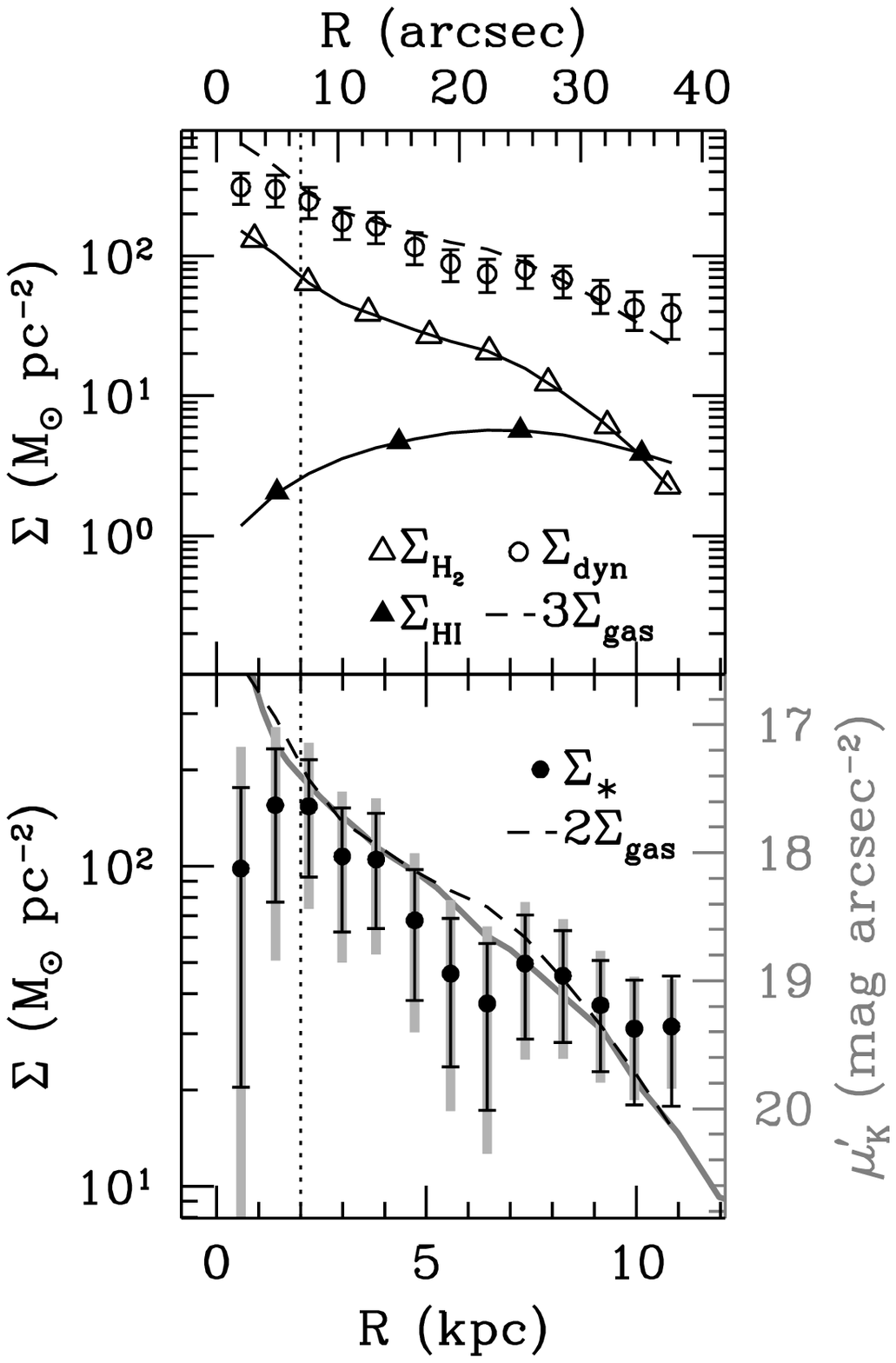}
\caption{
Mass surface density measurements for UGC 463.  The total dynamical ($\sddisk$;
{\it open circles}), \hone\ ($\sdhi$; {\it filled triangles}), \molh\ ($\sdmh$;
{\it open triangles}), and stellar ($\sds$; {\it filled circles}) mass surface
densities are plotted as a function of radius.  Random errors ({\it black error
bars}) are plotted for both $\sddisk$ and $\sds$; systematic errors ({\it
light-gray error bars}) are plotted only for $\sds$.  As described in the text,
dashed lines provide $3\sdg$ and $2\sdg$ in the top and bottom panels,
respectively.  The vertical dotted line marks the radius beyond which the
S{\'e}rsic profile describing the intrinsic central light concentration (\sect
\ref{sec:muk}) contributes less than 3\% to $\mu^{\prime}_K$.  We overplot
$\mu^{\prime}_K$ ({\it thick gray line}) in the bottom panel with a scaling such
that $\sds$ data falling directly on the surface-brightness measurements have
$\mlskdisk \sim 0.2$.
}
\label{fig:sdd}
\end{figure}

We use the scale length fitted to $\mu^{\prime}_K$ to calculate $q$ and $h_z$ as
provided in Table \ref{tab:expdisk}; the band used to define $h_R$ is
insignificant to the result.  Figure \ref{fig:sdd} plots the resulting $\sddisk$
calculations along with the measurements of $\sdhi$ and $\sdmh$ from \sects
\ref{sec:vla} and \sect \ref{sec:xco}, respectively.  Random (including
contributions from $\vsys$, $m_K$, $i$, $h_R$, $\azmean{\sstar}$, $\alpha$, and
$\beta$) and systematic (including contributions from $q$, $k$, and $D$) errors
are calculated separately; the former are dominated by the error in the SVE and
the latter have roughly equal contributions from $k$ and $q$ ($\sim 20-25$\%).
For clarity, Figure \ref{fig:sdd} shows neither the systematic error in
$\sddisk$ nor any errors in $\sdhi$ and $\sdmh$; the systematic errors in
$\sddisk$ are comparable to the random errors.  Computing the total gas mass
surface density, we find a very reasonable fit to $\sddisk$ by scaling $\sdg$ by
a factor of 3, akin to the correlation of $\sdg$ with $\mu^{\prime}_K$ shown in
Figure \ref{fig:sdgas}; $\sdhi$ and $\sdmh$ have been interpolated to the radii
of the $\azmean{\sstar}$ measurements for this comparison.

\subsection{Stellar Disk-Mass Surface Density, $\sds$}
\label{sec:sds}

Calculations of $\mlskdisk$ require a decomposition of $\sddisk$ into stellar
and non-stellar components.  Five mass components contribute to $\sddisk$:
stars, atomic gas, molecular gas, dust, and other non-stellar matter.  However,
we account for only the first three listed; the others are subsumed in the
``stellar'' mass surface density, $\sds$, for the following reasons.  First, our
data do not provide dynamically differentiable measurements of the
stellar-remnant and other DM (baryonic or otherwise) mass.  Second, the {\it
total} dust mass of normal star-forming (or even star-bursting) spiral galaxies
is expected to be less than 10\% (and more typically $\sim1$\%) of the total
hydrogen mass \citep{2007ApJ...663..866D}; therefore, it should not be
dynamically important to the mass surface density at scales relevant to our
analysis.  

Our measurements of $\sds = \sddisk - 1.4(\sdhi + \sdmh)$ are presented in the
bottom panel of Figure \ref{fig:sdd}.  As with $\sddisk$, we calculate random
and systematic errors in $\sds$ separately; compared with $\sddisk$, $\sds$
includes additional random error from $\imi$ and $\sdhi$ and systematic error
from the $\imi/\ico$ calibration and $\xco$.  Given the modest errors in $\imi$
and $\sdhi$, the random error in $\sds$ is dominated by the errors in $\sigz$;
and, despite the large systematic uncertainties in $\imi/\ico$ and $\xco$
(roughly 30\% for both), the systematic errors in $\sds$ are dominated by the
errors in $k$ and $q$.  We plot our $\sds$ measurements against $2\sdg$ and
$\mu^{\prime}_K$ (\sect \ref{sec:muk}); the scale of the plot is such that all
$\sds$ measurements falling on the $\mu^{\prime}_K$ profile have
$\mlskdisk\sim0.2$.  We find that $\sds(R)$ is roughly consistent with both
$2\sdg$ and $\mu^{\prime}_K$ at the level of the random errors.

\subsection{Global Stability, $Q$}
\label{sec:Q}

We calculate the global stability, $Q$, of the gaseous and stellar disks
separately and for the composite disk using our mass-surface-density and
velocity-dispersion measurements.\footnote{
Subscripts of $Q$ differentiate between each derivation and disk component.
}  These calculations are of interest to our dynamical study of UGC 463 given
our measurement of a systematically low $\sddisk$ with respect to a maximal disk
(see \sect \ref{sec:mass}):  A dependence between spiral-arm multiplicity and
disk maximality is expected \citep{1981seng.proc..111T, 1987A&A...179...23A}
based on stability arguments within the context of swing-amplification theory
--- see reviews by \citet{1984PhR...114..319A} and \citet{2010arXiv1006.4855S}.
Qualitatively, one expects a higher spiral-arm multiplicity for lower surface
density disks of a fixed rotation curve.  This expectation is in line with our
measurement of a submaximal disk for and the three-arm multiplicity of UGC 463
(see Figure \ref{fig:maps}); however, it should be noted that we find
submaximal disks for all 30 galaxies we have studied so far
\citep[\pV;][]{TPKMPhD}, regardless of their spiral structure.  Stability issues
will be discussed at greater length based on our full survey stemming from the
simple analysis given as an example here for UGC 463.

Assuming a razor-thin disk, \citet{1964ApJ...139.1217T} derived
\begin{eqnarray}
\qtg & = & \frac{\kappa\ c_s}{\pi\ G\ \sdg} \label{eq:qtg}\\
\qts & = & \frac{\kappa\ \sigr}{3.36\ G\ \sds}, \label{eq:qts}
\end{eqnarray}
where
\begin{equation}
\kappa^2 = 2\frac{\vc}{R}\left( \frac{\vc}{R} + \frac{d\vc}{dR}\right)
\label{eq:kap}
\end{equation}
is the epicyclic frequency and $c_s$ is the sound speed in the gas; such disks
should be stable if $Q_{\rm T} > 1$.  Following \citet{1984ApJ...276..114J} by
treating both the gaseous and stellar disks as fluids,
\citet{2001MNRAS.323..445R} derived a combined stability criterion
\begin{equation}
\frac{1}{\qr} = \frac{2}{\qrs}\ \frac{\omega}{1+\omega^2} + \frac{2}{\qrg}\
\frac{r\omega}{1+r^2\omega^2},
\label{eq:qr}
\end{equation}
where $\qrs/\qts = 3.36/\pi$, $\qrg = \qtg$, $\omega = \kl\sigr/\kappa$, $\kl =
2\pi/\lambda$ is the wavenumber of the instability, and $r= c_s/\sigr$; this
multi-component, razor-thin disk is stable if $\qr > 1$.  The finite thickness
of disks systematically increases the stability over that estimated via the
equations above; however, finite-thickness corrections are small
\citep{1992MNRAS.256..307R}.

\begin{figure}
\epsscale{1.1}
\plotone{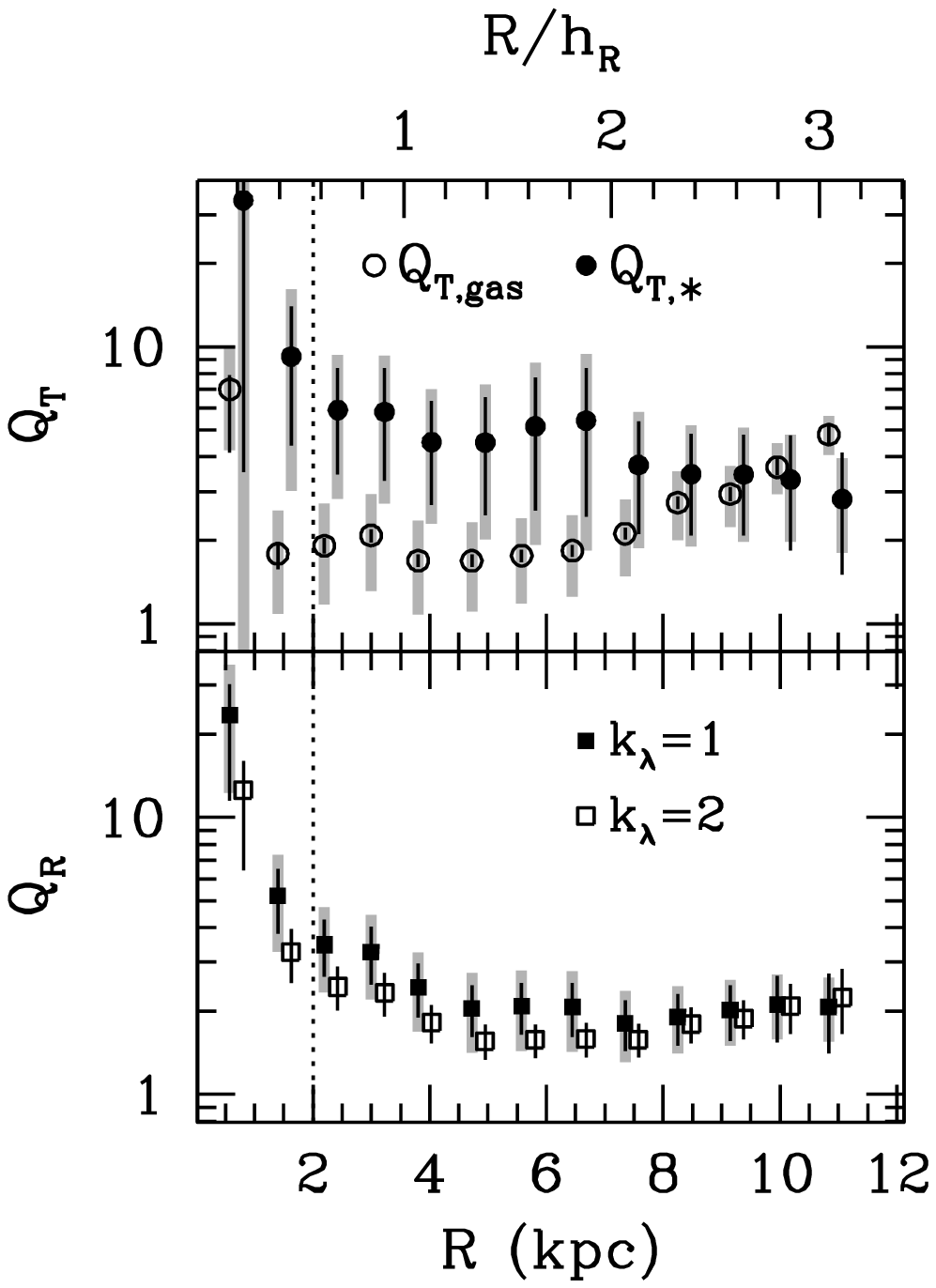}
\caption{
Disk stability measurements for the isolated gaseous ({\it open circles}) and
stellar ({\it filled points}) disks following from equations derived by
\citet[][$Q_{\rm T}$, {\it top}]{1964ApJ...139.1217T} are plotted in the top
panel.  The multi-component disk stability from \citet[$Q_{\rm R}$, {\it
bottom}]{2001MNRAS.323..445R} for wavenumbers of $\kl = 1$ ({\it black squares})
and $\kl = 2$ ({\it gray squares}) are provided in the bottom panel.  Values of
$\qtg$ and $\qts$, and $\qr$ for different $\kl$, are slightly offset in radius
for clarity.  Random errors are plotted in black; systematic errors, not
provided for $\qr$ when $\kl=2$, are plotted in light gray.  The vertical dotted
line is the same as plotted in Figure \ref{fig:sdd}.
}
\label{fig:q}
\end{figure}

Figure \ref{fig:q} provides our measurements of $Q$ for the disk of UGC 463; all
measurements demonstrate global stability.  To calculate $\kappa$, we use the
derivative of equation \ref{eq:vgcor} to calculate $d\vc/dR$, adopting the model
gas rotation curve from Figure \ref{fig:beamc} and the model $\sgas$ and
$\delta_P$ functions discussed in \sect \ref{sec:rcsdisp}.

Given that turbulence dominates the gas kinematics (\sect \ref{sec:rcsdisp}), we
calculate $\qtg$ by replacing $c_s$ with our measurements of $\azmean{\sgas}$
from the ionized gas.\footnote{
Reliable measurements of the \hone\ velocity dispersion are difficult given both
the spatial and spectral beam smearing of our UGC 463 data.
}  Therefore, our measurements of $\qtg$ may represent upper limits; the
measured $\allmean{\sgas} = 16.6\pm1.1$ \kms\ for UGC 463 is a factor of two or
more greater than the typical turbulent motions seen in the \hone\ or \molh\ gas
observed in local spiral galaxies \citep{2009A&A...495..795H,
2009AJ....137.4424T}.  Figure \ref{fig:q} shows that $\qtg$ increases
monotonically for $R\gtrsim4$ kpc.  In the range 2 kpc $\lesssim R\lesssim 7$
kpc ($0.6\lesssim R/h_R\lesssim 2.0$), we find\footnote{
Quoted errors in mean quantities here and below are (1) the random error in the
mean and (2) the quadrature sum of the systematic error and the standard
deviation in the quantity.
} $\allmean{\qtg} = 1.83\pm0.04^{+0.79}_{-0.72}$.  Thus, in the absence of the
stellar disk, the gas disk would be globally stable, but only by $\sim1.2$ times
the error.  A factor of two decrease in $\azmean{\sgas}$ would produce an
unstable gas disk, when in the absence of the stellar component.

The calculation of $\qts$ is determined directly from the data shown in Figures
\ref{fig:rcaxisym} and \ref{fig:sdd}, where we calculate $\sigr = (1.59\pm0.17)
\azmean{\sstar}$ using our measurements of the SVE axial ratios (\sect
\ref{sec:sve}).  We find $\allmean{\qts} = 5.2\pm2.1^{+3.1}_{-2.2}$ between 2
kpc $\lesssim R\lesssim 7$ kpc, decreasing to $\sim 0.6 \qtg$ at the radial
limit of our data.  The isolated stellar disk appears to be extremely stable, in
stark contrast to nominal expectations ($Q\sim2$) based on empirical
\citep[e.g.,][]{1993A&A...275...16B} or theoretical
\citep[e.g.,][]{1984ApJ...282...61S} arguments.  Given the mass of the gaseous
disk, a measurement of the composite stability is physically more meaningful.

The multi-component-disk stability measurement $\qr$ asymptotically decreases to
$\qr\sim2$, more in line with the theoretical expectations.  In detail, we find
$\allmean{\qr} = 2.1\pm0.4\pm0.7$ at $R\gtrsim h_R$ for $\kl=1$ indicating a
globally stable disk;\footnote{
A total stability calculation following $Q^{-1}_{{\rm gas}+\ast} = \qrg^{-1} +
\qrs^{-1}$ \citep{1994ApJ...427..759W} is 30\% smaller than $\qr$.
} measurements assuming $\kl = 2$, also provided in Figure \ref{fig:q}, reduce
$\qr$ by no more than its error.  A factor of two reduction in $\azmean{\sgas}$
decreases $\allmean{\qr}$ at $R\gtrsim h_R$ by 4\% and 20\% for $\kl=1$ and 2,
respectively, such that the composite disk should remain stable in this limit.

\subsection{$K$-Band Mass-to-Light Ratios, $\mldynkdisk$ and $\mlskdisk$}
\label{sec:ml}

We calculate dynamical ($\mldynkdisk$) and stellar ($\mlskdisk$) mass-to-light
ratios using our measurements of $\sddisk$ and $\sds$, respectively, and the
$K$-band surface-brightness
\begin{eqnarray}
I^{\rm disk}_K & = & {\rm dexp}[-0.4(\mu^{\prime}_K - A^i_K - {\mathcal K} -
M_{\odot}^K - 21.57)] \times \nonumber \\
& & \left(1-\frac{I_K^{\rm dust}}{I_K^{\rm obs}}\right)\cos i
\label{eq:disksb}
\end{eqnarray}
in $\lksol {\rm pc}^{-2}$, where $I_K^{\rm obs}$ is the observed emission after
correcting for Galactic extinction, $A^i_K$ is the internal dust extinction in
magnitudes (\sect \ref{sec:dustext}), ${\mathcal K} = 0.035$ mag is the
${\mathcal K}$-correction \citep{1995AJ....109...87B}, $I_K^{\rm dust}/I_K^{\rm
obs} = 0.02\pm0.01$ is the fraction of dust emission in the $K$-band (\sect
\ref{sec:dustem}), $i=27\arcdeg\pm2\arcdeg$ is the inclination (\sect
\ref{sec:inclination}), and $M_{\odot,K} = 3.30\pm0.04$ is the absolute
magnitude of the Sun.\footnote{
Our value and error for $M_{\odot,K}$ are, respectively, the mean and standard
deviation of measurements compiled from the following literature sources:
\citet{1994ApJS...95..107W, 1998gaas.book.....B, 2001ApJ...550..212B,
2003ApJS..149..289B}; and \citet{2004MNRAS.347..691P}.
}  Equation 10 from \citetalias{2010ApJ...716..234B} did not include some terms
in equation \ref{eq:disksb}, the most significant of which ($0.125\pm0.019$
\muu) is the face-on correction ($\cos i$); such terms negligibly affect our
error budget.

\begin{figure*}
\epsscale{0.9}
\plotone{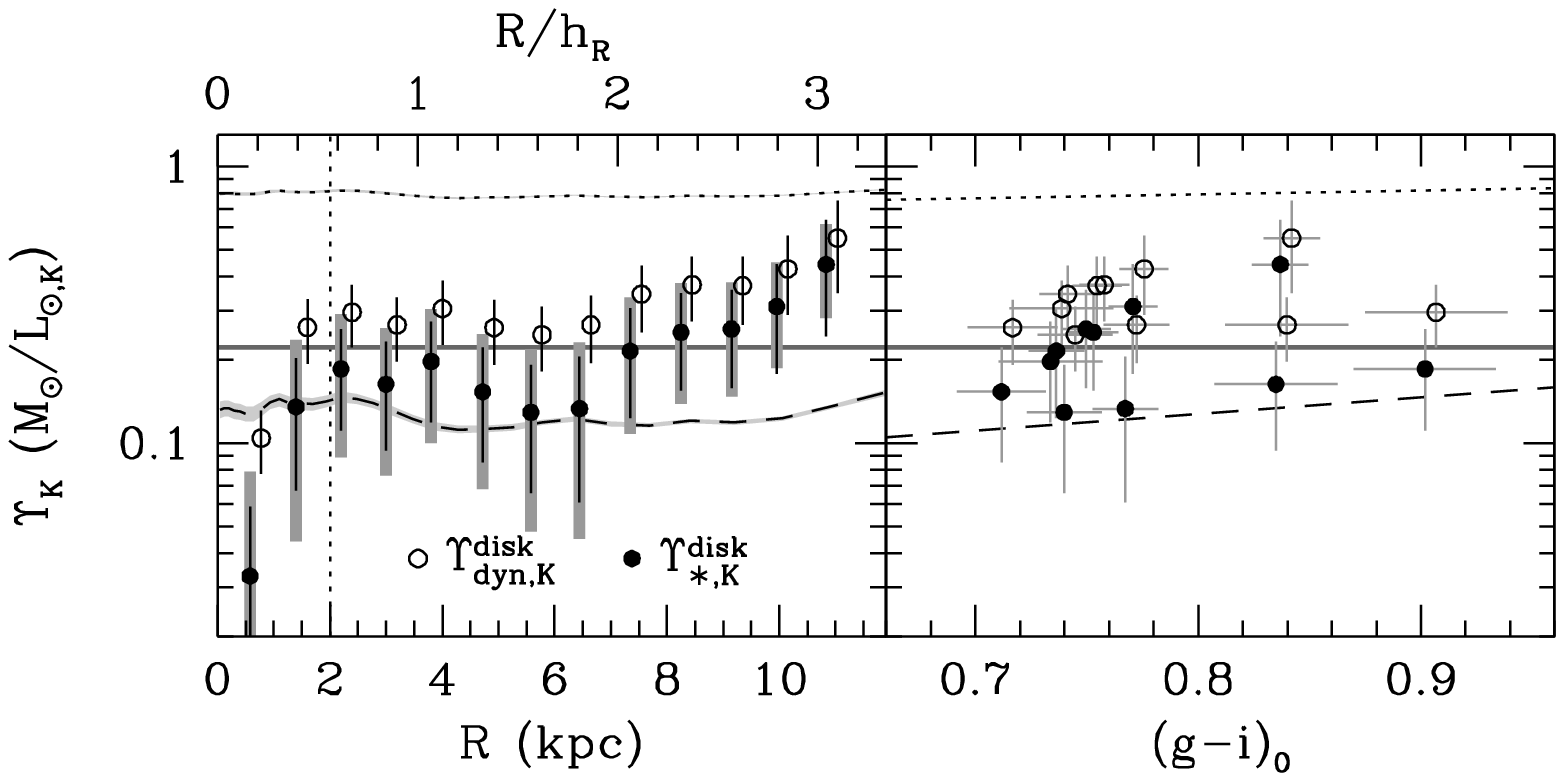}
\caption{
Measurements of $\mlskdisk$ ({\it filled points}) and $\mldynkdisk$ ({\it open
circles}) as a function of radius ({\it left}) and $(g-i)_0$ color ({\it
right}); the $\mldynkdisk$ measurements are slightly offset in radius from the
$\mlskdisk$ measurements for clarity.  Random errors are plotted in black;
systematic errors, only plotted for $\mlskdisk$ in the left panel, are light
gray.  The vertical dotted line delineates the radius at which the central light
concentration contributes $\sim3$\% to the total flux ($R=2$ kpc); data within
this radius are not shown in the right panel.  The solid dark-gray line provides
the mean $\mlskdisk$ at $R>2$ kpc.  The $(g-i)_0$ colors are used to predict
$\mlsksps$ based on the work of \citetalias{2003ApJS..149..289B} ({\it dotted
line}) and \citetalias{2009MNRAS.400.1181Z} ({\it dashed line}); random errors
due to the photometry are shown by the surrounding gray regions in the left
panel.
}
\label{fig:ml}
\end{figure*}

Figure \ref{fig:ml} provides our measurements of $\mldynkdisk = \sddisk/I_K^{\rm
disk}$ and $\mlskdisk = \sds/I_K^{\rm disk}$, both in units of $\msol/\lksol$;
the difference illustrates the effect of the gas-mass correction.  These data
are plotted both as a function of radius and $(g-i)_0$ color.  Given our
dynamical assumptions (e.g., negligible radial forces), $\sddisk$ measurements
at small radius may be systematically in error, especially in the presence of a
pressure-dominated bulge (as possible given the central light concentration;
\sect \ref{sec:muk}), a weak bar (\sect \ref{sec:axisym}), or a massive halo
(\sect \ref{sec:mass}).  The vertical dashed lined in Figure \ref{fig:ml} marks
where our model of the central light concentration contributes less than 3\% to
the total light; $\sddisk$ measurements beyond this radius (i.e., $R\geq2$ kpc)
should not be strongly affected by either the ``bulge'' or the dynamical
assumptions.  The two measurements of $\mldynkdisk$ and $\mlskdisk$ within this
radius are not plotted as a function of $(g-i)_0$.  Measurements of $(g-i)_0$
are interpolated from our SDSS photometry (\sect \ref{sec:sbprof}) after
accounting for internal reddening (\sect \ref{sec:dustext});
instrumental-smoothing- and ${\mathcal K}$-corrections have not been applied to
the $g$- or $i$-band data.  We compare our dynamical measurements with SPS model
calculations in Figure \ref{fig:ml} by plotting
\begin{equation}
\mlsksps = {\rm dexp}[a_K + b_K (g-i)_0],
\label{eq:brml}
\end{equation}
where the coefficients $a_K$ and $b_K$ are taken from \citet[][$a_K=-0.211$ and
$b_K=0.137$, hereafter \citetalias{2003ApJS..149..289B}]{2003ApJS..149..289B}
and \citet[][$a_K=-1.379$ and $b_K=0.604$, hereafter
\citetalias{2009MNRAS.400.1181Z}]{2009MNRAS.400.1181Z}; these models roughly
represent the extrema of similar SPS modeling done by, e.g.,
\citet{2001ApJ...550..212B} and \citet{2004MNRAS.347..691P}.  Consistent with
our photometry in \sect \ref{sec:optnir}, \citetalias{2003ApJS..149..289B} and
\citetalias{2009MNRAS.400.1181Z} assume $K$-band measurements are in Vega
magnitudes and $g$- and $i$-band measurements are in AB magnitudes.  We discuss
our results below after generating approximate probability distributions for
$\mlskdisk(R)$ and $\allmean{\mlskdisk}$.

\subsubsection{Probability Distribution}
\label{sec:mlsys}

We create an approximate probability distribution (combining random and
systematic components) for our measurements of $\mlskdisk(R)$ and
$\allmean{\mlskdisk}$ by Monte Carlo (MC) sampling individual probability
distributions for each component of our $\mlskdisk$ calculation.  In contrast
to the random error contributors, systematic errors --- particularly in $q$,
$k$, and $\xco$ --- may not be normally distributed, providing the primary
motivation for this test.  Our simulation is limited by the exclusion of any
parameter covariance, such as might be manifest in a refitting of $\alpha$ and
$\beta$ after adjusting $i$; however, we do not expect parameter covariance to
dramatically change the fundamental conclusions drawn from this MC simulation.

Random-error contributors --- {\it all} quantities with measurement errors
contributing to the random error in $\mlskdisk$ --- are assigned Gaussian
probability distributions according to their derived $\epsilon$.  We also assign
a Gaussian distribution for $D$ (combining the random and systematic error in
quadrature) and $\imi/\ico$ (using the $\pm0.11$ dex error in equation
\ref{eq:24_co_fit}).  Information on the probability distribution for $\xco$,
particularly for spiral galaxies like UGC 463, is limited; therefore, we simply
assume a uniform distribution with the range $\xco = (2.7 \pm 0.9) \times
10^{20}$ cm$^{-2}$ (K \kms)$^{-1}$ (\sect \ref{sec:xco}).  We have derived an
empirical probability distribution for $q$ in \citetalias{2010ApJ...716..234B}
(see Figure 1, and references, therein).  For our MC simulation, we smooth the
growth curve of this empirical distribution by a low-order polynomial to avoid
the discrete-measurement quantization noise.

A robust probability distribution for $k$ is elusive, lacking an empirical
measurement.  For UGC 463, the measurements of a relatively massive gas disk
(\sect \ref{sec:totgas}) and DM halo (\sect \ref{sec:mass}) are particularly
relevant to the value of $k$.  In an extreme scenario, the stellar disk is
vertically exponential and the gas disk is razor thin, yielding an effective
30\% decrease in $k$ for our measured values of $\sdg$ and $\sds$.  Despite the
resulting increase in $\sddisk$, we would still infer a massive DM halo that
increases $k$ by $20 - 30$\% (\sect \ref{sec:rhorat}) and, therefore, roughly
offsets the effect of the massive gas disk.  In view of further complications
introduced by finite-thickness gas disks, multi-component stellar disks, and
triaxial halos, we have decided to take a simple approach and assume $k$ is
quantized and equally distributed among the exponential ($k = 1.5$), sech
($k=1.71$), and sech$^2$ ($k=2.0$) cases.

\begin{figure*}
\epsscale{0.9}
\plotone{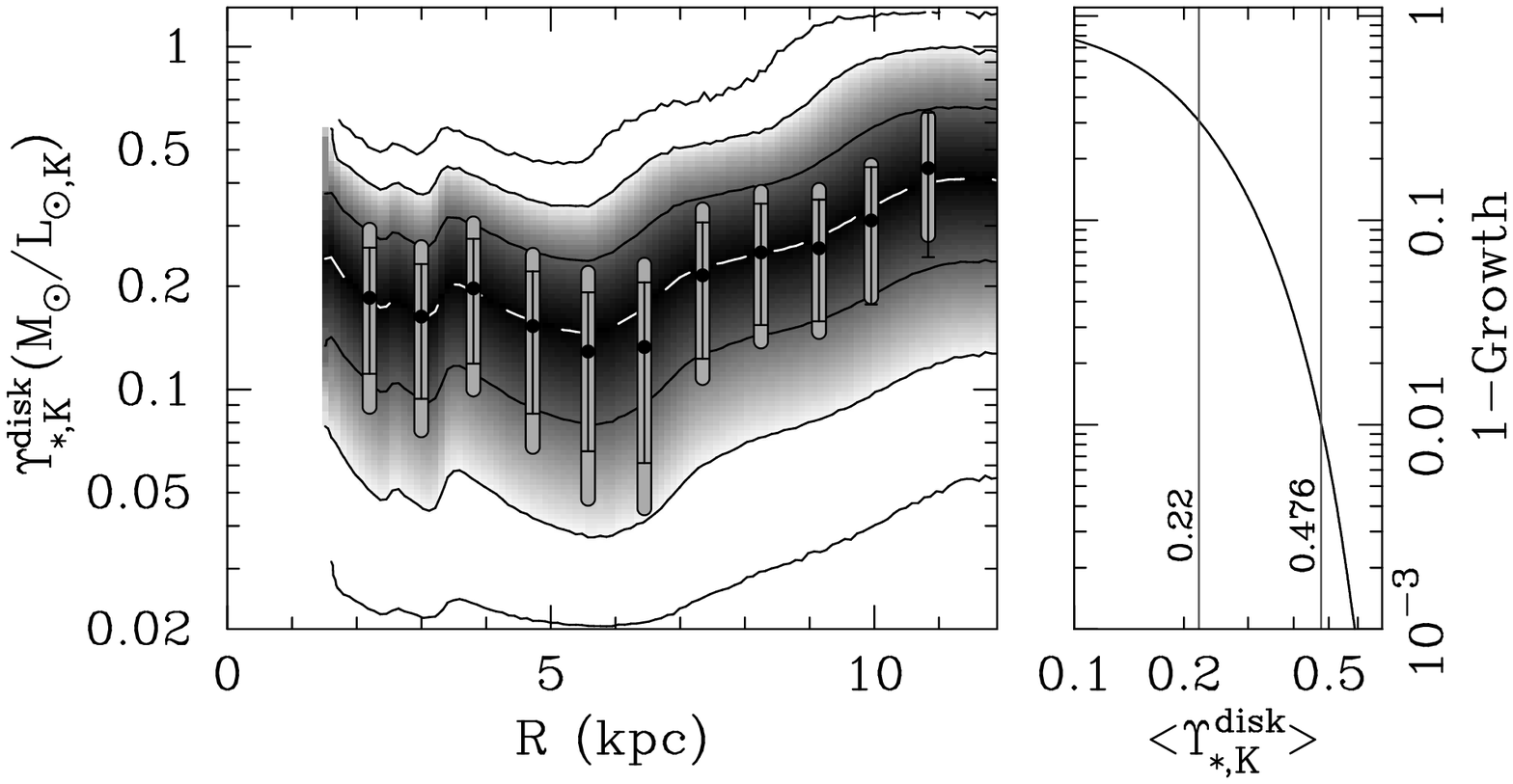}
\caption{
Result of the Monte Carlo simulation outlined in \sect \ref{sec:mlsys}.  The
left panel shows the gray-scale image of the folded growth curve in $\mlskdisk$
at a given radius. The median $\mlskdisk(R)$ lies at the peak intensity ({\it
black}) as traced by the dashed white line; black contours mark the 68\%, 95\%,
and 99\% confidence intervals.  The nominal measurements of $\mlskdisk(R)$ from
Figure \ref{fig:ml} are overplotted ({\it black points}) with random and
systematic errors in black and gray, respectively; data at $R<2$ kpc are
omitted.  The growth curve of $\allmean{\mlskdisk}$ is plotted in the right
panel. Gray lines mark $\allmean{\mlskdisk} = 0.22$ and 0.476; less than 1\% and
0.1\% of the simulations produce $\allmean{\mlskdisk} > 0.476$ and
$\allmean{\mlskdisk} > 0.59$, respectively.
}
\label{fig:mlrsim}
\end{figure*}

The results of our MC simulation are shown in Figure \ref{fig:mlrsim} based on
$10^6$ recalculations of $\mlskdisk$ for each measurement of $\azmean{\sstar}$
at $R\geq2$ kpc.  Each recalculation is binned in the two-dimensional ($R$,
$\mlskdisk$) plane; $R$ is binned in physical units, incorporating the MC
sampling of $D$.  For each radial bin, we create a growth curve for $\mlskdisk$
such that higher intensity (darker) cells in the left panel of Figure
\ref{fig:mlrsim} represent more probable measurement of $\mlskdisk$.  We
overplot 68\%, 95\%, and 99\% confidence contours for $\mlskdisk(R)$, as well as
a contour following the median value.  We also overplot the nominal measurements
of $\mlskdisk$ from Figure \ref{fig:ml} for reference, again differentiating
between random and systematic error.  Figure \ref{fig:mlrsim} also provides the
growth curve of $\allmean{\mlskdisk}$, calculated for each of the $10^6$
recalculations of $\mlskdisk(R)$.

\subsubsection{Discussion}
\label{sec:mldisc}

Based on our nominal calculation (Figure \ref{fig:ml}), we find
$\allmean{\mldynkdisk} = 0.34\pm0.09\pm0.15$ and $\allmean{\mlskdisk} =
0.22\pm0.09^{+0.16}_{-0.15}$ at $R>2$ kpc, in units of $\msol/\lksol$.  In
contrast, \citetalias{2003ApJS..149..289B} and \citetalias{2009MNRAS.400.1181Z}
predict $\allmean{\mlsksps} = 3.6 \allmean{\mlskdisk}$ and $\allmean{\mlsksps} =
0.56 \allmean{\mlskdisk}$, respectively.  From our derived probability
distribution we find a median of $\allmean{\mlskdisk} = 0.17^{+0.12}_{-0.09}$
such that the \citetalias{2009MNRAS.400.1181Z} prediction is within our 68\%
confidence interval; however, \citetalias{2003ApJS..149..289B} predict a
measurement that occurs for less than 1 in $10^5$ recalculations of
$\allmean{\mlskdisk}$.  The median value of our probability distribution is
below our nominal measurement because the median value for $k$ is above our
nominal value of $k=1.5$.  Within the parameter space probed by our MC
simulation, we find a {\it maximum} measurement of $\allmean{\mlskdisk}=0.82$.

The prescriptions adopted for the many ingredients of SPS modeling --- such as
the initial mass function (IMF), star-formation and chemical-enrichment history,
dust content, and the treatment of specific phases of stellar evolution --- play
an important role in setting $\mlsksps$ and its trend with color
\citep{2004MNRAS.347..691P, 2009ApJ...699..486C, 2010ApJ...708...58C,
2010ApJ...712..833C}.  Indeed, in their discussion of the differences between
their $\mls$--color relations and those from \citetalias{2003ApJS..149..289B},
\citetalias{2009MNRAS.400.1181Z} isolate their treatment of the star-formation
histories and thermally pulsating asymptotic giant branch (TP-AGB) phases of
stellar evolution from intermediate-age populations as the primary culprits.
The latter particularly effects differences in SPS predictions of $\mls$ in the
NIR bands \citep{2005MNRAS.362..799M}.  Aside from the factor of $\sim7$
difference in the mean \citetalias{2003ApJS..149..289B} and
\citetalias{2009MNRAS.400.1181Z} predictions for UGC 463, it is also important
to keep in mind that there is an additional factor of 2 or more {\it internal}
variation in the $\mlsksps$ calibration associated with each study as determined
by their search of the SPS modeling parameter space.  It is encouraging that the
advancement in SPS modeling, as represented by the
\citetalias{2009MNRAS.400.1181Z} study, are more consistent with our dynamical
measurements.  However, Figure \ref{fig:ml} shows that neither the zeropoint nor
the trend of $\mls$ with $(g-i)$ color from these models are a good match to our
measurements; therefore, we cannot conclude that this specific SPS treatment ---
either for the TP-AGB phase or star-formation histories --- is correct or
applicable to our entire sample.  An absolute calibration of $\mls$ using a
multi-color approach for the full DMS Phase-B sample will be presented in
forthcoming papers.

As expected from the discussion in \sect \ref{sec:hsigma}, Figure \ref{fig:ml}
shows $\mldynkdisk$ and $\mlskdisk$ generally increase with radius.  Considering
only the data at $R=5.6$ and 10.8 kpc, $\mldynkdisk$ and $\mlskdisk$ increase by
a factor of $2.2\pm1.0^{+1.2}_{-1.1}$ and $3.4\pm2.3^{+2.8}_{-2.5}$,
respectively.  Figure \ref{fig:mlrsim} shows that the radial trend holds for the
median of the probability distribution in $\mlskdisk(R)$.  Variation in
$\mlsksps$ is expected in galaxy disks given the observed arm/inter-arm and
radially averaged color gradients.  Yet, in terms of the latter, Figure
\ref{fig:ml} shows that the variation in $\mlskdisk$ for UGC 463 is consistent
with neither of the plotted $\mlsksps$ predictions.  Given both the errors in
our measurement and the errors in the $\mlsksps$ calibration, it is difficult
for us to conclude that our measurements are inconsistent with a radially
invariant $\mlskdisk$.  Indeed, a radially independent $\mlskdisk$ is consistent
with our 68\% confidence limits derived in Figure \ref{fig:mlrsim}.

Moreover, the more shallow decline of $\sigz^2$ with respect to the
surface-brightness profile, which is the primary driver for our measurement of a
radially varying $\mlskdisk$, may also be interpreted as a flaring of the
stellar disk (\sect \ref{sec:hsigma}).  A flared stellar disk has been measured
for the Galaxy \citep{2002A&A...394..883L, 2003A&A...409..523R,
2006A&A...451..515M} and one might expect disks to be flared due to, e.g.,
interactions with dark and/or luminous satellites \citep{2006PASJ...58..835H,
2008ASPC..396..321D, 2008MNRAS.389.1041R, 2009ApJ...700.1896K}.  Indeed,
\citet{2009ApJ...693L..19H} suggest their measurements of a nearly constant
$\sigz$ at large radii in M 83 and M 94 provide evidence for such interactions.
However, the onset radius for disk flaring is expected to be beyond the region
relevant to our dynamical measurements for UGC 463.  Robust photometric evidence
for stellar-disk flares in edge-on galaxies remains elusive:  The empirical and
theoretical foundation for the vertical structure of galaxy disks with radially
independent scale heights developed by \citet{1981A&A....95..105V,
1981A&A....95..116V, 1982A&A...110...61V, 1982A&A...110...79V} remains the
current paradigm due to repeated confirmations of little to no variation in
scale height measurements from surface photometry, particularly for galaxies of
similar Hubble type to UGC 463 (SABc) \citep[e.g.,][]{1997A&A...320L..21D,
2002A&A...389..795B}.  However, claims of factors of 2 or more increase in scale
height within the optical extent of some stellar disks exist in the literature
\citep[e.g.,][]{2002A&A...390L..35N, 2009MNRAS.396..409S}.  For our UGC 463
data, the measured increase in $\mlskdisk$ with radius is consistent with these
claims; however, we cannot claim a stellar-disk flare exists in UGC 463 based
solely on our data.  Therefore, barring more detailed information on the disk
structure of UGC 463, we simplify our mass decomposition in \sect \ref{sec:mass}
by largely focusing on results that assume $\mlskdisk$ and $h_z$ are constant
for the entire disk.

Finally, we note that our data provide a few, limited assessments of the
presence of a ``DM disk'' in UGC 463.  Such a structure has been predicted by
recent simulations \citep[e.g.,][]{2009MNRAS.397...44R} and modeling of the
Galaxy by \citet{2003ApJ...588..805K} \citep[cf.,][]{2010ApJ...724L.122M}.  DM
disks are expected to be more extended both radially and vertically than stellar
thin disks; \citet{2003ApJ...588..805K} fit a DM disk that has a scale length
and scale height that are, respectively, three and 10 times larger than for the
stellar thin disk.  Our stellar kinematic data are expected to trace the thin
disk mass distribution only such that we can place an upper limit on any DM
distributed identically to this structure as follows:  Assuming the $\mlsksps$
prediction from \citetalias{2009MNRAS.400.1181Z} is exactly correct, our
$\mlskdisk$ measurements suggest a thin DM disk that has $\sim80$\% of the
stellar mass surface density (or $\sim30$\% of the baryonic mass surface
density).  One can increase the mass surface density of a DM disk in UGC 463 by
proportionally increasing its scale height with relatively moderate effects on
our calculation of $\sddisk$.  Assuming no influence on $\sddisk$, the scale
height of the DM disk would need to be roughly the same as the thin-disk scale
length to reach the mass ratio that \citet{2003ApJ...588..805K} measure for the
Galaxy.  Although one may accommodate such a disk within the current
understanding of DM disks, such a structure is mostly conjectural with respect
to our data.  Given the uncertainty in both the expected vertical distribution
of a DM disk and the SPS modeling results for $\mlsksps$, our mass decomposition
assumes no DM disk exists in UGC 463, which is consistent with our observations.

\section{Mass Budget}
\label{sec:mass}

We produce a detailed mass budget of UGC 463 via a traditional rotation-curve
mass decomposition \citep[e.g.,][]{1985ApJ...294..494C, 1985ApJ...295..305V,
1991MNRAS.249..523B}.  We assume
\begin{equation}
\vc^2 = \sum_j V_j^2,
\label{eq:vcirc}
\end{equation}
where $V_j$ is the circular-speed of a test particle associated with
potential-density pair $\Phi_j$ and $\rho_j$ for each mass component $j$; all
potential-density pairs are considered independent and separable, neglecting any
covariance among the $j$ components \citep[cf.,
\citetalias{1986RSPTA.320..447V};][]{2010A&A...519A..47A}.  We calculate $V_j$
for each baryonic mass component based on our mass-surface-density measurements
using {\tt rotmod}, a program within the {\it GIPSY}\footnote{
Groningen Image Processing System; \url{http://www.astro.rug.nl/~gipsy/}.
} software package that calculates $V_j$ for oblate and spherical density
distributions \citep[following][]{1983MNRAS.203..735C}.

Studies of our Galaxy suggest the potential-density structure of UGC 463 may be
very complex.  For simplicity, we assume here that the total gravitational
potential is composed of four unique, axisymmetric density distributions,
yielding the following circular speeds: (1) $\vhalo$ for the spherical halo, (2)
$\vbulgestar$ for the stellar bulge, (3) $\vdiskstar$ for the stellar disk, and
(4) $\vdiskgas$ for the gaseous disk.  We attribute the central light
concentration to a bulge \citep[cf.][who propose such surface density peaks may
be attributed to the disk]{2009MNRAS.396..121D}; however, our use of the term
``bulge'' does not distinguish between a (disk-like) pseudo-bulge
\citep{2004ARA&A..42..603K} or (spherical) classical bulge.  We neglect any
significant contribution from, e.g., an inner halo, thick stellar disk, or
flattened dark-matter component for two reasons: (i) Our spectroscopic and
imaging data present no evidence for significant contributions of such
components to either the measured kinematics or the stellar light profile; and,
therefore, (ii) if present, the influence of such mass components is negligible
with respect to our analysis of the gravitational potential in UGC 463.
Furthermore, we assume the halo is dominated by DM (i.e., $\vhalo \approx \vdm$)
and the vertical distributions of the atomic and molecular gas are identical
($\vdiskgas$ is determined by $\sdg$ directly).  We define the baryonic circular
speed, $\vbary^2 = (\vbulgestar)^2 + (\vdiskstar)^2 + (\vdiskgas)^2$, such that
we can isolate the DM mass contribution via $\vdm^2 = \vc^2 -\vbary^2$.  The
fundamental advantage of our rotation-curve mass decomposition over previous
studies is that $\vbary$ is {\it uniquely} defined by our observations.

In summary, we calculate the circular speeds of all baryonic mass components in
\sect \ref{sec:baryrc}; we measure $\vdm(R)$ and the DM-halo volume-density
profile, $\rhodm(R)$, in \sect \ref{sec:rhodm}; and we discuss the relative
contributions of the baryonic matter and DM components to the total mass budget
of UGC 463 in \sect \ref{sec:dvb} finding that DM dominates at $R>h_R$.

\subsection{Baryonic Mass and Circular Speed}
\label{sec:baryrc}

\begin{figure}
\epsscale{1.1}
\plotone{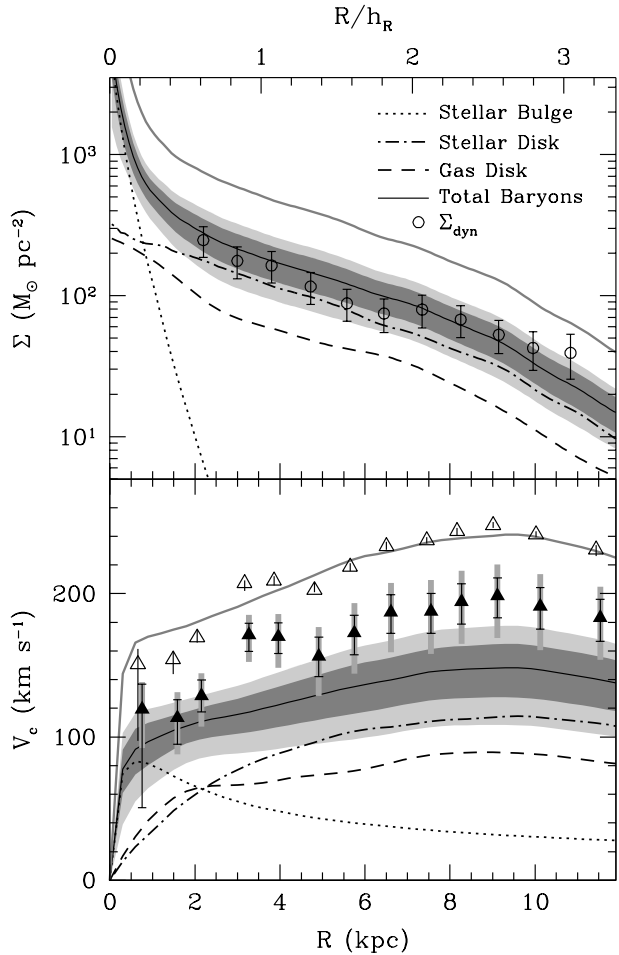}
\caption{
Mass-surface-density ({\it top}) and circular-speed ({\it bottom}) distributions
for the stellar bulge ({\it dotted line}), stellar disk ({\it dot-dashed line}),
gaseous disk ({\it dashed line}), and all baryonic matter ({\it solid black
line}) in UGC 463 assuming a constant $\mlsk(R)$.  Our measurements of $\sddisk$
at $R>2$ kpc are overplotted as open symbols in the top panel for reference,
which are directly traced during our calculations assuming a variable $\mlsk(R)$.
Dark- and light-gray regions illustrate, respectively, the random and systematic
error in $\sdbary$ and $\vbary$.  The solid gray lines assume $\mlsk = \mlsksps$
as predicted by \citetalias{2003ApJS..149..289B}.  Measurements of the DM-only
circular speed ($\vdm$; {\it black triangles}) are calculated by subtracting
$\vbary$ from $\vc$ ({\it open triangles}) in quadrature.  Random and systematic
errors in $\vdm$ are shown as black and gray error bars, respectively. 
}
\label{fig:rcdecomp}
\end{figure}

Following the bulge-disk decomposition discussed in \sect \ref{sec:muk}, we
calculate the mass-surface-density distribution for each stellar component
assuming $\mlsk(R) = \allmean{\mlskdisk} = 0.22\pm0.09^{+0.16}_{-0.15}$.  We use
a single $\mlsk$ for the disk and bulge components, which is reasonable
considering the marginal change in $\mlsksps$ between the disk- and
bulge-dominated regions in Figure \ref{fig:ml}.  The results are shown in Figure
\ref{fig:rcdecomp} along with our measurements of $\sdg$.  We calculate masses
enclosed within $R = 15$ kpc ($R = 4.2 h_R$) for all baryonic components in
Table \ref{tab:massbudget} by integrating each mass-surface-density profile;
these results are further discussed in \sect \ref{sec:massbudget}.  We note that
the instrumental-smoothing correction to $\mu^{\prime}_K$ (\sect \ref{sec:muk})
amounts to a marginal 3\% increase of the stellar-bulge mass.

\begin{deluxetable}{ l l c c }
\tabletypesize{\small}
\tablewidth{0pt}
\tablecaption{UGC 463 Enclosed Mass at 15 kpc (4.2$h_R$)}
\tablehead{ & & \multicolumn{2}{c}{Mass Fraction} \\ \cline{3-4} & \colhead{Mass} & \colhead{Baryonic} & \colhead{Total} \\ \colhead{Component} & \colhead{($10^{10}\msol$)} & \colhead{(\%)} & \colhead{(\%)} }
\startdata
Stellar Bulge      & $0.22\pm0.09\pm0.16$                  &  5.2 &  1.2 \\
Stellar Disk       & $2.6\pm1.1^{+1.9}_{-1.8}$             & 62   & 15 \\
Total Stars        & $2.8\pm1.2^{+2.0}_{-1.9}$             & 67   & 16 \\
\hline
Atomic Hydrogen    & $0.24\pm0.02$                         &  5.7 & 1.4 \\
Molecular Hydrogen & $0.76\pm0.04^{+0.34}_{-0.31}$         & 18   & 4.3 \\
Total Gas          & $1.40\pm0.07^{+0.47}_{-0.43}$         & 33   & 7.9 \\
\hline
Baryonic Matter    & $4.2\pm1.1^{+2.1}_{-1.9}$             &      &  24 \\
Dark Matter        & $13.5^{+1.9}_{-2.4}\ ^{+3.2}_{-4.2}$  &      &  76 \\
Total Mass         & $17.7^{+2.2}_{-2.7}\ ^{+3.8}_{-4.6}$  &      &
\enddata
\label{tab:massbudget}
\end{deluxetable}

Circular-speed calculations for each baryonic component use the
mass-surface-density profiles in Figure \ref{fig:rcdecomp} and an assigned
three-dimensional, axisymmetric density distribution.  We assume any truncation
of each mass element occurs well beyond our last dynamical measurement.  We
assume $\sdg$ is distributed in a razor-thin disk; and, consistent with our
previous assumptions, we adopt an exponential vertical distribution for the
stellar disk with a constant scale height of $h_z = 0.44$ kpc.  For the stellar
bulge (or central mass concentration), our nominal approach is to assume a
spherical distribution.

Due to our imposition of a constant $\mlsk$ for the mass decomposition
illustrated in Figure \ref{fig:rcdecomp}, the surface density of all baryonic
mass components, $\sdbary$, does not exactly follow our $\sddisk$ measurements.
We have also performed a more direct mass decomposition that adopts $\sdbary =
\sddisk$ by applying a smooth interpolation for $\mlsk(R) = \mlskdisk(R)$ at
$2<R<11$ kpc; we extrapolate by simply extending the $\mlskdisk$ measurements at
the two endpoints of this range to small and large radius.  This approach does not
produce statistically different values for the quantities discussed below.  For
completeness, Table \ref{tab:dmp} provides the implied DM properties for both a
constant and variable $\mlsk(R)$; however, we simplify the discussion below by
primarily focusing on the results obtained by assuming a constant $\mlsk(R)$.
This approach is justified by the consistency of our measurements with a
radially invariant $\mlskdisk$ as discussed in \sect \ref{sec:mldisc}.

In addition to the constant $\mlsk$ based on our dynamical measurements, Figure
\ref{fig:rcdecomp} also provides the calculation of $\sdbary$ and $\vbary$
assuming $\mlsk = \mlsksps = 0.79$, based on the prediction of
\citetalias{2003ApJS..149..289B}; adopting the \citetalias{2009MNRAS.400.1181Z}
prediction produces $\vbary$ within the systematic errors of our dynamical
measurements.  The stellar and baryonic mass are, thereby, increased by a factor
of 3.6 and 2.7, respectively, resulting in a maximal disk (i.e., $\vbary \sim
\vc$).  Therefore, we can account for the measured rotation velocity within
$R\leq 3.2 h_R$ by simply scaling $\mu^{\prime}_K$ by a constant $\mlsk$.  This
result is not unexpected given that the ``diet Salpeter'' IMF, used by
\citetalias{2003ApJS..149..289B}, was chosen by \citet{2001ApJ...550..212B} to
accommodate the ``maximum-disk'' rotation-curve mass decompositions produced by
\citet{1997PhDT........13V}.  Moreover, this is consistent with the expectation
from most rotation-curve mass decompositions in the literature performed within
similar radial regimes \citep[see discussion by][]{2004IAUS..220..233S},
including the recent study of dwarf galaxies performed by
\citet{2011ApJ...729..118S}.  However, we have shown in \sect \ref{sec:ml} that
the assumption $\mlsk = \mlsksps = 0.79$ is rejected by our $\mlskdisk$
measurements at $\gg99$\% confidence.  That is, while it is {\it possible} to
account for the rotation curve of UGC 463 within $R\leq 3.2 h_R$ without
invoking DM, DM is effectively {\it required} by our treatment of the observed
stellar kinematics and, in fact, dominates the mass budget (\sect
\ref{sec:massbudget}).  Moreover, the implied DM-mass distribution must be
substantially less oblate than the baryonic-mass (disk) distribution traced by
our stellar kinematics to simultaneously explain our measurements of $\vc$ and
$\azmean{\sstar}$, as discussed in \sect \ref{sec:mldisc}.

We note here that the DM properties we infer below are robust against many of
the assumptions made above concerning the detailed baryonic mass decomposition.
First, our characterization of the intrinsic central light concentration is a
marginal consideration for calculating $\vbary$.  For the innermost measurement
of $\vc$ ($R = 0.7$ kpc), $\vbary$ is 10\% (9.1 \kms) higher for our nominal
measurement than if we were to adopt the stellar-disk density distribution at
all radii; the random errors are 15\%.  This difference increases to 25\% (23.3
\kms) if we also omit the instrumental-smoothing correction to $\mu^{\prime}_K$;
however, the difference at the radius of the second measurement of $\vc$ (at $R
= 1.5$ kpc) in this case is only 2\% (3.5 \kms).  Second, we find statistically
equivalent measurements of the DM rotation speed, $\vdm$, when simply
calculating $\vbary$ directly from $\sdbary = \sddisk$ and assuming the nominal
oblateness of the disk.  The detailed baryonic mass decomposition presented here
is meant to couch our rotation-curve mass decomposition within the traditional
construct found in the literature and to provide insight into the {\it baryonic}
mass budget of UGC 463.

\subsection{Dark-Matter-Halo Circular Speed, Mass, and Volume Density}
\label{sec:rhodm}

\begin{figure}
\epsscale{1.1}
\plotone{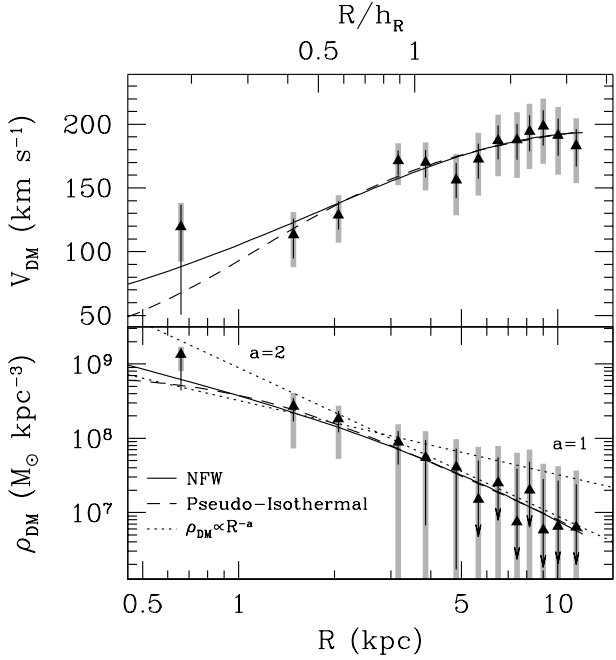}
\caption{
The DM-only circular speed ($\vdm$; {\it top}) and volume-density profile
($\rhodm$; {\it bottom}) for UGC 463 assuming a constant $\mlsk(R)$; random and
systematic errors are shown in black and gray, respectively.  Upper limit arrows
are used for data with random errors of $\epsilon(\rhodm) > \rhodm$.  The
best-fitting  NFW ({\it solid lines}) and pseudo-isothermal ({\it dashed lines})
parameterizations of the DM-halo are shown against the data.  For reference,
dotted lines in the bottom panel provide simple power-law density distributions
following $\rhodm \propto R^{-a}$ for $a=1$ and 2; zero-points are set to
approximately match the trend at small and large radius for $a = 1$ and $a = 2$,
respectively.
}
\label{fig:rhodm}
\end{figure}

Figure \ref{fig:rcdecomp} provides measurements of $\vdm^2 = \vc^2 - \vbary^2$
assuming a constant $\mlsk(R)$; the enclosed mass of the presumed spherical halo is
$\mhalodm = (8.9^{+1.3}_{-1.6}\ ^{+2.1}_{-2.9})\times 10^{10} \msol$ within the
radial range of our measurements ($R \leq 11.4$ kpc).  Measurements of $\vdm$
also provide the spherical-halo volume-density profile via
\begin{equation}
\rhodm = \frac{\vdm^2}{4\pi GR^2} \left( 1 + \frac{d\ln\vdm^2}{d\ln R} \right)
\label{eq:rhodm}
\end{equation}
\citep{2001ApJ...552L..23D, 2003ApJ...583..732S} as shown in Figure
\ref{fig:rhodm}, where we calculate $d\vdm^2/dR = d\vc^2/dR - d\vbary^2/dR$
based on the model values for $\vc$ and $\vbary$.  The calculation of $d\vc/dR$
is the same as used in \sect \ref{sec:Q} to calculate $\kappa$, and the
calculation of $d\vbary/dR$ is done via finite-differencing of the $\vbary$ data
in Figure \ref{fig:rcdecomp}.  Measurements of $\rhodm$ are upper limits at
$R\gtrsim 4$ kpc ($R\gtrsim 1.1h_R$) because the random error,
$\epsilon(\rhodm)$, becomes larger than the measured value.

We model the $\vdm$ data in Figure \ref{fig:rhodm} using a spherical ``NFW''
halo \citep{1996ApJ...462..563N, 1997ApJ...490..493N} and a pseudo-isothermal
sphere \citep{1979IAUS...84..441O, 1985IAUS..106...75S, 1986AJ.....91.1301K}.
The former is parameterized by the total halo mass, $\mhalodm$, and its
concentration $c=R_{200}/R_s$, where $R_{200}$ is the radius at which the halo
reaches 200 times the critical density ($\rho_{\rm crit} = 3 H_0^2/8\pi G$) and
$R_s$ is the characteristic scale of the density profile; as above, we adopt
$H_0 = 73$ \kms~Mpc$^{-1}$.  The pseudo-isothermal sphere is parameterized by
its central density, $\rho_0$, and core radius, $R_c$.  Best-fitting DM halo
parameters are determined by $\chi^2$-minimization, where $\chi^2$ is defined
using only the random errors.  Bootstrap simulations are used to calculate the
random error in each the parameter.  Systematic errors are based on 100
simulations of refitting the DM halo after MC sampling of the systematic errors
in $\vdm$, again minimizing a $\chi^2$ incorporating only the random errors.
The best-fitting DM halo models are plotted against our measurements in Figure
\ref{fig:rhodm}; we provide the best-fitting parameters for both a constant and
variable $\mlsk(R)$ in Table \ref{tab:dmp}.  Statistically, the NFW and
pseudo-isothermal halos are equally suitable descriptions of the DM halo of UGC
463, and there is only a marginal difference between the results when using a
constant or variable $\mlsk(R)$.  While the reduced $\chi^2$ is lower when
adopting a constant $\mlsk(R)$, the systematic deviations of the data about the
model is such that this difference is not statistically meaningful.  For
comparison, Figure \ref{fig:rhodm} also shows simple power-law density profiles
following $\rhodm\propto R^{-a}$ for $a = 1$ and 2; zero-points have been set
by-eye in each case.

\begin{deluxetable*}{ l l l l }
\tabletypesize{\small}
\tablewidth{0pt}
\tablecaption{UGC 463 DM-Halo Properties}
\tablehead{ \multicolumn{2}{c}{Parameter} & \colhead{Constant $\mlsk$} & \colhead{Variable $\mlsk$} }
\startdata
Pseudo-Isothermal: & $\log\rho_0$   & $8.85\pm0.21\pm0.13$                       & $8.83\pm0.16\pm0.09$                     \\
                   & $R_c$          & $1.06\pm0.21\pm0.01$                       & $1.16\pm0.21\pm0.04$                     \\
		   & $\chi^2_{\nu}$ & 0.64                                       & 1.20                                     \\
\hline
NFW Halo:          & $\log\mhalodm$ & $11.51\pm0.09\pm0.15$                      & $11.60\pm0.09\pm0.07$                    \\
	           & $c$            & $19.2\pm2.7\pm1.6$                         & $18.9\pm2.7\pm1.2$                       \\
		   & $\chi^2_{\nu}$ & 0.61                                       & 1.14                                     \\
\hline
                   & $\Fbary(R_e)$    & $0.81^{+0.48}_{-0.15}\ ^{+0.24}_{-0.35}$ & $0.78^{+0.43}_{-0.13}\ ^{+0.17}_{-0.22}$ \\
                   & $\Fbary(2.2h_R)$ & $0.61^{+0.07}_{-0.09}\ ^{+0.12}_{-0.18}$ & $0.53^{+0.05}_{-0.07}\ ^{+0.09}_{-0.12}$ \\
                   & $\Fdisk(2.2h_R)$ & $0.46^{+0.09}_{-0.12}\ ^{+0.15}_{-0.22}$ & $0.36^{+0.08}_{-0.10}\ ^{+0.11}_{-0.15}$
\enddata
\tablenotetext{~}{{\bf Notes.} Units of $\rho_0$, $R_c$, and $\mhalodm$ are
$\msol~{\rm kpc}^{-3}$, kpc, and $\msol$, respectively.}
\label{tab:dmp}
\end{deluxetable*}

We measure the concentration of the NFW halo to be $c=19^{+3}_{-2}$ in
reasonable agreement with the expectation for low-redshift DM halos of similar
mass \citep{1997ApJ...490..493N, 2001MNRAS.321..559B}.  Thus, consistent with
our measurement of a relatively low-mass disk, it appears that the baryons have
had little effect on the structure of the DM halo.  However, it is difficult to
assess the physical meaning of this concentration index given the simplicity of
our halo fitting; specifically, we do not include adiabatic contraction, which
should be a relatively small effect given the expected gravitational influence
of the baryons.  It is interesting that the innermost $\rhodm$ datum indicates a
steeper slope than provided by either of the DM-halo parameterizations,
demonstrating $a\sim 2$ as opposed to unity.  However, the error in this
measurement is large and is highly subject to our treatment of the central mass
concentration.  The full DMS sample will provide stronger statistical
constraints on the inner halo profile.

\subsection{The Dominant Gravitational Influence of Dark Matter}
\label{sec:dvb}

As discussed in \sect \ref{sec:intro}, assessments of the relative contribution
of dark and baryonic matter to disk-galaxy mass budgets have been limited by the
disk-halo degeneracy \citep{1985ApJ...295..305V}.  We have measured these
contributions directly for UGC 463.  Although allowing for a novel investigation
of the DM mass distribution, it is useful to cast our analysis also in terms of
a traditional approach such that we can compare with previous studies.

In particular, a common praxis in rotation-curve mass decomposition is the
so-called ``maximum-disk'' hypothesis \citepalias{1986RSPTA.320..447V},
producing an extremum of the mass budget.  Application of the ``maximum-disk''
hypothesis amounts to maximizing a radially independent mass-to-light ratio
while adjusting any DM-halo parameters to fit the observed circular speed;
however, exact implementations have varied.  All baryonic components have been
approximated by a single exponential disk \citep[as in][]{1985ApJ...295..305V},
values of $\mls$ have been distinct \citep[as in, e.g.,][]{1986AJ.....91.1301K,
1987AJ.....93..816K} or identical (as done above) for the bulge and disk
components; and the gas disk has been subsumed into or isolated from the stellar
component(s).  Thus, one should keep in mind that, although termed the
``maximum-{\it disk}'' hypothesis, the direct association of this hypothesis
with the stellar disk, in particular, can be tenuous in its practical
implementation.  Regardless, {\it all} implementations minimize the DM
contribution to the mass budget at small radius.  Galaxies adhering to the
``maximum-disk'' hypothesis are often said to have maximal disks; however, this
definition remains inchoate.

The ``maximality'' of a galaxy is often assessed via the stellar-disk mass
fraction, $\Fdisk = \vdiskstar/\vc$ \citepalias[equation 11
in][]{2010ApJ...716..234B}.  This quantity is traditionally measured at
$2.2h_R$, the radius at which the circular speed peaks for razor-thin, radially
exponential disks \citepalias[see the generalization to oblate disks
in][]{2010ApJ...716..234B}.  In the idealized case of a two-component galaxy
with a spherical DM halo and an exponential stellar disk, $\Fdisk(R=2.2h_R)$
uniquely quantifies the influence of the DM on the mass budget {\it at all
radii} for a given disk scale length, oblateness, and DM-halo density
parameterization.  This idealized case provides a useful fiducial model with
which to compare observations, as we discuss below.  Real galaxies deviate from
the idealized case due to (1) the inclusion of other baryonic components with
generally different mass distributions, such as bulges and gaseous disks, and
(2) perturbations of the stellar-disk mass-surface-density profile away from the
nominal exponential, as inferred from surface-brightness variations.  Thus, as
explicitly associated with the stellar disk, $\Fdisk(R=2.2h_R)$ has an intrinsic
distribution for maximal disks: \citet{1997ApJ...483..103S} adopted
$\Fdisk(R=2.2h_R) = 0.85\pm0.10$ as an appropriate definition for maximal disks
in galaxies of similar Hubble type to the Milky Way (Sb to Sc), also
representative of the DMS Phase-B sample.  We directly compare this definition
to our measurements in UGC 463; however, we note that the literature studies
upon which this definition was based do not remove the molecular gas component
from the total disk mass distribution as we do for UGC 463.

As roughly synonymous throughout the discussion by
\citetalias{1986RSPTA.320..447V}, the ``maximum-disk'' hypothesis could also be
termed the ``maximum-baryon'' or ``minimum-dark-matter'' hypothesis.  In this
respect, it is also useful to calculate the baryonic mass fraction, $\Fbary =
\vbary/\vc$.\footnote{
This should not be confused with the {\it total} baryon fraction, ${\mathcal
F}_{\rm bar} = \mtotbar/\mtot$, discussed in \citetalias{2010ApJ...716..234B}.
} As implemented by \citetalias{1986RSPTA.320..447V}, $\Fbary \approx \Fdisk$ at
least in the sense that their mass-to-light ratios incorporated {\it all} mass
distributed similarly to the luminous disk and they assumed the stellar disk was
by far the most massive baryonic component.  Our detailed accounting of multiple
baryonic components in UGC 463 with different mass distributions, particularly
with regard to the massive gas disk, means that (1) $\Fbary \neq \Fdisk$ and (2)
quoting $\Fdisk (2.2 h_R)$ has a more limited bearing on the relative influence
of the baryonic and DM mass on the total mass budget than described above.
Therefore, it is useful to consider the {\it radial functions} $\Fdisk(R)$ and
$\Fbary(R)$ and to define multiple fiducial radii for $\Fbary$ based on the
expectation of that each baryonic component can dominate the mass budget in
distinct radial regimes.  

One expects $\Fbary$ to decrease with radius as DM increasingly dominates the
mass budget.  The ``maximum-disk'' hypothesis effectively states $\Fbary\sim 1$
at small radius, regardless of whether or not the {\it disk} or bulge dominates the
baryonic mass \citep[see the implementation of the ``maximum-disk'' hypothesis
by, e.g.,][]{1986AJ.....91.1301K, 1987AJ.....93..816K}.  The application of
$\Fbary\sim 1$ at small radius to Sb--Sc galaxies has shown $\Fdisk (2.2 h_R) =
0.85$, which may be considered a lower limit for $\Fbary$ at these radii.
Finally, it is expected that $\Fbary\to 0$ at large radius because rotation curves
have been shown to remain nearly constant up to the radial extent of \hone\
disks.  The recent study by \citet{2011arXiv1101.1622D} has shown that this
expectation for $\Fbary(R)$ is likely a limited picture of the range in galaxy
properties.  By combining strong lensing analysis and stellar kinematics,
\citeauthor{2011arXiv1101.1622D} have shown $\Fbary(R_e) = 0.99$ and
$\Fbary(2.2h_R) = 0.67$ for the late-type lens galaxy SDSS J2141-0001, which is
at a redshift of 0.14; $R_e$ is the effective (half-light) radius of the bulge
(\sect \ref{sec:muk}).  Thus, while adhering to the fundamental tenant of the
``maximum-disk'' hypothesis, SDSS J2141-0001 exhibits a disk that is relatively
less massive than the disks of local Sb--Sc galaxies, such that it would be
considered ``submaximal'' under the definition proposed by
\citet{1997ApJ...483..103S}.

For UGC 463, we assess the ratio of the dark and baryonic mass components via
the mass fractions $\Fdisk$ and $\Fbary$ in \sect \ref{sec:fdisk}.  We quote
$\Fbary$ at $R_e = 0.5$ kpc and $2.2h_R = 7.8$ kpc as to compare with both the
expectations from the ``maximum-disk'' hypothesis and the results from
\citet{2011arXiv1101.1622D}.\footnote{
In galaxies with extended gas disks, one might also quote $\Fbary$ at a radius
where the gas disk dominates the baryonic circular speed; however, Figure
\ref{fig:rcdecomp} shows that there is no such radius for UGC 463 within the
radial regime of our measurements.
} These radii are close to the radii at which the circular speeds of the stellar
components are maximized (Figure \ref{fig:rcdecomp}).  We also quote
$\Fdisk(2.2h_R)$ for a direct comparison with the definition of a maximal disk
proposed by \citet{1997ApJ...483..103S}.  Further assessments are made using the
ratio of the mid-plane volume densities $(\rhodm/\rhobary)_{z=0}$ in \sect
\ref{sec:rhorat} and the enclosed-mass budget in \sect \ref{sec:massbudget}.  In
all of these sections, we compare our measurements for UGC 463 with an
idealized, maximal-disk galaxy; this model provides the equivalent quantities
for a maximal disk that has been embedded in the NFW and pseudo-isothermal halos
fitted to our measurements of $\vdm$ (\sect \ref{sec:rhodm}).  This fiducial
maximal disk has the scale length and oblateness as measured for UGC 463, is
purely exponential in both $R$ and $z$ and has $\Fd = 0.85$ at $R=2.2 h_R$.
Here, our notation $\Fd$, as opposed to $\Fdisk$, purposely leaves the nature of
the matter in the disk undefined, as done by \citetalias{1986RSPTA.320..447V} in
their implementation of the ``maximum-disk'' hypothesis.  The combination of
this fiducial maximal disk and our fitted DM halos produces a substantially
higher circular speed than we measure for UGC 463, and the shape of the rotation
curve does not adhere to the disk-halo conspiracy
\citepalias{1986RSPTA.320..447V}; we have shown in \sect \ref{sec:baryrc} that
application of the ``maximum-disk'' hypothesis to UGC 463 essentially results in
a marginal DM-halo mass within the radial regime probed by our data.

Before continuing, we note that \citet{1993A&A...275...16B, 1997A&A...328..517B}
used kinematic measurements similar to our own and found $\Fd(2.2h_R) =
0.63\pm0.10$ for a sample of 12 late-type galaxies; Bottema did not
differentiate between stellar and gas components in this decomposition, hence
our use of $\Fd$.  In detail, Bottema's assumptions are not exactly the same as
our own; however, we can cast UGC 463 in terms of his measurement to find
$\Fd(2.2 h_R) = 0.6$, which is compatible with these previous results.  We also
note that, using planetary nebulae as kinematic tracers,
\citet{2009ApJ...705.1686H} reported submaximal disks for four of the five
galaxies they studied.

\subsubsection{Stellar-Disk and Baryonic Mass Fractions, $\Fdisk$ and $\Fbary$}
\label{sec:fdisk}

\begin{figure}
\epsscale{1.1}
\plotone{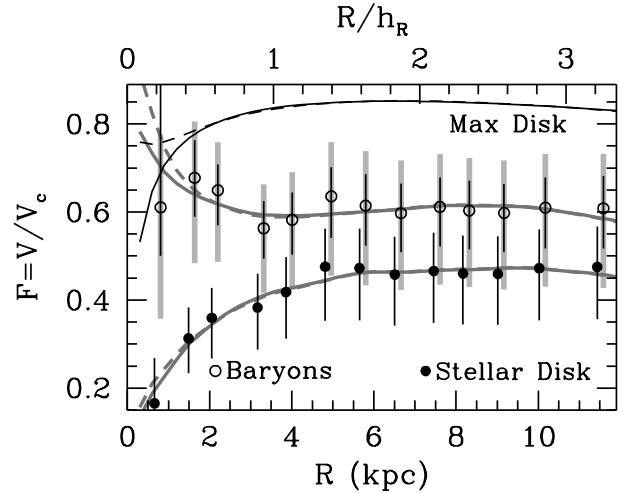}
\caption{
Mass fractions ($\mathcal{F} = V/\vc$) of the stellar disk ($\Fdisk$; {\it
filled points}) and all baryonic matter ($\Fbary$; {\it open circles}) as a
function of radius assuming a constant $\mlsk(R)$.  The discrete measurements
use the measured values of $\vc$; random errors are shown in black for both
quantities, whereas systematic errors are shown in gray only for measurements of
$\Fbary$.  Gray lines show the result of calculating $\vc^2 = \vdm^2+\vbary^2$
using the NFW ({\it solid gray lines}) and pseudo-isothermal ({\it dashed gray
lines}) DM halo models from Figure \ref{fig:rhodm}.  Black lines show $\Fd$ for
a fiducial maximal disk ($\Fd = 0.85$ at $R=2.2 h_R$) that has been embedded
in the fitted DM halos (see text).
}
\label{fig:fdisk}
\end{figure}

Figure \ref{fig:fdisk} provides $\Fdisk$ and $\Fbary$ as a function of radius
assuming a constant $\mlsk(R)$.  Individual data points use our direct
measurements of $\vc$, whereas the gray lines calculate $\vc^2 =
\vbary^2+\vdm^2$ using $\vbary$ from Figure \ref{fig:rcdecomp} and $\vdm$ from
the NFW and pseudo-isothermal halos fitted in Figure \ref{fig:rhodm}.  Both
$\Fdisk(R)$ and $\Fbary(R)$ are effectively constant between $1.0 \lesssim R/h_R
\lesssim 3.5$; the roughly constant $\Fbary-\Fdisk$ at these radii reflects the
similarity between the radial distribution of the gas and stars.  Figure
\ref{fig:fdisk} also provides the expected $\Fd$ for the fiducial maximum disk,
which is the same for the NFW and pseudo-isothermal halos at $R \gtrsim 0.5
h_R$.  The shape of $\Fdisk(R)$ is very similar to that calculated for a maximal
disk, which is essentially a statement that $\mu^{\prime}_K$ is very close to an
exponential after subtracting the central light concentration; however, the
normalization is very different.  At small radius, $\Fbary$ and $\Fdisk$ diverge
largely due to the exclusion of the central mass concentration in the
calculation of the latter.

Table \ref{tab:dmp} gives $\Fbary(R_e)$, $\Fbary(2.2h_R)$, and $\Fdisk(2.2h_R)$
for both the constant and variable $\mlsk(R)$ assumptions.  Our measurement of
$\Fbary(R_e)$ has significant uncertainty due to the error in measurement of
$\vc$ at this radius (largely due to centering errors), the uncertainty in the
baryonic rotation speed, and the difference in the extrapolation when based on
either the NFW or pseudo-isothermal halo.  To the contrary, both $\Fbary$ and
$\Fdisk$ at $2.2h_R$ are relatively well constrained.  Following the definition
proposed by \citet{1997ApJ...483..103S} and assuming a constant $\mlsk(R)$,
these quantities demonstrate that UGC 463 has a substantially submaximal disk
by a factor of $\sim(0.85/0.46)^2\approx3.4$ in mass; this reduces to 1.9 if one
instead defines $\Fbary (2.2h_R) = 0.85$ as a maximal disk.  These factors are
consistent with our previous discussion in \sects \ref{sec:mldisc} and
\ref{sec:baryrc} regarding the comparison of our dynamical measurements with the
SPS predictions from \citetalias{2003ApJS..149..289B}.  Adopting a variable
$\mlsk(R)$ results in a mass profile of the disk that yields a peak rotation
speed at or beyond the limit of our calculation ($R=15$ kpc), whereas the
constant $\mlsk(R)$ disk has a peak rotation at $2.7 h_R$.  However, in both
cases the stellar (and baryonic) disk remains submaximal due to the relatively
constant value of $\Fdisk$ (and $\Fbary$) at $R>h_R$.  Although it is possible
that the baryonic mass is close to satisfying the $\Fbary(R_e)\sim 1$ given the
uncertainty, our data suggest that the circular speed may have substantial
contributions from DM even within the bulge region.

\subsubsection{Mid-Plane Volume-Density Ratio, $(\rhodm/\rhobary)_{z=0}$}
\label{sec:rhorat}

\begin{figure}
\epsscale{1.1}
\plotone{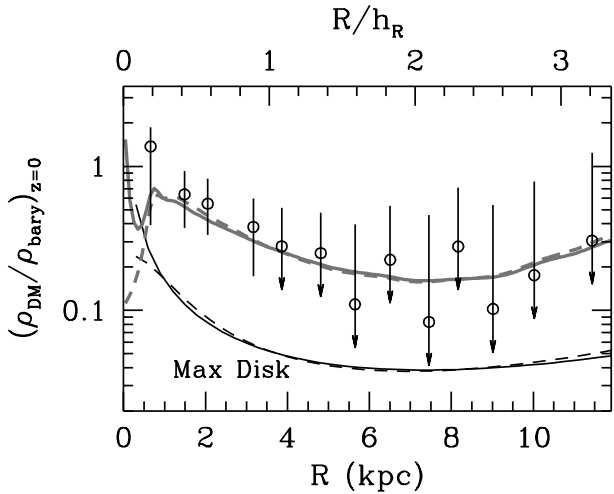}
\caption{
Mid-plane density ratio $(\rhodm/\rhobary)_{z=0}$ of all baryonic matter as a
function of radius assuming a constant $\mlsk(R)$.  The calculation of
$(\rhobary)_{z=0}$ is discussed in the text.  Data points use the $\rhodm$
measurements from Figure \ref{fig:rhodm}, whereas gray lines use the NFW ({\it
gray solid lines}) and pseudo-isothermal ({\it gray dashed lines})
parameterizations of $\rhodm$.  Only random errors are shown; upper limit arrows
are used when the error is larger than the measured value.  The results for the
fiducial maximal disk are plotted as black lines for both the NFW ({\it solid
black lines}) and pseudo-isothermal ({\it dashed black lines}) DM halo
parameterizations.
}
\label{fig:rhoh2d}
\end{figure}

Our mass-surface-density and $\rhodm$ measurements from Figures
\ref{fig:rcdecomp} and \ref{fig:rhodm}, respectively, provide the mid-plane
volume-density ratio between the dark and baryonic matter,
$(\rhodm/\rhobary)_{z=0}$.  We assume that the stratification of the gaseous and
stellar disks combine to produce an exponential vertical density distribution
with a single scale height $h_z$ and that the bulge is spherical.  Thus, the
baryonic volume density at the disk mid-plane is $(\rhobary)_{z=0} =
(\sds+\sdg)/2h_z + \rhobulge$.  If the gas is more confined to the plane, this
calculation results in upper limits.  Figure \ref{fig:rhoh2d} provides the
mid-plane volume-density ratio when assuming a constant $\mlsk(R)$, where the
discrete measurements use the non-parametric calculation of $\rhodm$ and the
gray lines use the NFW and pseudo-isothermal sphere parameterizations.  We also
plot the expectation for the fiducial maximal disk discussed above.  For
reference, \citet{2006A&A...446..933B} find $(\rhodm/\rhobary)_{z=0} = 0.14$ in
the solar neighborhood assuming a spherical halo.

At $R \gg z$, the change in $\rhodm$ with $z$ is much smaller than the change in
$\rhobary$ such that $(\rhodm/\rhobary) \propto \exp(|z|/h_z)$.  Averaging the
data at $1<R/h_R<3$ in Figure \ref{fig:rhoh2d}, we find that
$(\rhodm/\rhobary)_{z=0} \sim 0.2$, which is a factor of 5 larger than the
expectation for a maximal disk (only 40\% larger than the Milky-Way value).  The
mass volume density of UGC 463 is, therefore, dominated by dark matter at $|z|
\gtrsim 1.6 h_z$ at $R\gtrsim h_R$.  This result is particularly important for
our understanding of out-of-plane motions in the disk of UGC 463: The derivation
of $\sddisk = \sigz^2/\pi k G h_z$ assumes an isolated, plane-parallel, infinite
disk.  Deviations from these assumptions, such as embedding the disk in a very
massive DM halo, introduces systematic errors in the calculation, as briefly
discussed in \citetalias{2010ApJ...716..234B}.  The result for
$(\rhodm/\rhobary)_{z=0}$ in UGC 463 suggests that such effects may be
significant for this galaxy.

\citet{1993A&A...275...16B} has discussed the influence of a massive DM halo on
$\sigz$ in disk stars, continuing the work of \citet{1984ApJ...276..156B}.
These authors find that $\sigz$ should be {\it inflated} relative to an isolated
disk when embedded in a massive, spherical halo; the degree of the inflation is
proportional to $(\rhodm/\rhobary)_{z=0}$, as shown in Figure 15 from
\citet{1993A&A...275...16B}.  For the fiducial maximal disk shown in our Figure
\ref{fig:rhoh2d}, \citet{1993A&A...275...16B} would predict a less than 5\%
increase in the $\sigz$ over an isolated disk, whereas our measurements for UGC
463 from Figure \ref{fig:rhoh2d} suggest $\sigz$ could be increased by
$10-15$\%.  This means that our calculation of $\sddisk \propto \sigz^2$ could
{\it overestimate} the mass of the disk by $20-30$\%.

Ideally, one would calculate $\sddisk$ by first assuming an isolated disk and
then converging to a solution that incorporated the effects of the DM halo.
However, we have not done so here for UGC 463 because (1) the random error in
our isolated-disk measurements of $\sddisk$ are of the same order as this
systematic correction; (2) there is substantial error in our measurement of
$(\rhodm/\rhobary)_{z=0} \sim 0.2$, even if we adopt the parameterized solutions
for $\rhodm$; and (3) there are equally unknown competing systematics that work
in the opposite direction, such as the inclusion of a massive, razor-thin gas
disk.  The continued study of the effects of a massive halo on the {\it velocity
dispersion}, as opposed to just the rotation curve \citep{2006MNRAS.373.1117H,
2008ApJ...679.1232W}, is worthy of a dedicated effort.  However, given
ambiguities regarding the three-dimensional structure of galaxies and the
vertical stratification of disks (see discussion in \sect \ref{sec:mlsys}), a
detailed understanding of the influence of the DM halo on the disk is
complicated.  Here, we simply note that the impact of a relatively massive DM
halo in UGC 463 works to further lower the maximality of an already submaximal
disk.

\subsubsection{Enclosed-Mass Budget}
\label{sec:massbudget}

\begin{figure}
\epsscale{1.1}
\plotone{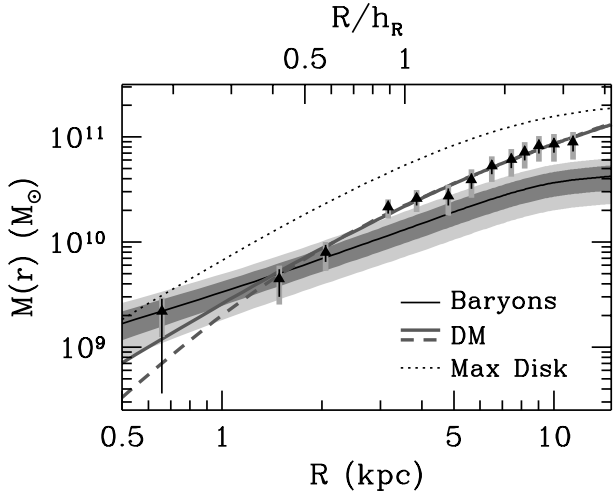}
\caption{
Mass growth curves for baryonic mass ({\it solid black line}), NFW ({\it gray
solid line}) and pseudo-isothermal ({\it gray dashed line}) DM halo
parameterizations, and the fiducial maximal disk ({\it dotted line}).  The
baryonic mass profile assumes a constant $\mlsk(R)$; the dark- and light-gray
regions represent the random and systematic errors, respectively.  Individual
measurements of $\vdm$ from Figure \ref{fig:rhodm} are converted to masses
assuming a spherical halo and plotted as black triangles; black and gray bars
represent the random and systematic errors, respectively.
}
\label{fig:mass}
\end{figure}

As discussed in \citetalias{2010ApJ...716..234B}, the {\it total} baryon
fraction ${\mathcal F}_{\rm bar} = \mtotbary/\mtot$ is ill-defined; however, the
surface density and $\vdm$ measurements from Figure \ref{fig:rcdecomp} allow for
a robust calculation of the enclosed-mass budget to a finite radius, assuming
the halo is spherical.  The resulting mass budget is presented in Table
\ref{tab:massbudget} and the mass growth curves are shown in Figure
\ref{fig:mass}.  We end the calculation at $R=15$ kpc ($R=4.2h_R$; $\sim
R_{200}/10$ for our fitted NFW halo), well within the limiting radius of our
$\mu^{\prime}_K$ measurements (Figure \ref{fig:sbprof}) but extrapolating beyond
our dynamical data.  We calculate the DM mass at 15 kpc using the fitted halo
parameterizations and the percentage errors from the measured data.  The mass
growth curve of the fiducial maximal disk discussed above is overplotted in
Figure \ref{fig:mass} for reference.  Although Table \ref{tab:massbudget} only
provides the results when assuming a constant $\mlsk(R)$, the results assuming a
variable $\mlsk(R)$ are insignificantly different.

Figure \ref{fig:mass} shows that the integrated baryonic mass of UGC 463 is, at
most, equal to the integrated DM mass at $R\sim h_R$.  Beyond this radius, the
integrated DM mass quickly begins to dominate, such that $\mb/\mdm\sim 0.3$ at
15 kpc.  To the contrary, the integrated mass of the fiducial maximal disk {\it
always} dominates over the DM mass; the minimum mass ratio $\mdisk/\mdm\sim1.4$
is at $R=15$ kpc.

\section{Summary}
\label{sec:summary}

In this paper, we have presented a detailed case study of the dynamics and
implied mass budget of the low-inclination, SABc galaxy UGC 463.  We find the
galaxy to be dominated by DM at nearly all radii, a classification usually
reserved for low-surface-brightness galaxies whereas UGC 463 is $\sim1$
magnitude above the \citet{1970ApJ...160..811F} mean central surface brightness
\citepalias{2010ApJ...716..198B}.  The stellar disk of UGC 463 is submaximal by
a factor of $\gtrsim3$ in mass.  Submaximal disks have been both indirectly
\citep[e.g.][]{1999ApJ...513..561C} or directly \citep{1993A&A...275...16B,
2009ApJ...705.1686H} measured in the past, and our results are consistent with
these measurements.  At present, this general result should be unaffected by
systematic error, despite the albeit large number of assumptions.  A summary of
our analysis follows.

Our analysis of UGC 463 draws from nearly all of our survey data for this galaxy
as collectively described in \sect \ref{sec:data}.  We derive a distance of $D =
59.67 \pm 0.01 \pm 4.15$ Mpc using the flow-corrected systemic velocity and $H_0
= 73 \pm 5$ \kms~Mpc$^{-1}$.  We derive optical and NIR surface brightness
profiles from, respectively, archival SDSS and 2MASS data; we use the latter to
calculate a total $K$-band magnitude of $m_K = 9.32\pm0.02$.  Our photometry is
corrected for Galactic extinction; internal extinction and dust emission are
marginal considerations for our results.  We also correct for instrumental
smoothing of the surface brightness profile at small radius.  Ionized-gas and
stellar kinematics are derived using SparsePak and PPak IFS.  Ionized-gas
kinematics combine results from atomic emission lines near \halp\ and the
\oiii$\lambda5007$ line; kinematics are measured using single or double Gaussian
line fits \citep[as in][]{2008ApJ...688..990A}.  Stellar kinematics are derived
from absorption lines with rest wavelengths between $\sim 492-522$nm using \dct\
\citepalias{2011ApJS..193...21W} and a single K1 III template star (HD 167042
for SparsePak and HD 162555 for PPak); template mismatch is $\lesssim5$\%
\citepalias{2010ApJ...716..234B}.  All kinematics are corrected for instrumental
broadening.  We measure \hone\ mass surface densities and the \hone\ velocity
field using 21cm aperture synthesis imaging from the VLA.  Finally, we estimate
the \molh\ mass surface density by combining 24$\mu$m {\it Spitzer} imaging and
a $\imi/\ico$ calibration based on our reanalysis of data tabulated by
\citetalias{2008AJ....136.2782L}.  Errors in our measurements of $\sdmh$
incorporate the systematic error in this calibration and the error in our
adopted value of $\xco = (2.7 \pm 0.9) \times 10^{20}$ cm$^{-2}$ (K
\kms)$^{-1}$.

In \sect \ref{sec:geom}, we determine the detailed on-sky geometric projection
of the disk of UGC 463, including extensive tests of the inclination.  These
efforts are important to subsequent analysis of the measured kinematics due to
the substantial influence of inclination errors on the fundamental calculations
in this study \citepalias{2010ApJ...716..234B}.  Inclinations are measured both
kinematically --- using an algorithm explained in \citet{2008ApJ...688..990A}
that assumes circular motion and a single, coplanar disk --- and by inverting
the TF relations from \citetalias{2001ApJ...563..694V}.  Kinematic inclinations
are most consistent between all ionized-gas and stellar tracers when adopting a
$\partial\vlos/\partial i$ weighting scheme, as explained in Appendix
\ref{app:ikin}.  We find $\ikin = 25\fdg1 \pm 2\fdg5$ and $\itf =
29\arcdeg\pm2\arcdeg$; a combined measurement of $i=27\arcdeg\pm2\arcdeg$
represents our best estimate for the inclination and is used throughout all
subsequent analysis.  We derive the dynamical center of each SparsePak
observation and, for the PPak data, we affix the dynamical center to the
morphological center determined from a reconstructed continuum map; our
two-dimensional maps in Figure \ref{fig:maps} demonstrate that the dynamical and
morphological centers of UGC 463 are identical to within the errors of our
measurement ($\sim 1\arcsec$).

Using the derived geometry, we create azimuthally averaged kinematics in \sect
\ref{sec:spk+ppk}.  To combine kinematic measurements using different
instruments and different wavelength regimes, we apply beam-smearing corrections
to both our ionized-gas and stellar kinematics; beam-smearing corrections to the
\hone\ data are described by \citet{TPKMPhD}.  The beam-smearing corrections
employ model surface-brightness, velocity, and velocity-dispersion distributions
to create a synthetic dataset that is compared to our observations; the
corrections are small (less than a few percent) except for data near the
dynamical center.  Figure \ref{fig:beamc} demonstrates that both the ionized-gas
and stellar kinematics measured separately by SparsePak and PPak are very well
matched.  We provide a cursory assessment of the kinematic axisymmetry of the
rotation curves and velocity dispersion profiles by overlaying $180\arcdeg$
azimuthally averaged kinematics for the receding and approaching sides in Figure
\ref{fig:beamc}; with respect to the errors, only moderate differences are
present.  Therefore, we impose axisymmetry by measuring the gas
(ionized$+$neutral) and stellar velocity and velocity dispersions only as a
function of in-plane galaxy radius.  We correct the gas rotation curve to the
circular speed using measurements of $\sgas$ and $\sdg$ following
\citet{2010ApJ...721..547D} to produce the axisymmetric radial profiles in
Figure \ref{fig:rcaxisym}.

Based on our collection of azimuthally averaged properties, we determine
physical properties of the disk of UGC 463 in \sect \ref{sec:disk}.  We find
that the stellar velocity dispersion profile, $\azmean{\sstar}(R)$, has only
minor deviations from a pure exponential; the fitted $e$-folding length
($h_{\sigma}\sim 2.6h_R$) suggests either $\mldyn$ or $h_z$ increases by a
factor of $\sim2.3$ over the radial range of our data, if the other quantity is
radially invariant.  Using the measured circular speed, stellar rotation curve,
and stellar velocity dispersion profile, we calculate the observable function
$\dad(R)$.  This function is directly related to the shape of the SVE such that
we find $\alpha = 0.48\pm0.09$ and $\beta = 1.04\pm0.22$, when assumed to be
constant over the entire disk.  Therefore, we find the conversion factors
$\sigz/\langle\sstar\rangle_{\gaz} = 0.76\pm0.09$ and
$\sigr/\langle\sstar\rangle_{\gaz} = 1.59\pm0.17$, which we use in our
calculations of $\sddisk$ and the disk stability, $Q$.  Our calculations of
$\sddisk$ (Figure \ref{fig:sdd}) use equation 9 from
\citetalias{2010ApJ...716..234B}, which assumes the oblateness from their
equation 1.  We find $\sddisk$ is well fit by 3$\sdg$, from which it follows
that $\sds\sim2\sdg$.  Using $\sdg$, $\sgas$, $\sds$, and $\sigr$, we calculate
individual and multi-component stability coefficients following from,
respectively, \citet{1964ApJ...139.1217T} and \citet{2001MNRAS.323..445R}.  We
find the disk to be globally stable, with the multi-component stability
asymptotically decreasing with radius to a value of $\qr\sim2$.  Based on
combining stability arguments with swing-amplification theory
\citep{1981seng.proc..111T, 1987A&A...179...23A}, disks of a fixed rotation
curve should exhibit higher spiral-arm multiplicity when the disk mass is
decreased.  This expectation is qualitatively consistent with our measurements
of a submaximal disk and three-arm multiplicity in UGC 463; however, \pV\ and
\citet{TPKMPhD} find submaximal disks regardless of spiral-arm multiplicity.

We calculate $\mlskdisk$ using $\sds$ and $\mu^{\prime}_K$ to find
$\allmean{\mlskdisk} = 0.22\pm0.09^{+0.16}_{-0.15}\ \msol/\lksol$ at $R>2$ kpc.
Our measurements of $\allmean{\mlskdisk}$ are systematically lower than SPS
model predictions ($\mlsksps$) from \citetalias{2003ApJS..149..289B} by a factor
of $\sim 3.6$ and larger than the SPS modeling of
\citetalias{2009MNRAS.400.1181Z} by a factor of $\sim 1.8$.  Based on an MC
sampling of probability distributions assigned to each quantity in the
calculation, we generate a composite (random$+$systematic error) probability
distribution for $\mlskdisk(R)$ and $\allmean{\mlskdisk}$.  We find that the
\citetalias{2009MNRAS.400.1181Z} prediction are within our 68\% confidence
interval for $\allmean{\mlskdisk}$; in contrast, measurements of
$\allmean{\mlskdisk}$ consistent with the \citetalias{2003ApJS..149..289B}
prediction occur for less than 1 in $10^5$ MC samples.
\citetalias{2009MNRAS.400.1181Z} attribute the disparity between their
$\mlsksps$ predictions and those from \citetalias{2003ApJS..149..289B} to
different treatments of the star-formation history and TP-AGB phases of stellar
evolution.  We also find a factor of $\sim2$ increase in $\mlskdisk$ with
radius, which is not predicted by the SPS modeling.  This feature may reflect a
true increase in the $\mlskdisk$, a flaring of the stellar disk, or a change in
the relative dynamical influence of the halo, thick stellar disk, and/or
razor-thin gas disk.

We discuss the mass budget of UGC 463 out to 15 kpc (4.2 $h_R$) in \sect
\ref{sec:mass} using a traditional rotation-curve mass decomposition, which
benefits from our {\it unique} and direct measurement of $\mlskdisk$.  Our
primary discussion assumes a constant $\mlsk(R) = \allmean{\mlskdisk}$; however,
we also briefly discuss results obtained by assuming a variable $\mlsk(R) =
\mlskdisk(R)$.  Our mass decomposition also assumes that $h_z$ is constant with
radius and that the galaxy is composed of four separable potentials (halo,
stellar bulge, stellar disk, and gas disk).  The total mass budget obtained by
assuming a constant $\mlsk(R)$ is provided in Table \ref{tab:massbudget};
assuming a variable $\mlsk(R)$ produces results that are statistically
identical.

We calculate the circular speed of each baryonic component in \sect
\ref{sec:baryrc}.  We find that a maximal disk may be produced by adopting the
$\mlsksps$ prediction from \citetalias{2003ApJS..149..289B}, which amounts to
increasing the stellar (baryonic) mass by a factor of 3.6 (2.7) above our
measurements of $\allmean{\mlskdisk}$.  However, this result is effectively
excluded by our simultaneous measurements of $\vc$ and $\azmean{\sstar}$.  Using
our dynamical measurements, we produce $\vdm^2 = \vc^2 - \vbary^2$ and use these
measurements to calculate $\rhodm$; both $\vdm$ and $\rhodm$ assume a spherical
halo.  In \sect \ref{sec:rhodm}, we fit $\vdm$ with an NFW and pseudo-isothermal
DM halo, and we find both to be statistically suitable descriptions of our
measurements.  The concentration of the NFW halo is consistent with expectations
from DM-only simulations, implying that the halo structure has been relatively
unaffected by the collapsed baryons.  Our measurements of $\rhodm$ show a slope
that may be steeper than both $\rhodm \propto 1/R$ and the slope predicted by
either DM-halo parameterization; however, this result is highly dependent on the
error-prone assessment of the baryonic mass within the central kpc.  Results for
additional galaxies in our Phase-B sample are required to place better
statistical constraints on the shape of DM halos of local disk galaxies.

We discuss the dominant gravitational influence of DM over baryonic matter in
UGC 463 in \sect \ref{sec:dvb}, as summarized in Tables \ref{tab:massbudget} and
\ref{tab:dmp}.  Assuming a constant $\mlsk(R)$, we find the baryonic disk to be
substantially submaximal with $\Fbary(2.2h_R) = 0.61^{+0.07}_{-0.09}\
^{+0.12}_{-0.18}$.  Considering only the stellar disk and adopting $\Fdisk
(2.2h_R) = 0.85\pm0.10$ as the definition of a maximal disk
\citep{1997ApJ...483..103S}, UGC 463 is submaximal by a factor of
$\sim(0.85/0.46)^2=3.4$ in mass, consistent with our expectation based on the
difference between $\allmean{\mlskdisk}$ and the $\mlsksps$ prediction from
\citetalias{2003ApJS..149..289B} (\sect \ref{sec:baryrc}).  We also compare our
measurements of the baryonic component to a fiducial maximal disk (having $\Fd
(2.2 h_R) = 0.85$) resulting from embedding a purely exponential disk (with
$h_R$ and $h_z$ as measured for UGC 463) in our fitted NFW and pseudo-isothermal
DM halos.  In the disk mid-plane, we find the ratio $(\rhodm/\rhobary)_{z=0}$ is
a factor of five larger than expected by our fiducial maximal disk, which may
lead to an overestimate of $\sddisk$.  Additionally, we find that the
enclosed-mass of the galaxy is dominated by DM at $R\gtrsim h_R$, whereas the
fiducial maximal disk dominates the enclosed-mass budget at all radii sampled by
our observations.  Finally, assuming a constant $\mlsk(R)$, we find a
baryonic-to-DM mass ratio of $0.31\pm0.8^{+0.16}_{-0.14}$ for the mass enclosed
within the central 15 kpc.

Although our results are for a single galaxy, an analysis of 30 galaxies in our
sample demonstrate that {\it all} of these disks are comparably submaximal
\citep[\pV;][]{TPKMPhD}.  One can increase the maximality of these disks by
changing the assumptions concerning the vertical mass distribution (quantified
by the constant $k$) or the applied oblateness ($q$), as discussed in \sects
\ref{sec:sdd} and \ref{sec:mlsys}; however, $k$ and $q$ would have to take on
values that are effectively excluded by empirical constraints on the structural
parameters of disk galaxies based on edge-on systems \citep[e.g.,][]{KregelPhD}.
The recalibration of $\mls$ based on these results has significant consequences
for, e.g., our understanding of the baryonic mass of galaxies as a function of
redshift and for the gravitational interplay between baryonic and dark matter in
the process of galaxy formation.

\acknowledgments Support for this work was provided by the National Science
Foundation (NSF) via grants AST-0307417 and AST-0607516 (M.A.B., K.B.W., and
A.S.-R.), OISE-0754437 (K.B.W.), and AST-1009491 (M.A.B.\ and A.S.-R.).  K.B.W.\
is also supported by grant 614.000.807 from the Netherlands Organisation for
Scientiﬁc Research (NWO).  M.A.W.V.\ and T.P.K.M.\ acknowledge financial support
provided by NOVA, the Netherlands Research School for Astronomy, and travel
support from the Leids Kerkhoven-Bosscha Fonds. This work is based in part on
observations made with the {\it Spitzer} Space Telescope, which is operated by
the Jet Propulsion Laboratory, California Institute of Technology under a
contract with NASA.  R.A.S.\ and M.A.B.\ acknowledge support from NASA/{\it
Spitzer} grant GO-30894.  This work has made use of the SIMBAD,\footnote{
\url{http://simbad.u-strasbg.fr/simbad/}
} VizieR,\footnote{
\url{http://vizier.u-strasbg.fr/viz-bin/VizieR}
} NED,\footnote{
\url{http://nedwww.ipac.caltech.edu/}
} SDSS,\footnote{
\url{http://www.sdss.org/collaboration/credits.html}
} and 2MASS\footnote{
\url{http://www.ipac.caltech.edu/2mass/releases/allsky/faq.html\#reference}
} databases and data archives.




\appendix

\section{Surface-Brightness Calibration and Interpolation of IFU Data}
\label{app:maps}

The spectral-continuum surface-brightness maps in the \halp\ and \mgi\ regions
shown in Figure \ref{fig:maps} are calibrated against SDSS imaging data,
assuming that the dynamical center (Table \ref{tab:kcen}) is the same as the
morphological center.

The ``model'' flux in fiber $f$ measured from the CCD data is
\begin{equation}
C_{{\rm mod},f} = {\rm dlog}\left[ 0.4(Z_{\rm fib} - Z_{\rm CCD}) + \log I_{{\rm
CCD},f}\right] A_{\rm fib} - S_{\rm fib},
\label{eq:ccdcounts}
\end{equation}
where $Z$ represents a magnitude zero-point and $I_{{\rm CCD},f} = (C_{{\rm
CCD},f}-S_{\rm CCD})/A_{\rm CCD}$ represents the surface brightness in units of
DN arcsec$^{-2}$ --- determined by the total flux ($C$) within an aperture (of
area $A$) with a sky background ($S$).  The fiber aperture area, $A_{\rm CCD}
\sim A_{\rm fib} \equiv \pi D_{\rm fib}^2/4$, is known and the quantities
related to the CCD image are measured or provided by the SDSS calibration.  The
fiber-continuum zero-point, $Z_{\rm fib}$, and sky-level, $S_{\rm fib}$, are
free parameters; $S_{\rm fib}$ is fiber independent, adjusting the nominal
correction based on the average sky spectrum.  Equation \ref{eq:ccdcounts} is
fit to our IFS data by minimizing $(\sqrt{C_{{\rm fib},f}} - \sqrt{C_{{\rm
mod},f}})^2$, where $C_{{\rm fib},f}$ is the mean flux across the full spectral
range; data with erroneous negative flux are ignored.  We also limit the radial
region considered to avoid inflated errors were the sky subtraction of the IFS
is particularly problematic due to variations in the sky flux as measured by the
dedicated sky fibers.

We use the SDSS $g$-band ($1\farcs5$ seeing) and $r$-band ($1\farcs2$ seeing)
data to calibrate the \mgi-region and \halp-region IFS, respectively.  For the
PPak data, analysis of guide-camera images taken throughout each exposure show
that the average seeing was $1\farcs7$ \citep{TPKMPhD}; therefore, we match this
seeing by applying a Gaussian kernel with a FWHM of $0\farcs8$ to the $g$-band
image when fitting to the PPak data.  No such seeing measurements are available
for the SparsePak data.  Although we have allowed the seeing to be a fitted
parameter for SparsePak, seeing measurements were non-convergent in the sense
that there appears to be no substantial difference with the inherent seeing of
the SDSS images.  This is not surprising given the image quality quartiles at
WIYN and the large SparsePak fibers.  

\begin{figure*}
\centering
\leavevmode
\includegraphics[angle=-90,scale=0.7]{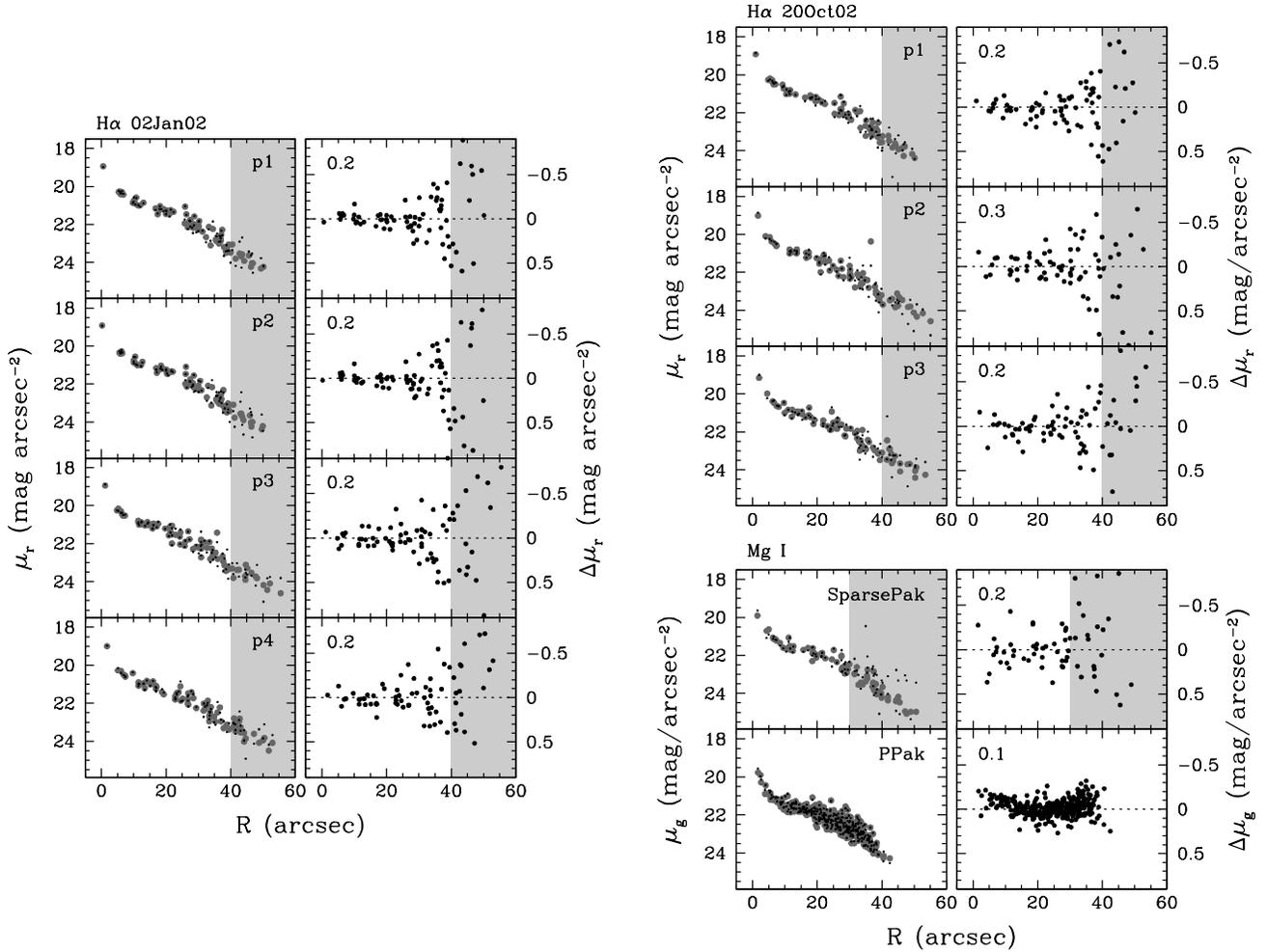}
\caption{
Photometric calibration of the IFS fiber continuum flux.  Three panel groups are
shown: the results for the four \halp\ 02Jan02 SparsePak pointings, the results
for the three \halp\ 20Oct02 pointings, and the results for the two \mgi\
pointings --- one from SparsePak and the other from PPak.  Each group has two
columns, overlaying the calibrated fiber fluxes ({\it black}) on the measured
aperture photometry from the SDSS images ({\it gray}) to the left and plotting
the residual, $\Delta\mu = \mu_{\rm fib} - \mu_{\rm CCD}$, to the right.  Radial
regions not considered during the calibration process are shaded gray.  The
pointing number or instrument is provide in the upper-right corner of the
left-column panels, and the RMS of the residuals are displayed in the upper-left
corner of the right-column panels.
}
\label{fig:fibphot}
\end{figure*}

The final calibration results are provided in Figure \ref{fig:fibphot} for all
data.  We mark each panel by the pointing number and include the residual RMS
within the fitting region.  The RMS values are typically 0.2 magnitudes, and the
PPak data have the smallest residual at 0.1 magnitudes.  Fits using no
additional seeing for the SparsePak data demonstrate good agreement with the
direct-imaging data at small radii, implying that the systematic errors due to
an inappropriate seeing match between the SDSS images and the SparsePak IFS are
inconsequential.  The results shown in Figure \ref{fig:fibphot} are used to
produce calibrated fluxes that are interpolated and converted to surface
brightness for Figure \ref{fig:maps}.

The interpolation of the continuum fluxes and kinematics from our IFS is
performed to fill interstitial fiber regions according to the following
algorithm:  Each kinematic measurement contributes to every pixel in the image,
weighted by a two-dimensional Gaussian function centered on the fiber aperture
with a FWHM equal to a multiple of the effective fiber diameter, $D_{\rm fib}$,
and inversely weighted by the measurement error; therefore, each image pixel is
weighted both by the quality and proximity of the surrounding fiber
measurements.  Mathematically, the weight of each fiber $f$ at pixel coordinate
($j$,$k$) is, therefore,
\begin{equation}
w_f(j,k) = \frac{1}{\epsilon_f^2}
\exp\left[\frac{-(x_j-x_f)^2-(y_k-y_f)^2}{(nD_{\rm fib}/2)^2 /\ln 2}\right],
\label{eq:wmom}
\end{equation}
where $\epsilon_f$ is the measurement error, ($x_j,y_k$) are the on-sky
coordinates of pixel ($j$,$k$), the fiber center has on-sky coordinates
($x_f,y_f$), and $n D_{\rm fib}$ is the FWHM of the circular Gaussian in
multiples of the fiber diameter.  Calibrated continuum fluxes include no
additional error weighting ($\epsilon_f$ is constant for all $f$). The
interpolated value at each pixel is then the weighted average, over all fibers,
of the continuum value or kinematic measurement in question.  For presentation
purposes, we limit the interpolation to only those regions with a
``significant'' contribution to the interpolated map.  For Figure
\ref{fig:maps}, we adopt the following representation of ``significant:''
\begin{equation}
\sum_{f=1}^{N_f} w_f(j,k) \geq \frac{w_{\rm min}}{N_x N_y}\sum_{j=0}^{N_x-1}\
\sum_{k=0}^{N_y-1}\ \sum_{f=1}^{N_f} w_f(j,k),
\label{eq:interplim}
\end{equation}
where $N_x$ and $N_y$ are the pixel dimensions of the image.  That is, the
summed weight of all fibers to a given pixel must be greater than $w_{\rm min}$
times the mean of all weights across the entire image.  This scheme is not ideal
given the dependence between the inclusion of an interpolated value and the
arbitrary size of the image; however, it can provide reasonable results as
demonstrated by Figure \ref{fig:maps}.  Values of $n$ and $w_{\rm min}$ for each
of the 5 interpolated maps in Figure \ref{fig:maps} are provided in Table
\ref{tab:mpar}.

\begin{deluxetable}{ c c c }
\tabletypesize{\small}
\tablewidth{0pt}
\tablecaption{Interpolation Parameters for Figure \ref{fig:maps}}
\tablehead{ \colhead{Quantity} & \colhead{$n$} & \colhead{$w_{\rm min}$}}
\startdata
$\mu_{{\rm H}\alpha}$ & 1.4 & 0.5 \\
$V_{{\rm H}\alpha}$   & 1.4 & 0.1 \\
$\mumg$   & 1.2 & 0.5 \\
$\vstar$              & 1.8 & 0.1 \\
$\sigma_{\ast}$       & 1.8 & 0.1
\enddata
\label{tab:mpar}
\end{deluxetable}

\section{Optimal Weighting Scheme for Kinematic Inclination Measurements}
\label{app:ikin}

In velocity-field modeling, parameter robustness and covariance can be,
respectively, improved and mitigated by introducing data-weighting schemes.  For
example, \citet{1989A&A...223...47B} introduced a cosine weighting scheme,
effectively weighting each datum by the derivative of the model LOS velocity
with respect to the projected rotation velocity, $w =
\partial\vlos/\partial\vrotproj = \cos\gaz$, thereby limiting the covariance
between $\vrot$ and $i$.  \citet{2003ApJ...599L..79A} mitigate this same
covariance by instead fitting the projected rotation curve directly.  Here, we
consider an optimal weighting scheme for measuring the kinematic inclination of
UGC 463 according to the approach described in \sect \ref{sec:ikin}.

\citet{2003ApJ...599L..79A} parameterized $\vrotproj(R)$ by a hyperbolic tangent
function and adopted a velocity-error weighting scheme, producing a face-on TF
relation that is well matched to samples of more inclined systems.  Their
error-weighting scheme combines, in quadrature, the measured velocity error with
a ``beam-smearing error'' and a ``stochastic error.''  The beam-smearing error
is based on a fiber-by-fiber measurement of the variance in $\vlos$ within the
fiber aperture, thereby reducing the effect of patchy emission on the fit.  The
stochastic error is a single error assessed for every velocity measurement that
reduces the influence of small-scale, incoherent non-circular motions on the
fit.

Here, we are primarily concerned with fitting inclination such that we test the
success/failure of a given weighting function, as applied to UGC 463, via the
correspondence/disparity of the inclinations determined from each of three
tracers: (1) \halp\ from SparsePak, $\iha$; (2) \oiii\ from PPak, $\iot$; and
(3) stars from PPak, $\ist$.  We apply four weighting schemes:  In addition to
the error-based and $\cos\gaz$ schemes described above, we include uniform
weighting and a weighting scheme defined by $w = \partial\vlos/\partial i$.  The
latter scheme affords those data with greater leverage on the fitted inclination
a greater influence on the goodness-of-fit statistic.

We omit data from consideration in our goodness-of-fit statistic in two steps.
First, we omit all data with velocity errors that are greater than 15 \kms,
eliminating 0.5\%, 0\%, and 1.0\% of the SparsePak \halp, PPak \oiii, and PPak
stellar data, respectively.  We note here that, after applying this omission,
the mean velocity measurement errors are 1, 3, and 6 \kms\ for the SparsePak
\halp, PPak \oiii, and PPak stellar data, respectively.  Second, we omit highly
discrepant velocities by first fitting the data using the error weighting scheme
and omitting data at high $\chi^2$.  Data are iteratively omitted while
adjusting the model and the stochastic error until the error-weighted
distribution of the data about the model follows a nominal Gaussian
\citep{KBWPhD}; in practice, no points are omitted with $\chi^2 = (V -
V_m)^2/\epsilon(V)^2 < 10$.  This omission stategy does not bias our results
toward, e.g., the initial guess parameters of the fit; instead, it serves to
eliminate a $\chi^2$-optimization bias driven by a few, highly discrepant
measurements.  Applying this procedure to our UGC 463 data eliminates an
additional 11.6\%, 14.0\%, and 5.2\% of the SparsePak \halp, PPak \oiii, and
PPak stellar data, respectively.  As described in \sect \ref{sec:ikin}, all
velocity-field parameters are simultaneously fit to the remaining data with the
dynamical center of the PPak data fixed to the morphological center.  Identical
data sets are fit by each weighting scheme for each tracer, and we use a set of
500 bootstrap simulations \citep[see \sect 15.6.2 of][]{NR3} to determine the
probability distribution for each fitted parameter.  The error-weighted standard
deviations of the velocities about the best-fitting models are typically 5, 6,
and 7 \kms\ for the SparsePak \halp, PPak \oiii, and PPak stellar data,
respectively, with only small ($5-10$\%) variations among results reached using
the different weighting schemes.

\begin{figure*}
\epsscale{1.1}
\plotone{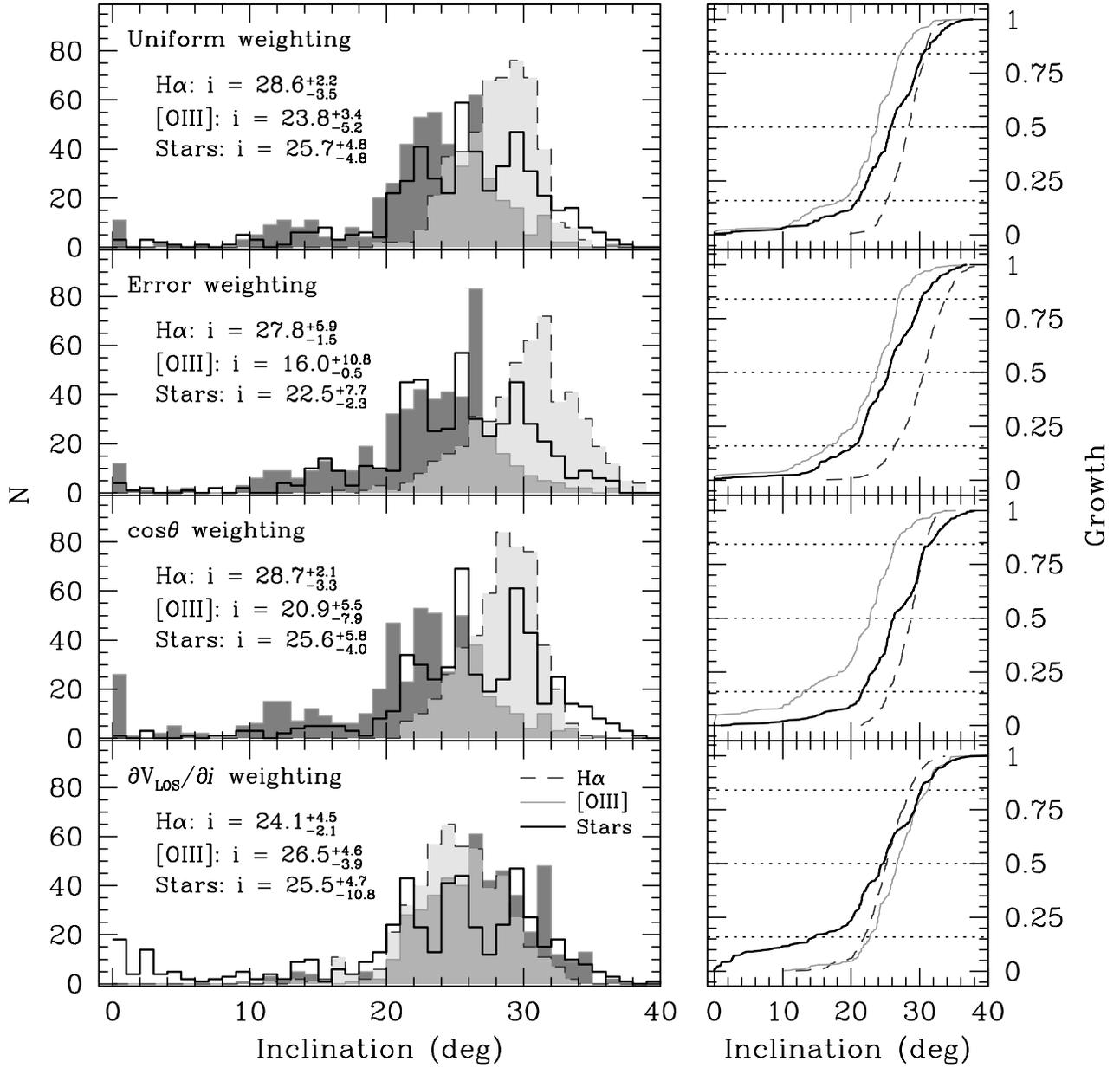}
\caption{
Inclination probability distributions determined from 500 bootstrap simulations
for each tracer using four different weighting schemes. {\it Left} --- Histogram
of the returned inclination values.  The fitted inclination for each tracer is
tabulated in the upper-left corner of each panel.  The line key for each
histogram is shown in the lower-left panel.  The SparsePak \halp\ histogram is
shaded in light (transparent) gray and the PPak \oiii\ histogram is shaded in
dark gray.  {\it Right} --- Growth curves of each histogram from the left
panels.  The line types are repeated from the left column.  The dotted lines
mark the median (0.5 growth) and the 68\% confidence interval ($0.16-0.84$
growth).
}
\label{fig:idist}
\end{figure*}

Figure \ref{fig:idist} provides the bootstrap-based probability distributions
and growth curves for the inclinations measured by each weighting scheme and
each tracer; the best-fit inclination and the 68\% confidence limits are
tabulated in the Figure.  For UGC 463, we find that weighting by
$\partial\vlos/\partial i$ produces inclination distributions that are the most
similar between the three kinematic tracers.  This is, therefore, the weighting
scheme we have adopted in \sect \ref{sec:ikin} to measure $\ikin$.  For all
results except those based on the error-weighting scheme, the error-weighted
mean of $\iha$, $\iot$, and $\ist$ is compatible with our final adopted
inclination of $i=27\arcdeg\pm2\arcdeg$ to better than the errors.

\end{document}